\documentclass[12pt]{article}
\usepackage{amsmath,amssymb}
\usepackage{arydshln} 
\usepackage[utf8]{inputenc}
\usepackage{natbib}
\usepackage{graphicx}
\usepackage{wasysym}
\usepackage{cool}
\usepackage{subcaption}
\usepackage{float}
\usepackage{adjustbox}
\usepackage{xcolor}
\usepackage{longtable}
\usepackage{booktabs}
\usepackage{comment}
\usepackage{ulem}
\usepackage[backref=page, colorlinks=true, linkcolor=blue, citecolor=blue, urlcolor=blue, filecolor=blue, pdfborder={0 0 1}]{hyperref}
\usepackage{multirow}
\usepackage{setspace}
\usepackage{threeparttable}
\usepackage{rotating}
\usepackage{pdflscape}
\usepackage{textcomp}
\usepackage{caption}
\usepackage{dsfont}
\usepackage{makecell}
\usepackage[margin=1in, left=1in, right=1in]{geometry} 
\emergencystretch=\maxdimen
\title{\singlespacing \vspace{-28mm} The Impact of Publicly Funded Small Business Advisory Services: Firm Take-up and Performance in the United States\thanks{We thank Eliane Barker, Thomas Ginn, Darrell Glaser, Dede Long, Doug Vanderwerken, and participants in the Naval Academy Economics Department Faculty Brown Bag Series, Liberal Arts Colleges Public and Labor Economics (LAC-PaL) Virtual Seminar Series and Annual Conference, the Society for Government Economists (SGE) Annual Conference, and the Small Business Administration Evidence and Evaluation Community of Practice workshop series for insightful comments. 
		We thank executives and employees apart of the Northern California Small Business Development Center (NorCal SBDC) Network Lead Center for access to administrative data and important information about the network's operations. 
		The NorCal SBDC Network is funded in part through a cooperative agreement with the US Small Business Administration (SBA) and through a grant with the Governor’s Office of Business and Economic Development. All opinions, conclusions, or recommendations expressed are those of the authors and do not necessarily reflect the view of the SBA, California Office of the Small Business Advocate or Cal Poly Humboldt Sponsored Programs Foundation. Special thanks to Jim Chrisman for providing supplementary budget and cost data for the NorCal SBDC Network. The views expressed in this paper are those of the authors and do not reflect the official policy position of the Department of Defense or the U.S. Government.
}}

\author{Scott Kaplan\thanks{Corresponding author. Assistant Professor, Department of Economics, United States Naval Academy. 121 Blake Rd, Annapolis, MD 21402. +1(410)293-6971. \textcolor{blue}{skaplan@usna.edu}.} \and Ryan Raimondi\thanks{Second Lieutenant, U.S. Marine Corps. \textcolor{blue}{ryan.raimondi@icloud.com}.} 
}
\date{July 2026
} 

\begin{document}

	\maketitle
	\vspace{-9mm}
	
	\begin{abstract}
		\singlespacing
		\vspace{-3mm}
		\noindent This paper studies the impact of geographic proximity to and utilization of publicly funded advisory services offered to US small businesses on firm take-up and performance. We leverage a novel administrative dataset from the Northern California Small Business Development Center (SBDC) Network covering all firm-center interactions from 2006–23. To address endogeneity in firm engagement with centers, we exploit exogenous variation in center-firm geographic proximity generated by center closures and openings. We instrument for paired center-firm consulting time with changes in distance resulting from these organizational shifts. A one standard deviation reduction in distance between a firm and corresponding center (20 miles) increases expected annual consulting time by 0.15 hours (7.5\%); each additional consulting hour raises average firm annual revenue and employment by 3.6--5.2\% and 1.6--2.9\%, respectively. Back-of-the-envelope calculations suggest advisory services are cost-effective. This study provides novel causal evidence on take-up and effectiveness of small business advisory services in the US using quasi-experimental variation in geographic proximity. Our findings highlight the importance of both physical distance and localized expertise in shaping small business outcomes.
	\end{abstract}
	
	\singlespacing
	
	\doublespacing
	
	\clearpage
	
	\section{Introduction}
	
	Economists have shown that firm productivity and economic development are tightly linked; countries with more productive firms tend to have higher GDP per capita and experience better economic outcomes (\citealt{bloom2010firms}). 
	A key factor behind differences in firm productivity has been shown to be good business management practices (\citealt{bloom2007measuring}). Business management practices consist of a wide variety of activities, including keeping records, budgeting and financial planning, and building professional networks. Studies in low- and middle-income countries have shown that firms---particularly sole proprietorships and those with only a few employees---using better business practices may experience higher profitability and growth (\citealt{bloom2013does}; \citealt{mckenzie2017business}), increased access to financing opportunities (\citealt{gonzalezuribe:leatherbee2017}), and higher sales and employment (\citealt{bruhn2018impact}; \citealt{iacovone2022improving}).
	
	
	Despite extensive work examining the provision of business training and advisory services in developing economies, there has been relatively little research in economics studying both take-up of business training opportunities as well as their impact on firm performance in wealthier countries. Some of this gap may be explained by higher baseline utilization rates of basic business practices among firms in more developed economies, which may limit expected benefits associated with provision of these services. At the same time, nearly every developed country offers partially or fully publicly funded business advisory services, which are intended to help businesses employ strong management practices (\citealt{oecd2021}). Therefore, assessing the extent to which these advisory services are utilized, their impact on firm outcomes, and whether they are cost-effective is important for economists and policy-makers to understand.
	
	To address this gap, this study identifies the impact of publicly funded business advisory services on important performance outcomes of participating small businesses, and the extent to which such investments are cost-effective. Because the empirical strategy leverages geographic shifts in where advisory services are offered, we also estimate the effect of spatial proximity to advisors on firm take-up. We examine a network of primarily publicly funded small business advisory centers present across the US, known as America's Small Business Development Center (SBDC) Network. In the mid-1970s, The Small Business Administration (SBA) 
	established this national network of centers to provide management and technical assistance to entrepreneurs and small business owners.\footnote{The SBA offers an \href{https://www.sba.gov/local-assistance/resource-partners/small-business-development-centers-sbdc}{online portal} providing information about the SBDC network for small businesses.} 
	The network has expanded significantly; SBDCs are now located in all 50 states and provide assistance to over one million clients across the US per year 
	(\citealt{sbdc:history2024b}). 
	
	The specific empirical setting for this research consists of all SBDCs and client firms in the Northern California (NorCal) regional SBDC network. Established in 2006, the NorCal network now serves 36 counties, spanning from the California-Oregon border to as far south as Santa Cruz (80 miles south of San Francisco)
	. As of 2024, the NorCal SBDC network employed over 500 individuals across 19 active centers, having served approximately 90,000 unique clients (businesses) over the last two decades. 
	
	We leverage a novel administrative dataset from 2006-2023 encompassing all firms to ever work with at least one NorCal SBDC during this period. The data contains time-invariant information about each business, including location of operation, associated industry, and establishment date. We also observe a complete log of each meeting between a firm and corresponding center, including contact time (hours spent in each consulting session between a center advisor and firm), preparation time (hours spent by a center advisor preparing for each consulting session), and the specific center offering services to each business, allowing us to study the evolution of center resources provided to each business over time. The data also includes timestamped annual revenue and employment values for each firm, although this information is collected less frequently than firm-center advisory interactions. 
	
	To measure the impact of public advisory services on firm revenue and employment outcomes, we use an instrumental variables strategy
	. Firm engagement with SBDCs is endogenous; for instance, more motivated firms may work more closely with SBDC advisors, and certain centers may have more available resources to support firms. To address this, we leverage plausibly exogenous variation in the geographic proximity between firms and SBDCs. Over the nearly 20 year time-frame the NorCal SBDC network has been in operation, there has been a number of center closures, openings, and re-locations, regularly adjusting the spatial distribution of the network. The constructed instrument is the change in distance between a treated firm and paired center induced by such changes; treated firms constitute those that experience at least one change in distance with a paired center during their tenure with the network.\footnote{In the event of a center closure, previous business clients of that center may be paired with an existing or newly-opened center. In the event of a center opening, previous business clients may move from an existing or recently-closed center.} Section \ref{Data} provides a detailed overview of center movement and other relevant context regarding the construction and validity of the instrument.

	The empirical analysis yields two primary findings. First, a one standard deviation reduction in the distance between a firm and corresponding center (20 miles) increased expected consulting time a firm receives by 0.15 hours per year, conditional on existing participation with the SBDC network. For context, the median firm in the data engages in 2 hours of consulting time annually. 
	Additionally, we find evidence that the importance of geographic proximity to firms varies meaningfully between rural and urban areas. A one standard deviation reduction in firm-center distance increases expected annual consulting time by 0.22 hours for firms in rural areas, compared with 0.09 hours for firms in urban locations. This pattern is consistent with institutional context: NorCal SBDC administrators report that a significantly larger share of advisory meetings for firms located in rural areas occurs in person, amplifying the importance of geographic proximity.
	
	The second set of findings identifies the impact of increased utilization of advisory services. We show that average annual revenue increases by approximately 3.6--5.2\% while average employment increases approximately 1.6--2.9\% for every 1 hour 
	increase in annual consulting time between centers and firms. The median firm in the data has an annual revenue (employment) of \$18,200 (1.5 employees), thus one additional consulting hour (corresponding to a 50\% increase) generates approximately \$655--\$946 in expected additional annual revenue (in 2023 USD) and 0.024--0.043 expected additional hires, respectively. The results suggest these public business advisory services are cost-effective. In 2023, nearly 60,000 advisor-hours were spent on NorCal SBDC clients at an average cost of \$230.60, only half of which is paid by federal tax dollars due to the SBDC matched funding model. In fact, this study may underestimate the economic impacts of public business advisory services, since it does not consider the external benefits SBDC clients may provide to other businesses, particularly within the local region.
	
	One limitation we face in the analysis relates to the frequency with which we observe firm annual revenue and employment. Because these outcomes are not recorded at every firm-center advisory session, we have limited within-firm variation available in the second-stage estimation. To address this, we aggregate and average firm annual revenue and employment at the industry-county-year level. Industry captures important commonalities across firms, while the county-year dimension preserves key identifying variation from the first stage, as changes in distance arise from local spatial adjustments over time. This approach mirrors existing studies that leverage richer, micro-level first-stage variation and aggregate predicted values to match coarser outcome data in the second stage (\citealt{frankel1999does}; \citealt{keller2013link}; \citealt{allcott2014gasoline}). Our findings are robust to alternative aggregation schemes and specifications, which inform the range of estimates in the previous paragraph.
	
	
	This research makes several important contributions. First, it adds to limited causal evidence assessing impacts of publicly funded business advisory services in developed economies. 
	\cite{fairlie2015behind} is the most closely related study in the U.S. context, leveraging a large-scale randomized program titled ``Project GATE" conducted by the US Department of Labor and SBA to evaluate the effects of offering free entrepreneurship training to interested individuals. While the paper documents substantial take-up of training and short-run impacts on business ownership and employment, it finds little evidence of long-run performance gains. In contrast, we exploit quasi-experimental variation arising from geographic changes in the SBDC advisory network to identify both take-up and performance effects. This approach allows us to estimate marginal changes in advisory intensity as a result of geographic proximity---rather than more general group-based comparisons via randomized assignment---and link these changes directly to firm outcomes. Beyond \cite{fairlie2015behind}, there exists a descriptive literature exploring the role of publicly-funded business development resources on entrepreneurship and firm outcomes in developed countries, with some studies finding significant growth in employment, sales, and research productivity (\citealt{siegel2003assessing}; \citealt{hart2003modelling}; \citealt{cumming2012publicly}), while others find no effect (\citealt{lambrecht2005evaluation}; \citealt{roper2005small}). 
	
	Second, we contribute to a broader literature examining business training and advisory interventions in developing economies (\cite{mckenzie2014we} and \cite{mckenzie2021small} offer extensive reviews of this literature). 
	While this work provides important insights, it largely focuses on experimental interventions in settings with low baseline adoption of managerial practices and varying take-up. Our study complements this work by examining an established advisory system in a developed economy that differs substantially in both institutional structure and underlying firm characteristics, and by leveraging continuous, geographically-driven variation in engagement. 

	Finally, we contribute to the literature assessing take-up of public services in the US, particularly as it pertains to geographic proximity. \cite{rossin2013wic} examine the opening and closing of Women, Infants, and Children (WIC) nutrition program centers, finding that having a center located in a mother's zip code increases her likelihood of food receipt by 6 percentage points. \cite{deshpande2019screened} study the effect of Social Security Administration office closures on disability applications, finding they cause a 16\% decline in the number of disability recipients in surrounding areas due to increased travel times and heightened congestion at neighboring offices. \cite{bitler2025intersection} finds that Supplemental Nutrition Assistance Program (SNAP) office closures led to a decrease in participation in surrounding census tracts in the two years following closure, while openings facilitated increased local participation. Our study uses similar identifying variation to assess changes in take-up of public business advisory services.

	The rest of the paper proceeds as follows. First, we provide background on the SBDC network and pertinent institutional context. Next, we describe the data and offer relevant descriptive statistics. We then detail the empirical strategy to evaluate firm outcomes and discuss the identifying assumptions necessary to interpret the findings. We present the results of the analyses, followed by a discussion and key policy implications. The final section provides concluding remarks and suggests opportunities for future research.

	\section{Background and Institutional Context}\label{SBDC_Background}
	
	\subsection{History of SBDCs}
	SBDCs provide assistance to over one million clients across the US each year, including new entrepreneurs and existing small business owners (\citealt{sbdc:history2024b}). The program is described by the SBA as the “Federal Government’s largest and most successful technical assistance program for small businesses.” According to a report published by America’s SBDC Association, long-term SBDC clients generated approximately \$6.6 billion in sales and added 80,995 new full-time jobs between 2022-23. In addition, each federal dollar spent on SBDC services generated \$23.91 in new capital to help businesses grow, \$1.59 in federal revenues, and \$2.40 in state revenues (\citealt{chrisman2024}).\footnote{America’s SBDC is a non-profit organization that represents America’s nationwide network of SBDCs. They work jointly with the SBA and United States Congress to ensure continued support for the SBDC program. Additionally, America's SBDC is responsible for a accrediting SBDC programs across the nation.}
	
	The SBDC program was founded in the early 1970s by William Flewellen and Reed Powell, both university educators and members of the SBA’s national advisory board. Flewellen and Powell believed the US would benefit from public investment in small businesses and partnerships between businesses, universities, and government (\citealt{sbdc:history2024b}). The first SBDC was established in 1976 at California State Polytechnic University at Pomona
	. This flagship center was announced under SBA’s University Business Development Center Program, whose goal was to utilize the resources and innovative atmosphere of higher education settings in small business development. In the subsequent months, seven additional US universities established similar programs
	.\footnote{These initial participating universities included California State University at Chico, The University of Georgia, The University of Missouri at St. Louis, The University of Nebraska at Omaha, Rutgers University, The University of Southern Maine and the University of West Florida.} 
	
	The success of these initial programs eventually led to the Small Business Development Act of 1980, which authorized federal funding to be matched one-for-one with non-federal funds (\citealt{smallbusiness:developmentact}).\footnote{The federal funds authorized for SBDCs are no longer matched one-for-one with non-federal funds. Federal funds are distributed on a \textit{pro rata} basis among the states. These funds are typically authorized through an Appropriations Act every 2-3 years (\citealt{levin2023}).} By 1991, as the network continued to expand outside of universities, every state in the nation operated an SBDC program
	. The US SBDC Network now consists of numerous state and regional conglomerates (like the NorCal SBDC Network) that receive federal, state, and private funding. It supports an infrastructure of approximately 1,000 centers and over 5,000 employees
	. 
	
	\subsection{The Northern California SBDC Network}
	This study examines the Northern California (NorCal) regional SBDC network, which covers nearly 12,000 square miles and 36 counties in the upper half of the state
	. The network has experienced major growth since 2006 when it was first established; as of 2018, it made up the 3rd largest regional SBDC program and operated 19 active centers with over 500 employees
	.\footnote{Previously, California had been broken into 6 different regions. In 2018, the Northwest SBDC region absorbed the Northeast region, adding new centers to the the NorCal SBDC Network we study, greatly expanding its size.} 
	According to the NorCal SBDC Network, in 2021 alone it provided one-on-one advising to 20,317 businesses, leading to over 10,000 new jobs created and nearly \$242 million in taxable revenue (\citealt{norcal:region2024}). 
	
\begin{figure}[h!]
		\centering
		\captionsetup{justification=centering}
		\caption{Top Industries by Share of Firms in NorCal SBDC and NorCal CBP Data}
		\label{fig:industry_histogram}
		\begin{minipage}[t]{0.48\textwidth}
			\centering
			\caption*{Panel A: SBDC Firms}
			\vspace{2mm}
			\includegraphics[width=\textwidth]{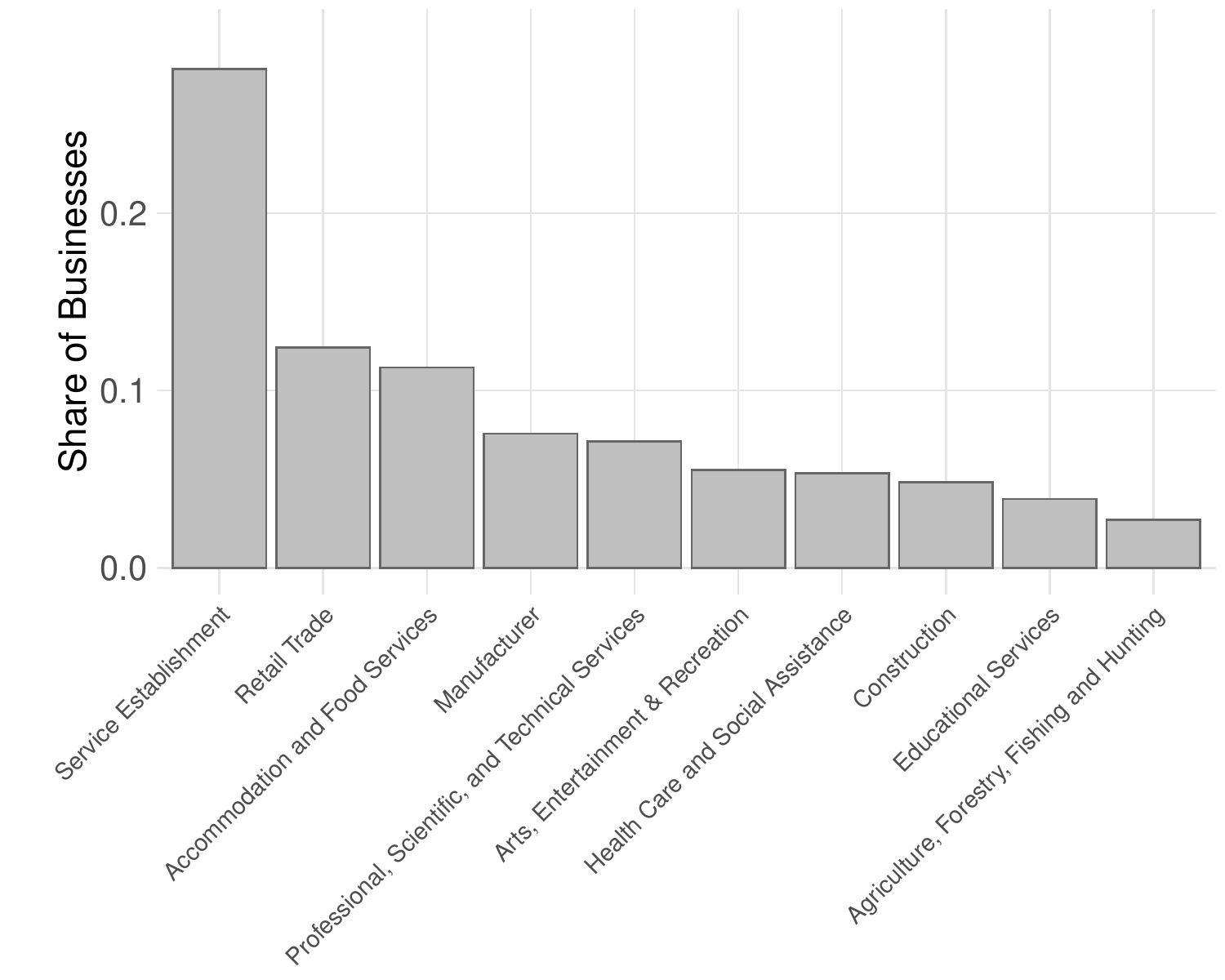}
		\end{minipage}
		\hfill
		\begin{minipage}[t]{0.48\textwidth}
			\centering
			\caption*{Panel B: CBP Firms}
			\vspace{2mm}
			\includegraphics[width=\textwidth]{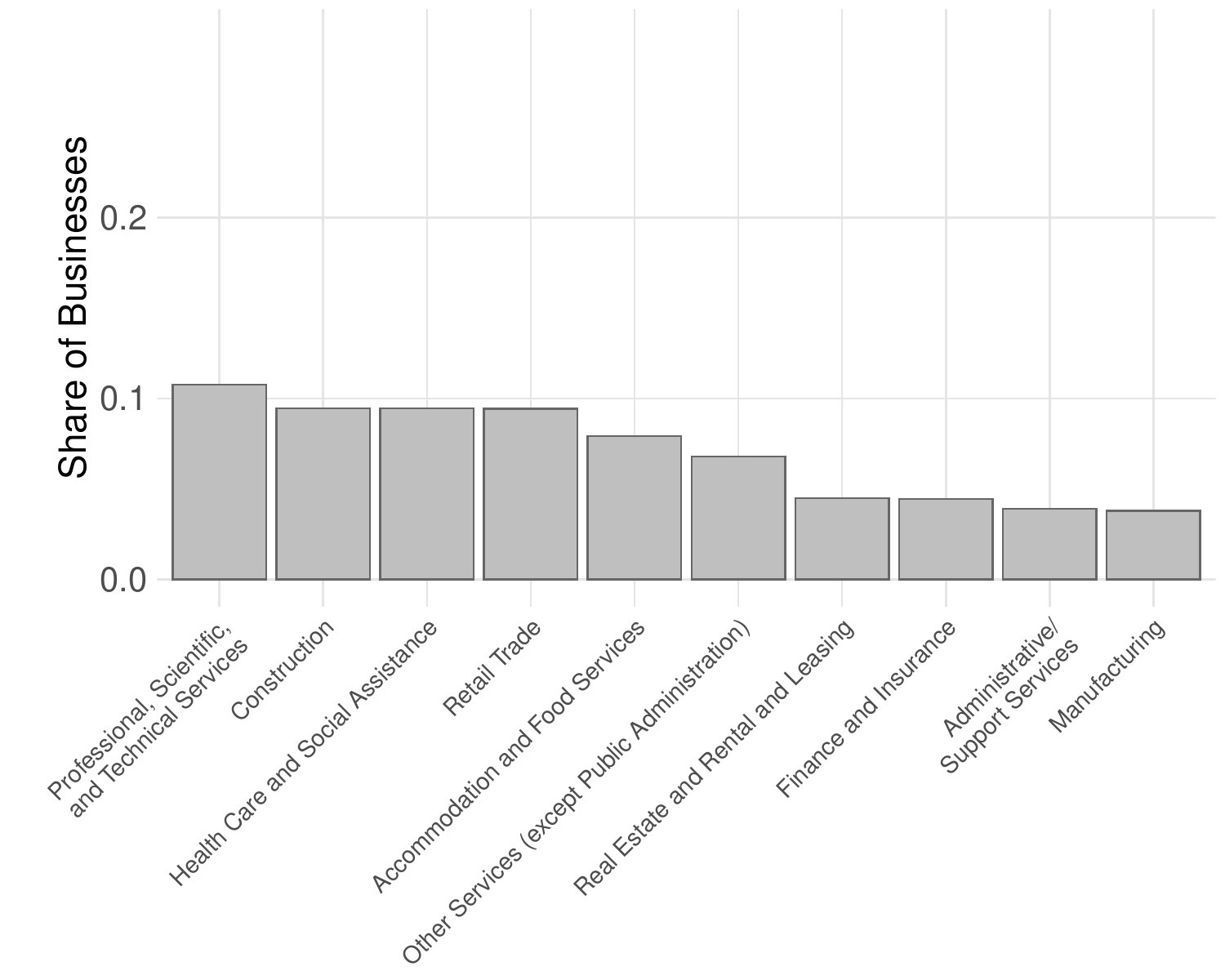}
		\end{minipage}
		\vspace{-4mm}
	\end{figure}

	Figure \ref{fig:industry_histogram} documents the top 10 industries, classified according to 2-digit North American Industry Classification System (NAICS) codes, by share of firms represented in the NorCal SBDC Network data (panel A) and US Census County Business Patterns (CBP) data (panel B) for the same counties and sample period. Panel A shows the variety of industries the NorCal SBDC Network has served over its existence. SBDC client firms tend to come from a wide range of highly competitive industries, including services, manufacturing and construction, and health and education. Moreover, while there is substantial overlap with top industries represented in the CBP data (i.e. Retail Trade; Accommodation and Food Services; Construction; Manufacturing; Professional, Scientific, and Technical Services; and Health Care and Social Assistance), the NorCal SBDC Network features a significantly higher share of firms in Service Establishments; Arts, Entertainment, and Recreation; Educational Services; and Agriculture, Forestry, Fishing, and Hunting.
	
	Each individual center in the NorCal SBDC Network is responsible for a geographic jurisdiction, typically encompassing 2-3 counties depending on whether the center is serving a more rural or urban area. 
	Centers are each associated with a fiscal host, examples of which include state universities, local chambers of commerce, and business councils
	. Hosts establish the SBDC program under their management, typically in conjunction with other business services they offer. They are in charge of providing facilities and fiscal management of centers, including managing grants and other funding received. Hosts are required to reach a certain threshold of funding to receive matched funds from the NorCal SBDC Network. 
	
	While hosts provide the physical and institutional infrastructure for SBDCs at a broad level, day-to-day operations at each center are run by a director employed by the SBDC Network. 
	This arrangement preserves a degree of continuity across the network, as directors are trained similarly to ensure that core practices and service delivery remains consistent from center to center. Directors are responsible for hiring business advisors to consult with small business clients. Advisors are hired on a contract basis and typically consist of local business executives and leaders in the region who work with centers on a part-time basis.\footnote{It should be noted that while the NorCal SBDC Network generally hires advisors as contractors, advisors in the majority of SBDC regional networks around the country are generally hired as full-time employees.} Advisors play a critical role in center services; their primary responsibility is to facilitate consulting sessions with client firms. Each consulting session is tailored to the specific business' needs and goals; more comprehensive needs and goals of businesses often require more consulting time. Crucially, advisors are typically hired from the community where the center is located, and thus tend to provide localized expertise. When centers close, open, or re-locate, advisor-firm pairings may or may not persist depending on feasibility.\footnote{Advisors often remain with the director whom they were hired by, though a new center location may induce changes to the advisor pool. This information was provided by informational discussions with several executives located at the NorCal SBDC Lead Center.}

	The SBDC advisory model is ``we teach and show, but we do not do." The approach aims to equip business owners with the skills and resources needed to operate independently, rather than provide ongoing support. Coupling this with the rapid evolution of small business needs and objectives, as well as the general observation that a substantial share of US small businesses do not survive longer than a few years, SBDC networks feature a relatively high firm turnover rate.\footnote{From 1994-2021, 32.1\% of new employer establishments did not survive more than two years and 50.8\% did not survive more than five years (\citealt{SBA2024FAQ}).} 
	
\begin{figure}[!h]
		\centering
		\captionsetup{justification=centering}
		\caption{Annual Firm Entry and Exit Counts in the NorCal SBDC Network}
		\label{fig:entry_exit_rates}
		\includegraphics[width=\textwidth,height=0.25\textheight]{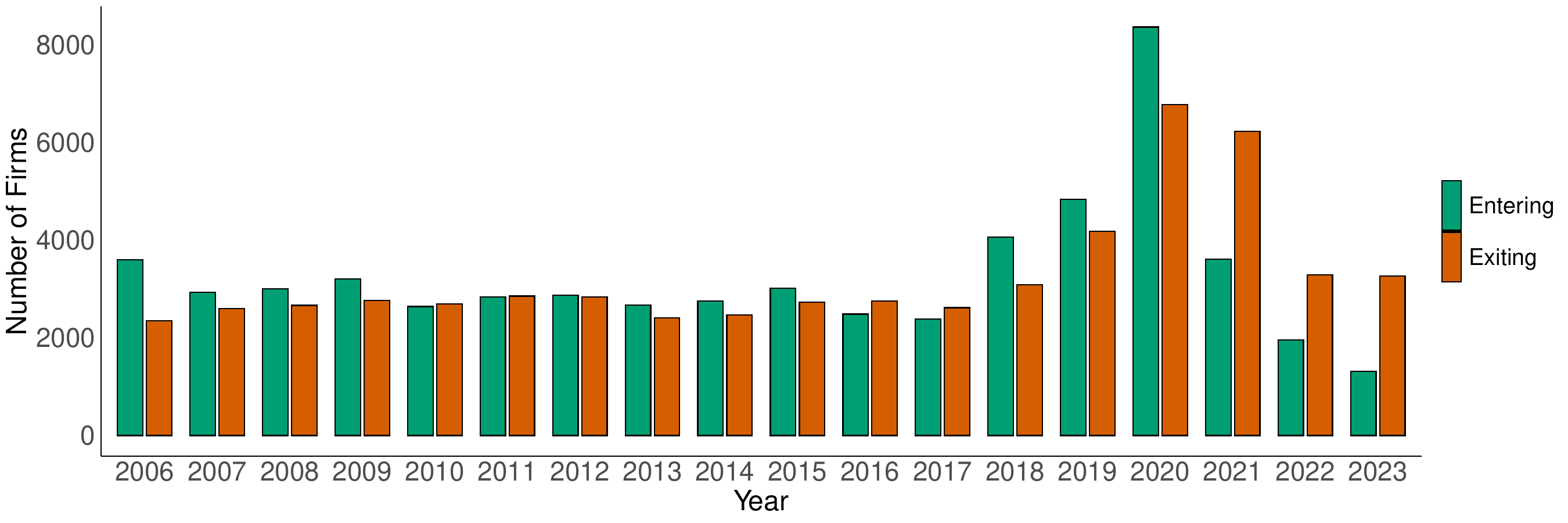} 
	\end{figure}

	Figure \ref{fig:entry_exit_rates} depicts these trends over the course of the NorCal SBDC Network's existence. One can see a significant uptick in entry and exit counts at the onset of the COVID-19 pandemic. This pattern mirrors the significant increase in small business applications in the US beginning in 2020, which was in large part due to the availability of additional financing opportunities from federal and state governments (e.g the Paycheck Protection Program).
	
	Oversight and coordination of this decentralized network is managed by the NorCal SBDC lead center, which is hosted by California State Polytechnic University at Humboldt in Arcata, CA. 
	They are responsible for maintaining standard practices across centers, expanding the network, and managing local centers in cooperation with their respective hosts. The lead center also operates a centralized intake process: when a business engages with the NorCal SBDC Network for the first time, it is typically through an online form processed by the lead center, who then matches businesses with an appropriate center. 
	
	Of particular importance for this study, the lead center generally oversees center movements, openings, and closures. These changes typically occur on a case-by-case basis, and are often due to fluctuations in funding or expiring leases.\footnote{Some hosts opt not to renew their lease agreements due to the increasing demands of managing a center.} One common mechanism for host changes is difficulty obtaining matching funds. If a host cannot reach the threshold of funds necessary for matching by the NorCal SBDC Network, a new host is sought out. According to discussions with NorCal SBDC network executives, when host changes occur, the lead center engages in a Request for Proposal (RFP) process to identify new potential hosts to facilitate the relocation or opening of a center. This process 
	identifies possible new hosts who can offer the most resources and best fit for SBDC clients.\footnote{When a center closure or relocation occurs, the lead center will continue to pay the contracted advisors at the previous center to continue working with business owners until the new center is established.}
	
	To introduce additional structure to the process of matching centers and hosts, the NorCal SBDC Network developed a regular Request for Proposals (RFP) process in 2020. In addition to case-by-case host changes, the network implemented the RFP process to be repeated every five years across all centers. This allows existing hosts to either renew or end their relationships with the network, and prospective hosts to submit proposals. The stated goal of this procedure is to ensure that each center host continues to serve as a strong and valuable partner.\footnote{The RFP process was recommended by America's SBDC, the nonprofit organization responsible for the NorCal Network's accreditation.} The first network-wide RFP cycle in the NorCal SBDC Network, held in 2020, led to major restructuring: 8 centers closed in 2021, and 8 new centers opened in 2022. This marked one of the largest shifts in the network’s configuration since its inception in 2006.
	
	These structural changes in the network's composition are noteworthy given the unique role each center plays in serving the specific needs of the local business community. 
	Differences in business needs across regions stems from a range of factors, including the makeup of local industries, regulatory landscape, access to capital, and broader socioeconomic conditions. 
	While many forces contribute to varying conditions across geographies, a consistent and impactful driver is the urban-rural divide. This consideration is particularly important in the Northern California region, which features both highly urban and highly rural areas. Such an environment underscores the importance of place-based business support models like the SBDC network, which can deliver services tailored to the differing needs of both urban and rural entrepreneurs.
	
	Additional detail regarding the NorCal SBDC network, particularly with respect to the process by which firms initially access the network, the services individual centers provide, and format of consulting sessions can be found in Appendix Section \ref{AdditionalDetailSBDC}.

	\section{Data}\label{Data}
	
	We leverage a novel administrative dataset from the NorCal SBDC network. Each observation examines a firm $i$ in year $t$ matched with one of 50 possible center locations spanning the NorCal network and active at some point between 2006-2023.\footnote{Centers may retain the same host when changing locations, but any change in host will necessarily be associated with a change in location. For simplicity and accuracy, we treat each location change as a new center. Figure \ref{activecenters} provides a map of active NorCal SBDC centers as of 2023.} 
	
	\begin{table}[h!]
		\centering
		\caption{Unique Counts of Key Panel Characteristics}
		\label{tab:table1_unique_count}
		\begin{tabular}{cccccc}
			\toprule
			Years & Client IDs & Treated Clients & Industries & Entity Types & Centers \\
			\midrule
			18 & 88,734 & 2,865 & 24 & 10 & 50 \\
			\bottomrule
		\end{tabular}
	\end{table}
	
	Table \ref{tab:table1_unique_count} provides general information about the structure of the dataset. Each firm is assigned a unique client ID by the NorCal network.\footnote{There were a small number of businesses with addresses outside of the Northern California region; we remove these businesses from the sample.} Treated clients include any business that experienced a closure, opening, or movement of a center they were matched with.\footnote{This does not include clients who were initially (or at some point) matched with the NorCal SBDC lead center.} Centers are uniquely identified by their specific postal address. Businesses are classified under 24 possible industries, each one defined by a 2-digit NAICS code, and 10 entity types (e.g. LLC, sole proprietorship, etc.). Other firm-specific characteristics include zip code and establishment date of the company.
	
	We also observe several time-varying characteristics at the firm level; Table \ref{tab:table2_summary_stats} displays summary statistics for these variables. Firms engage in an average of 3.42 consulting sessions per year with centers. Contact and preparation time are reported when a center and business engage in a session. Contact time measures the hours spent between an advisor and client during a consultation, while preparation time measures the hours spent by an advisor to prepare for a consulting session. These variables allow us to observe (1) different sessions for firms and (2) the center responsible for facilitating each session. As Table \ref{tab:table2_summary_stats} shows, firms spend an average of 3.73 hours annually in consulting sessions with advisors, while advisors spend nearly 1.5 hours annually in preparation per client. 
	Appendix Table \ref{tab:center_characteristics_compact} provides descriptive statistics on participating firms for each specific center in the data. Additionally, while firms apart of the NorCal SBDC Network are small by nature, they are highly representative of non-participating firms in the same geographic area over the same time horizon, according to US Census CBP data (Appendix Table \ref{tab:cbp_summary_stats}). In fact, more than half of all firms in the region during this time period feature 1-4 employees. 
	
	\begin{table}[h!]
		\centering
		\caption{Summary Statistics of Client Firms in the NorCal SBDC Network}
		\label{tab:table2_summary_stats}
		\hspace*{-0.75cm}
		\begin{threeparttable}
			\begin{tabular}{lllllll}
				\toprule
				Variable & \makecell{Number of \\ Observations} & Mean & SD & Median & \makecell{5\textsuperscript{th}\\Percentile} & \makecell{95\textsuperscript{th}\\Percentile}\\
				\midrule
				\# of Sessions & 91,832 & 3.42 & 4.02 & 2 & 1 & 11 \\
				Contact Time (hours) & 91,832 & 3.73 & 7.55 & 2 & 0.25 & 13 \\
				Preparation Time (hours) & 91,832 & 1.45 & 2.34 & 0.75 & 0 & 5.25 \\
				Distance (miles)
				& 48,089 & 13.2 & 20.0 & 7.0 & 0.9 & 46.1 \\
				$\Delta$ in Distance (miles)$^*$ & 3,409 & 0.33 & 43.1 & -0.17 & -47.7 & 43.0 \\
				\% Urban & 46,726 & 85.5 & 26.1 & 98.5 & 0 & 100 \\
				Firm Tenure w/ Network (years) & 58,576 & 1.81 & 1.75 & 1 & 1 & 5 \\
				Firm Age (years) & 30,306 & 9.30 & 12.20 & 5 & 0 & 33 \\
				Annual Revenue (\$1000s)$^{+}$ & 12,334 & 530.8 & 7,709.5 & 18.2 & 0.0 & 1,501.2 \\
				Employees$^\dagger$ & 10,954 & 4.01 & 15.2 & 1.5 & 0 & 14 \\
				\bottomrule
			\end{tabular}
			\begin{tablenotes}[flushleft]
				\footnotesize
				\singlespacing
				\vspace{-4mm}
				\item \textit{Note}: These statistics omit all businesses that engaged with the NorCal SBDC lead center. The lead center is often responsible for on-boarding businesses into the network, thus logging contact time not associated with the primary business advisory services.\\
				$^*$The number of observations for the ``$\Delta$ in Distance" variable consists of every unique change in business-center pairing, resulting from center openings, closures, or re-locations. Comparing this value to the number of treated businesses in Table \ref{tab:table1_unique_count} shows there are businesses who have experienced $>1$ changes in the center pairing. \\
				$^{+}$Annual revenue is given in 2023 USD terms. \\
				$^\dagger$Full-time and part-time employees are documented separately. We count each part-time employee as 0.5 full-time employees.
			\end{tablenotes}
		\end{threeparttable}
	\end{table}

\begin{figure}[h!]
		\centering
		\captionsetup{justification=centering}
		\caption{Distribution of Contact Time, Preparation Time, Revenue, and Employment}
		\label{fig:Summary_Histograms}
		\begin{minipage}[t]{0.48\textwidth}
			\centering
			\par
			\vspace{2mm}
			\includegraphics[width=\textwidth]{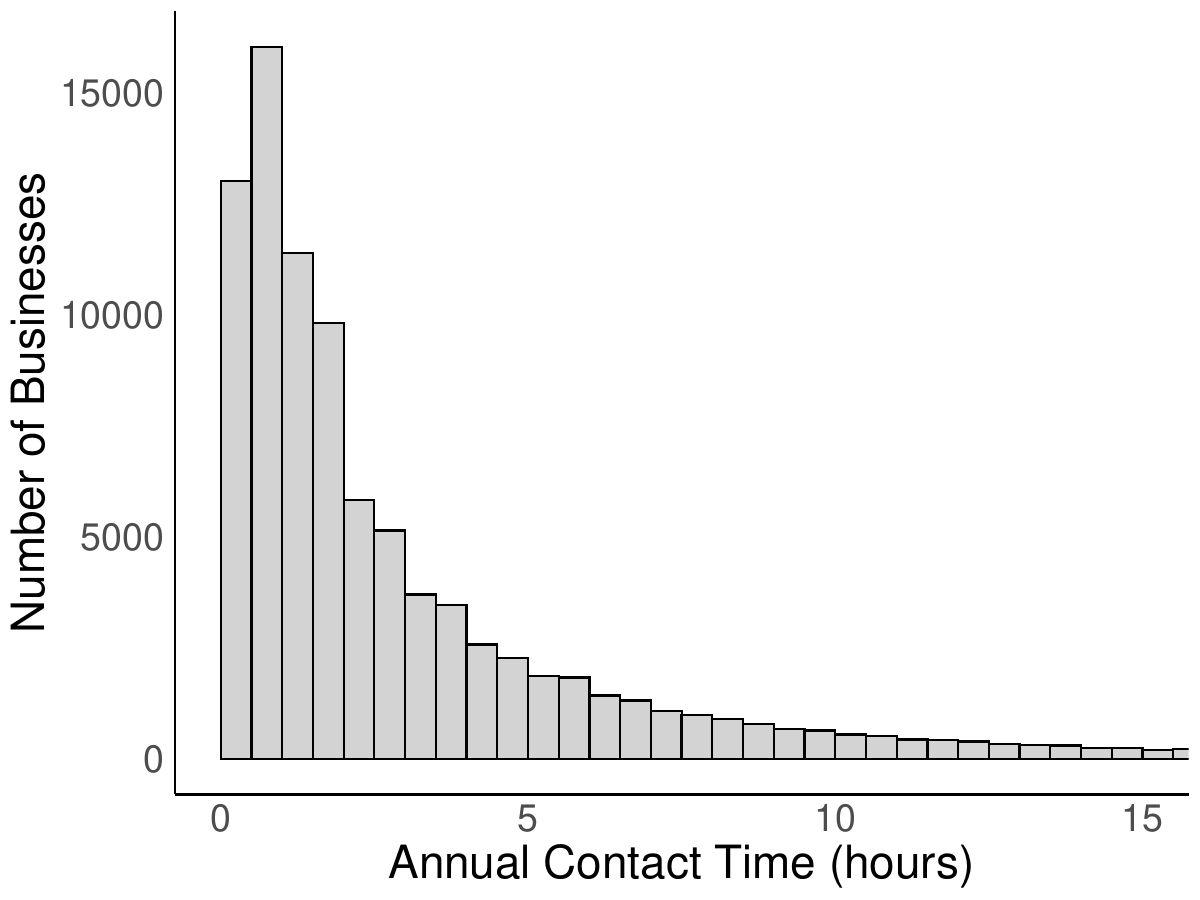}
		\end{minipage}
		\hfill
		\begin{minipage}[t]{0.48\textwidth}
			\centering
			\par
			\vspace{2mm}
			\includegraphics[width=\textwidth]{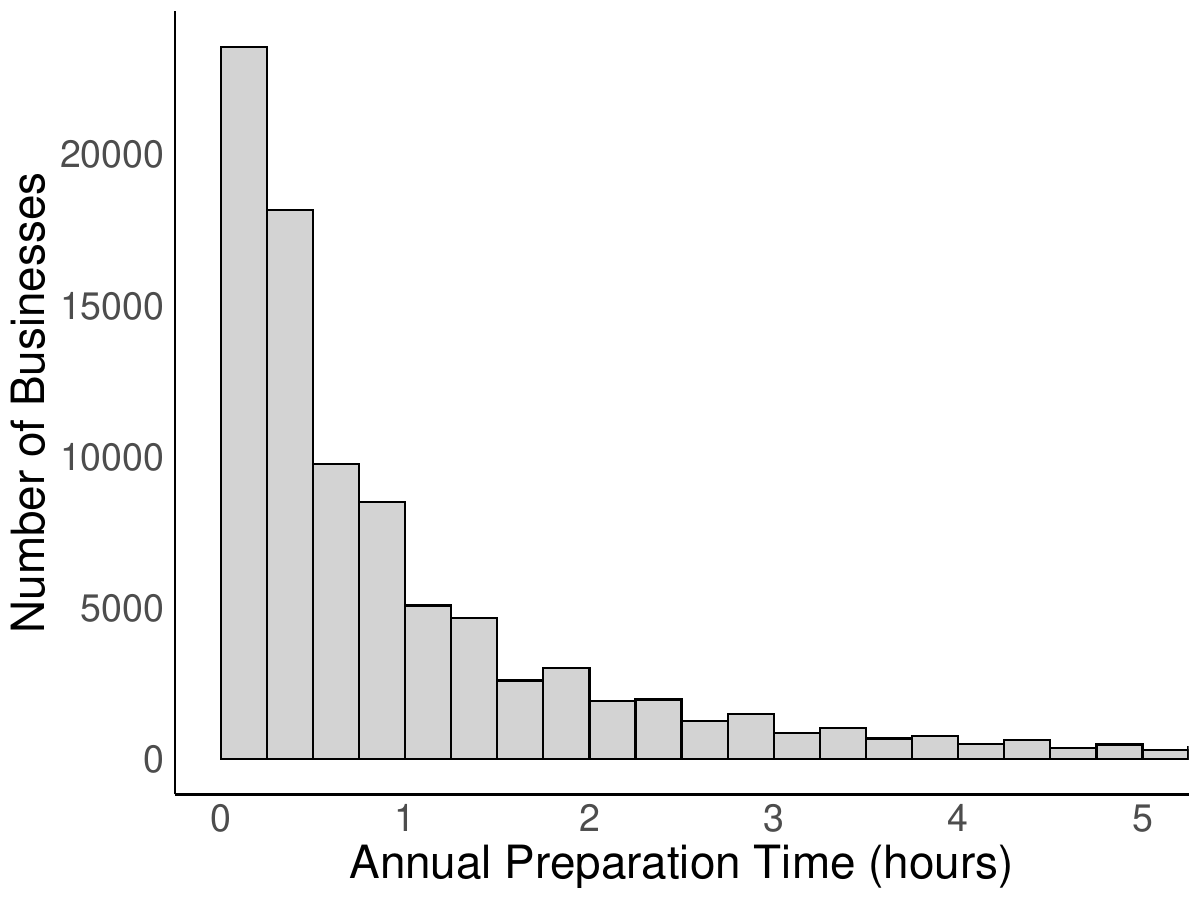}
		\end{minipage}
		\vspace{-4mm}
	\end{figure}

	Figure \ref{fig:Summary_Histograms} depicts the distributions of contact time, preparation time, annual revenue, and employment. The distributions of both contact and preparation time are heavily right-skewed, which is intuitive given that most firms engage in a few consulting sessions annually (each of which lasts approximately one hour), while a small number of firms engage in significantly more. Annual revenue appears approximately log-normal, while employment is also right-skewed.\footnote{The long right-tails of both employment and annual revenue inform the large standard deviation estimates found in Table \ref{tab:table2_summary_stats}.} These patterns are consistent with the composition of SBDC clientele, which consists primarily of small businesses with relatively few employees.
	
	\begin{figure}[h!]
		\centering
		\begin{minipage}[t]{0.48\textwidth}
			\centering
			\par
			\vspace{2mm}
			\includegraphics[width=\textwidth]{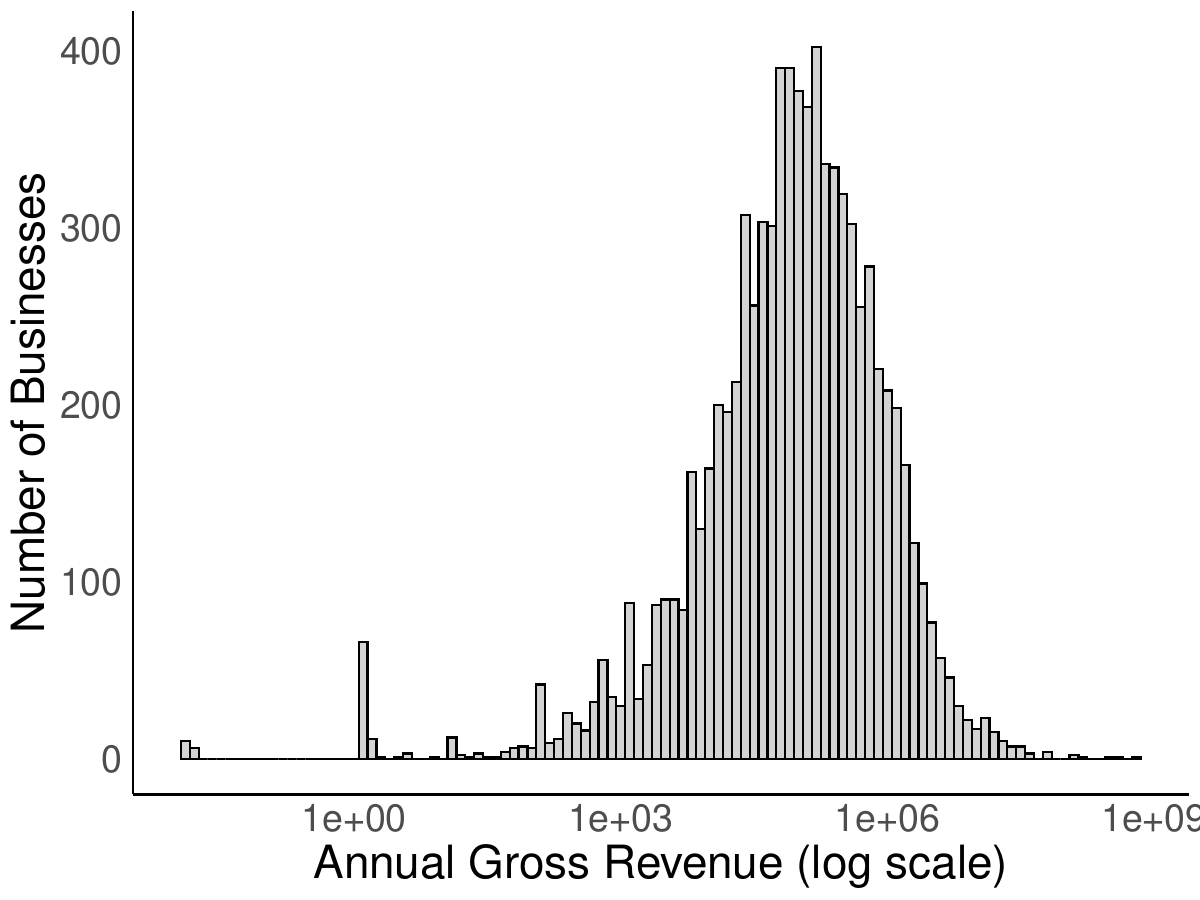}
		\end{minipage}
		\hfill
		\begin{minipage}[t]{0.48\textwidth}
			\centering
			\par
			\vspace{2mm}
			\includegraphics[width=\textwidth]{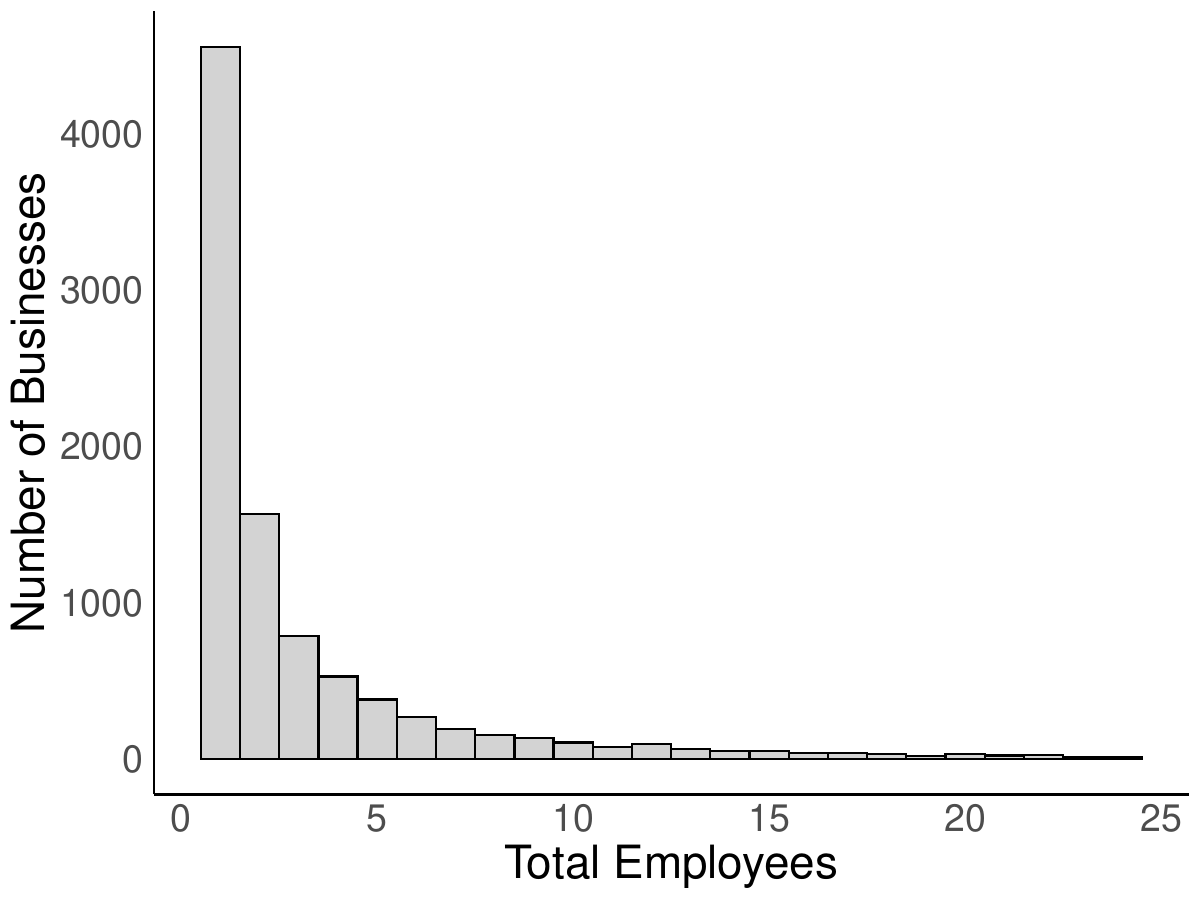}
		\end{minipage}
		\begin{threeparttable}
			\begin{tablenotes}
				\footnotesize
				\vspace{2mm}
				\item Note: This figure illustrates the distributions of the four labeled variables. The vertical axis displays the number of businesses that register a value for contact time, preparation time, total employees, or revenue in a given year. 
			\end{tablenotes}
		\end{threeparttable}
	\end{figure}

	An important limitation of the data is that revenue and employment are collected less frequently than session-level information. As a result, for most firms we observe only a single value for each of these outcomes and the associated time period in which it was recorded. A related concern is that firms are only observed during their tenure with the SBDC network, which may introduce selection based on survival or continued engagement. If firms that remain in the network differ systematically from those that exit, estimates may be biased in favor of outcomes of surviving firms. The direction of this bias is ambiguous; continued participation may reflect successful engagement and thus positive selection, or instead capture firms that required sustained advisory support due to ongoing challenges. Appendix Figures \ref{fig:firm_rev_emp_violinstats} and \ref{fig:firm_rev_emp_tenureyearofrecord} provide descriptive evidence on the distribution and timing of firm annual revenue and employment. While this descriptive evidence suggests that outcome measurement is not systematically related to firm tenure in the network, this concern cannot be fully ruled out.\footnote{The figures show that length of firm tenure with the NorCal SBDC Network does not predict moments of the annual revenue and employment distributions (Figure \ref{fig:firm_rev_emp_violinstats}), and is not associated with the timing of when data on these performance outcomes is collected from each firm (Figure \ref{fig:firm_rev_emp_tenureyearofrecord}).}
	
	Distances between firms and corresponding centers are computed using location data for each. To maintain confidentiality, business locations are provided at the 5-digit USPS ZIP code level, while center coordinates are geocoded from their full addresses. Using locations of centers and businesses at each encounter, we compute a straight-line measure of distance between a business and its corresponding center.\footnote{Distance is calculated using the Haversine Formula. 
	} 

	\begin{figure}[!h]
		\centering
		\captionsetup{justification=centering}
		\caption{Timeline of Center Closures, Openings, and Location Changes}
		\label{center_changes}
		\vspace{-2mm}
		\includegraphics[width=\textwidth]{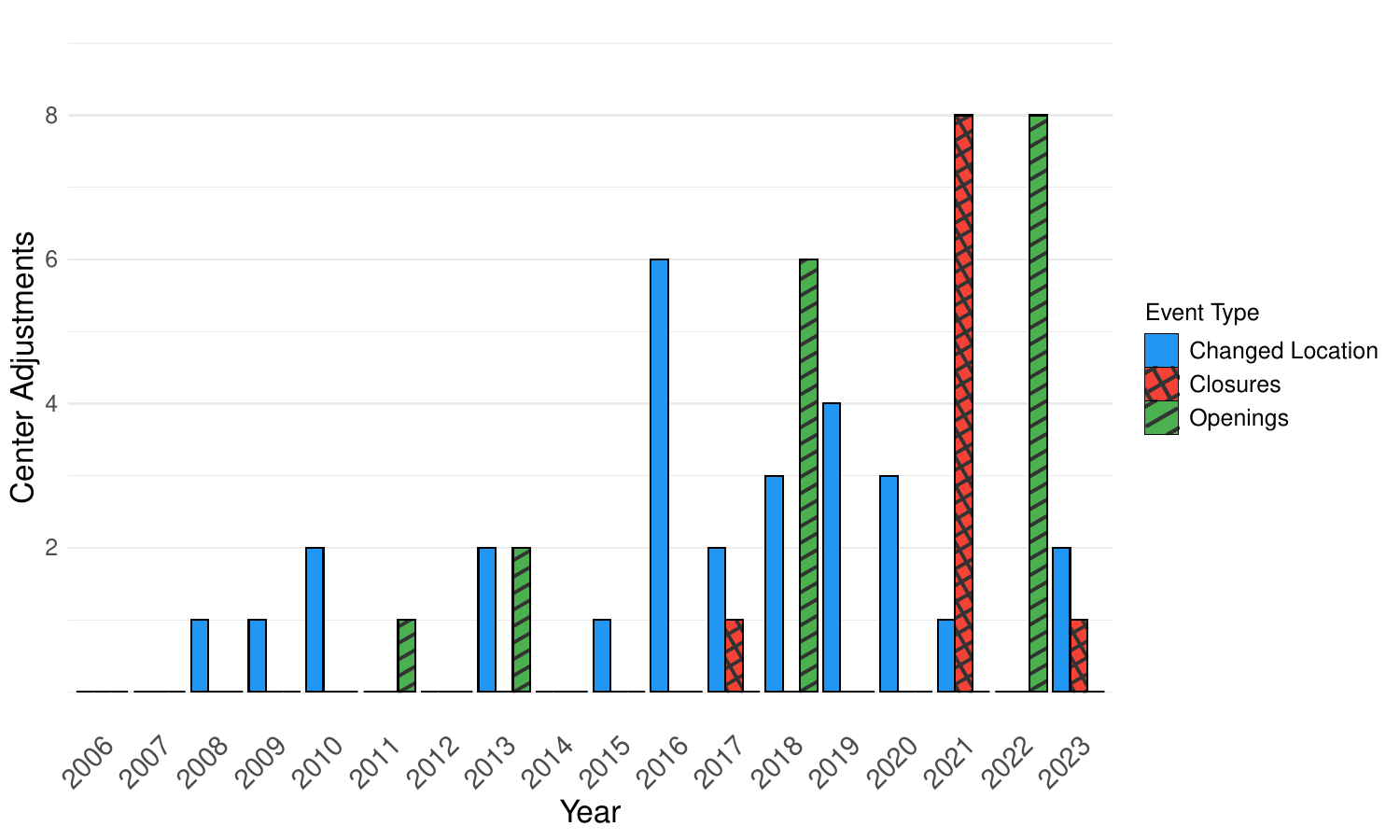} 
	\end{figure}

	A key source of variation in the data is the evolution of business-center pairings over time and space. There are three possible changes a pairing can experience, which we observe: a center closure, opening, or change in location. If a center closes, an associated business is typically re-assigned to the closest open center. If a center opens, certain businesses may switch from an existing center to this new center. If a center changes location, businesses may choose to move with the center, or be re-assigned to a different center. The data captures all business-center adjustments between 2006-2023. 
	
	Figure \ref{center_changes} illustrates a timeline of the count of center adjustments, indicating annual openings (green), closures (red), and location changes (blue). There were 11 initially active centers at the inception of the NorCal SBDC Network in 2006. One can see that in nearly every subsequent year there is at least one center adjustment event. Center openings and closures happen less frequently and in larger bunches, particularly in 2021 for closures and 2018 and 2022 for openings.\footnote{
		In 2018, the NorCal SBDC network we study expanded by absorbing the Northeast regional network of California; we classify these newly introduced centers as openings.}

\begin{figure}[h!]
		\centering
		\captionsetup{justification=centering}
		\caption{Geographic Distribution of NorCal SBDC Centers Over Time}
		\label{Map Timeline}
		\begin{minipage}[t]{0.40\textwidth}
			\centering
			\underline{2006} \par
			\vspace{2mm}
			\includegraphics[width=0.9\textwidth]{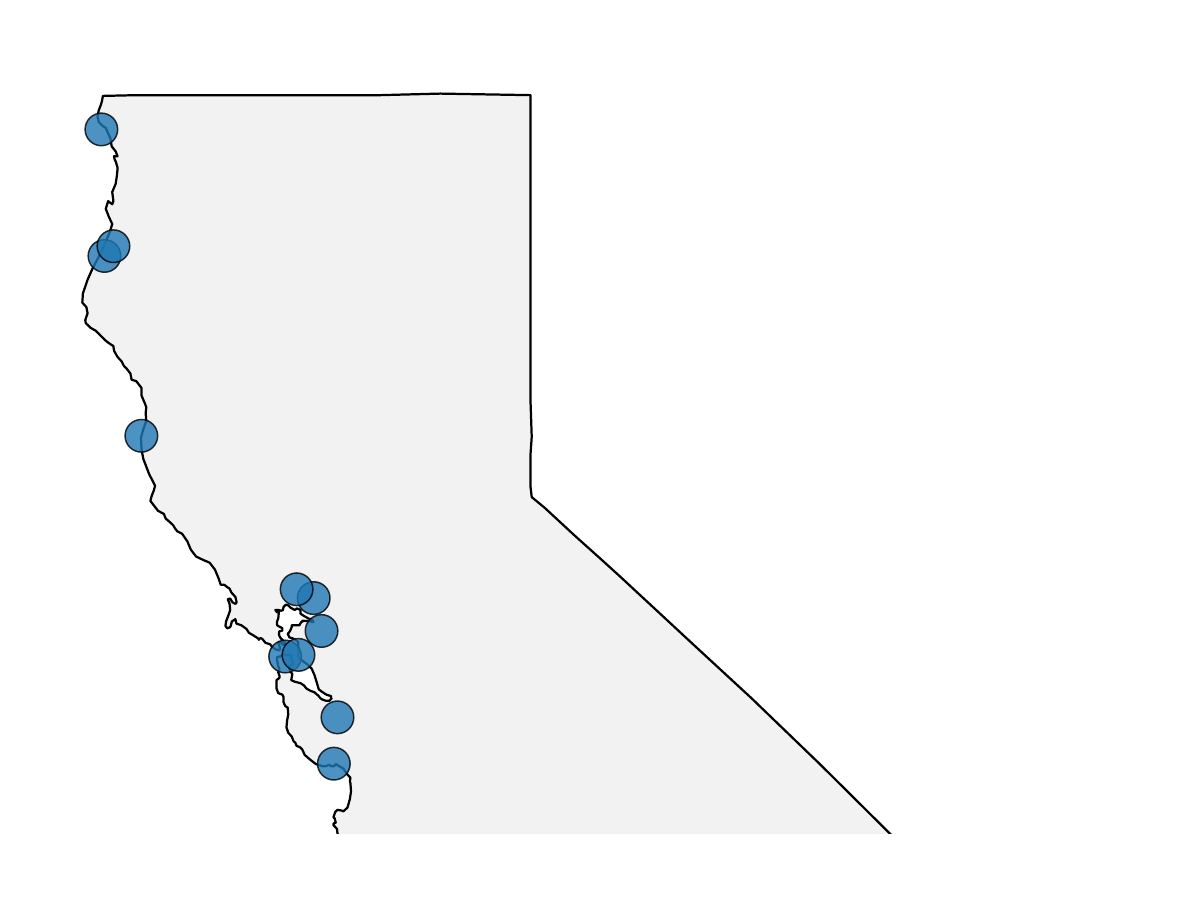}
			\vspace{2mm}
			Active Centers: 11
		\end{minipage}
		\hfill
		\begin{minipage}[t]{0.40\textwidth}
			\centering
			\underline{2013} \par
			\vspace{2mm}
			\includegraphics[width=0.9\textwidth]{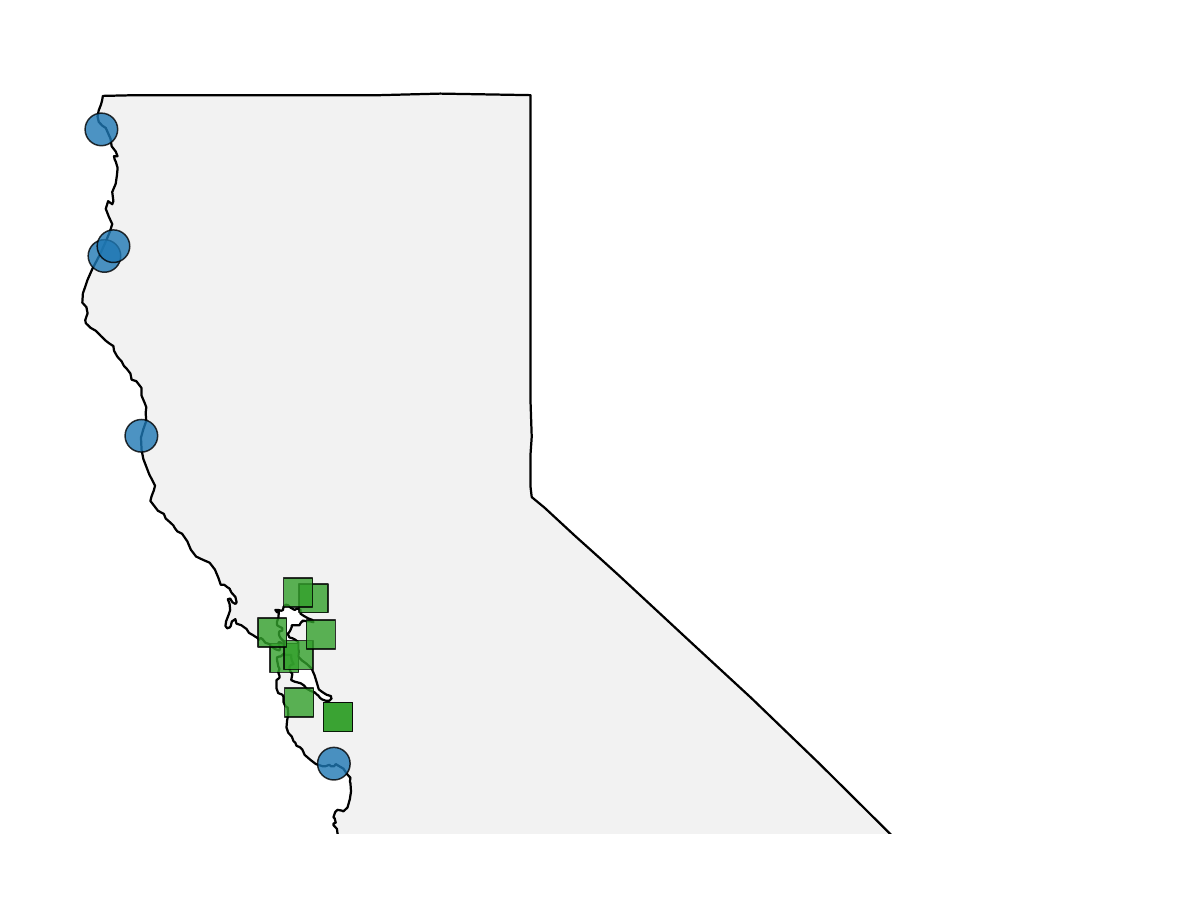}
			\vspace{2mm}
			Active Centers: 14
		\end{minipage}
	\end{figure}

	Figure~\ref{Map Timeline} shows how the geographic makeup of the NorCal SBDC network has evolved over time. It shows four key time periods: 2006, 2013, 2018, and 2023.\footnote{These four distinct periods account for over 90 percent of all center closures and openings between 2006-2023.} Each new shape introduced in a subsequent period indicates either a relocated or newly-established center compared to the previous period. The figure highlights the network’s substantial growth over the past 18 years; 
	the evolving spatial distribution generates important variation in the geographic proximity of businesses to paired centers. 
	Appendix Figure \ref{activecenters} shows the geographic distribution of and number of clients served across all active centers as of 2023.
	\footnote{Despite not being located in geographic proximity, some businesses will register contact time with the lead center (represented by the largest circle in the figure) in the first year they are observed as a result of the on-boarding process. Due to this administrative structure and because of the localized geographic variation we seek to exploit, we omit all businesses who register contact time with the lead center and subsequently experience a change in associated center.}

	\begin{figure}[h!]
		\centering
		\begin{minipage}[t]{0.40\textwidth}
			\centering
			\underline{2018} \par
			\vspace{2mm}
			\includegraphics[width=0.9\textwidth]{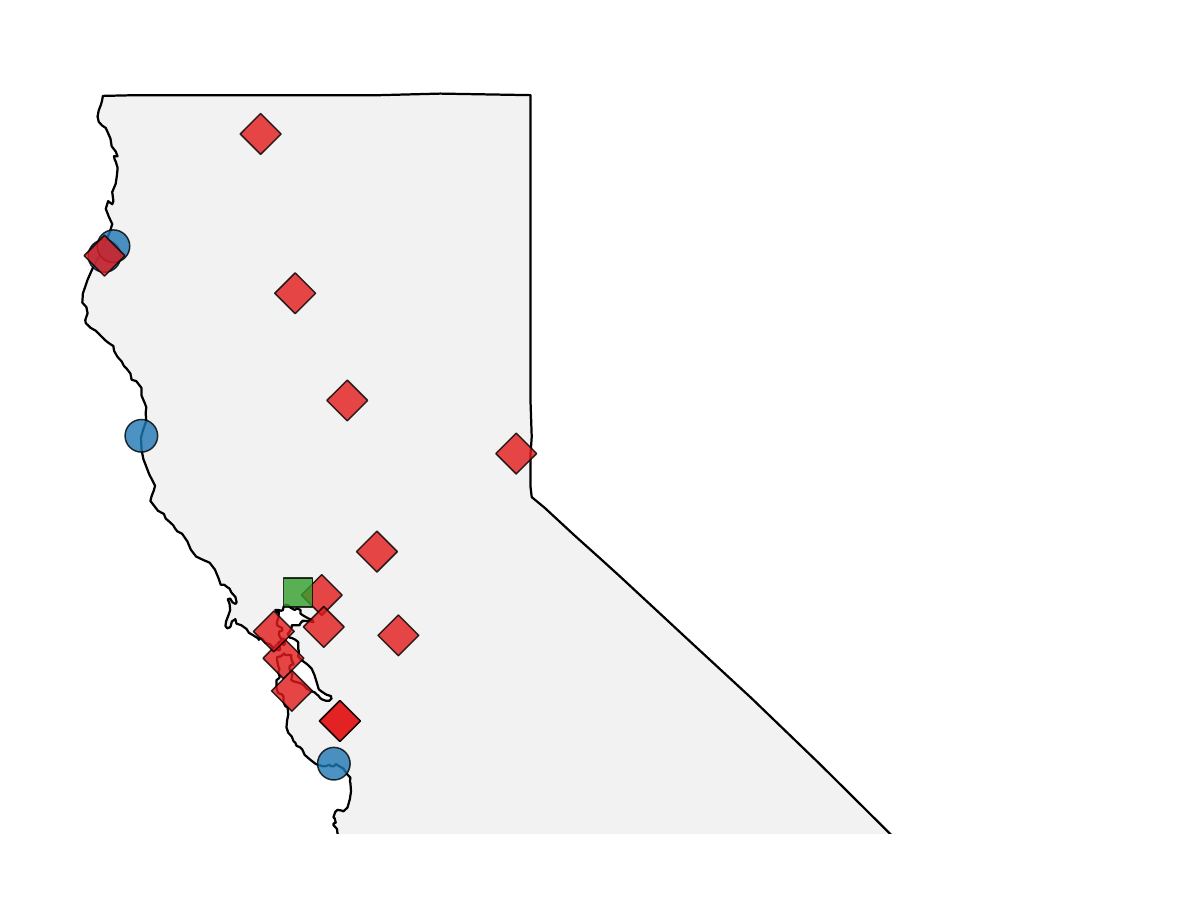}
			\vspace{2mm}
			Active Centers: 19
		\end{minipage}
		\hfill
		\begin{minipage}[t]{0.40\textwidth}
			\centering
			\underline{2023} \par
			\vspace{2mm}
			\includegraphics[width=0.9\textwidth]{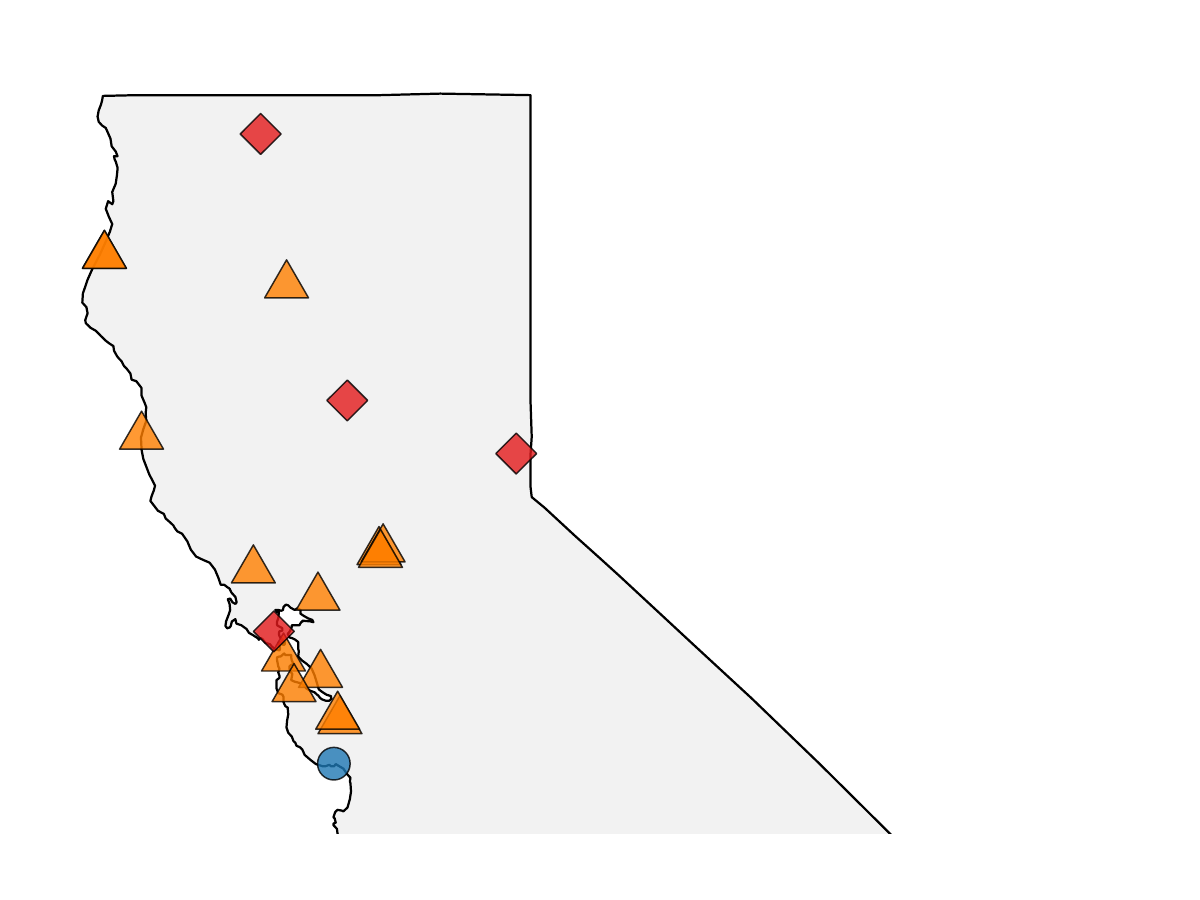}
			\vspace{2mm}
			Active Centers: 19
		\end{minipage}
		\begin{threeparttable}
			\begin{tablenotes}
				\footnotesize
				\vspace{2mm}
				\item Note: The four symbols include a blue circle (2006), green square (2010), red diamond (2018), or an orange triangle (2023). A new symbol appearing on the map indicates a center either changed location or was newly established in comparison to the previous period. Note that some centers visually overlap with one another on the map; this is due to multiple centers located in the same city or in close geographic proximity.
			\end{tablenotes}
		\end{threeparttable}
	\end{figure}

\begin{figure}[h!]
		\centering
		\captionsetup{justification=centering}
		\caption{Distribution of Change in Distance}
		\label{changeindistance}
		\vspace{-4.25mm}
		\includegraphics[width=0.8\textwidth]{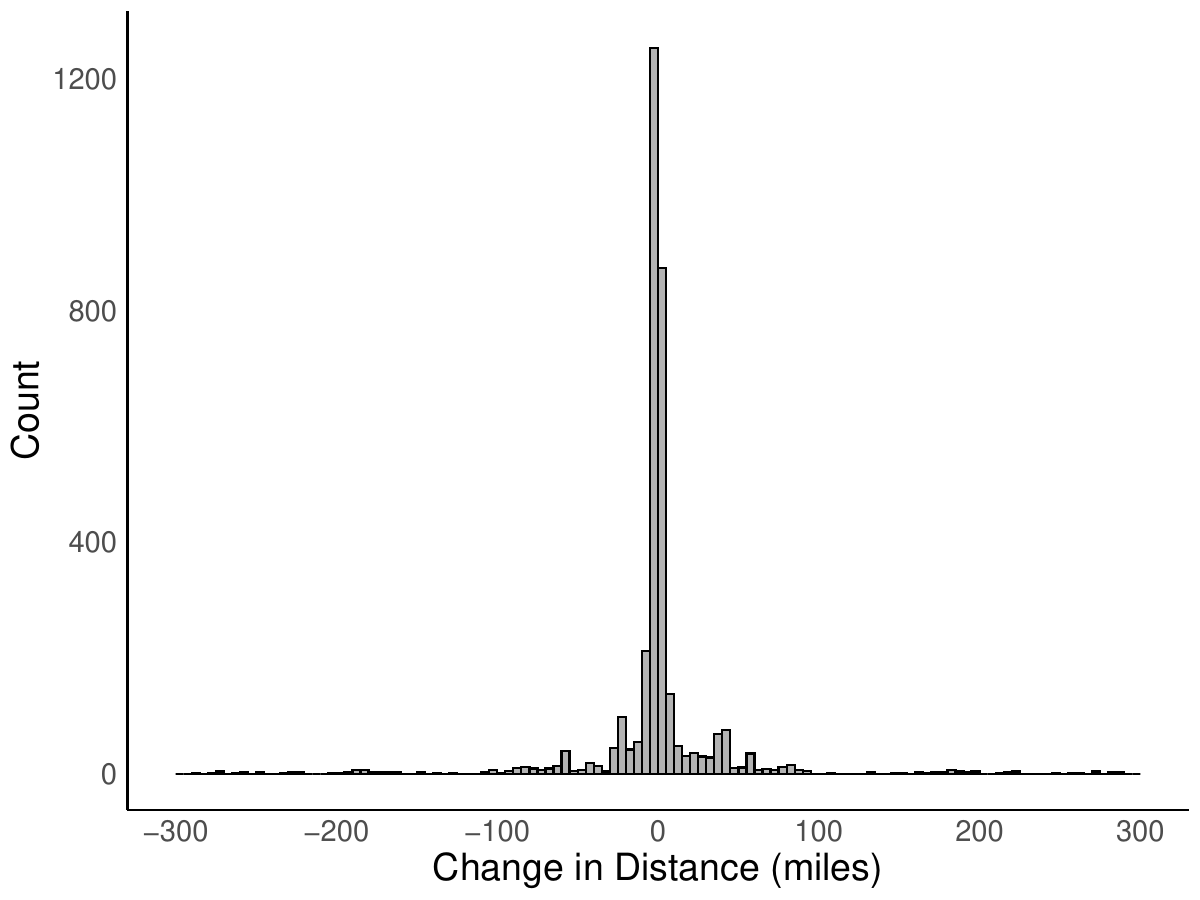} 
		\begin{tablenotes}
			\footnotesize
			\item \textit{Note}: This figure calculates the change in distance associated with any change in firm-center pairing. This means that negative change in distance values is associated with a center moving closer to a client, while positive change in distance values indicates a center moving farther away. For each subsequent year after treatment, the change in distance value remains the same. This figure omits any change in distance values greater than 300 miles or less than -300 miles.
		\end{tablenotes}
	\end{figure}

	The geographic redistribution of centers over time is central to our identification strategy. When a center closes, opens, or relocates, firms may be reassigned to a different center, introducing variation in distance between firms and matched centers. We exploit such variation to estimate the effect of geographic proximity on firm utilization of advisory services. Figure~\ref{changeindistance} depicts the distribution of change in distance values between treated businesses and matched centers.\footnote{Note that untreated businesses are simply businesses who do not experience a change in distance associated with a business-center pairing adjustment.} The average (standard deviation) change in distance is approximately 0.33 miles (43.1 miles) (Table \ref{tab:table2_summary_stats}). Figure~\ref{changeindistance} shows notable variation in change in distance, including substantial density among both positive (further distance between a business and center) and negative values. Most change in distance values range between -100 and 100 miles. The significant density near 0 reflects small geographic shifts often resulting from new hosts in the same local jurisdiction.
	
	\section{Empirical Strategy}
	We utilize an instrumental variables (IV) strategy to measure the impact of changes in geographic proximity between centers and firms on firm utilization of advisory services, and the resulting effects on firm annual revenue and employment. 
	Firm engagement with centers is endogenous; for instance, highly motivated firms are more likely to engage with centers and also exhibit stronger performance outcomes. The IV approach leverages changes in distance between firms and paired centers resulting from center closures and openings, which we show are orthogonal to pre-event firm performance outcomes and local economic conditions. 
	
	The first-stage estimating equation leverages panel data at the firm-year level and takes the following form:
	\begin{equation}
		\text{ContactTime}_{ijoct}= \alpha_0+\alpha_1\Delta\text{Distance}_{ijoct}+\gamma_i+\sigma_c+\delta_t+\varepsilon_{ijoct} \label{eq:first_stage}
	\end{equation}
	where $\text{Contacttime}_{ijoct}$ is the number of contact (consulting) hours received by firm $i$ classified in industry $j$ located in county $o$ from center $c$ during year $t$. $\gamma_i$, $\sigma_c$, and $\delta_t$ indicate firm-, center-, and year-fixed effects, respectively. The instrument, $\Delta\text{Distance}_{ijoct}$, also varies at the firm-year level. $\alpha_1$ indicates the average impact of changes in distance between a business and paired center on predicted annual consulting hours the business receives. 
	
	The second-stage estimating equation measures changes in firm outcomes due to changes in predicted contact time values estimated from the first-stage. We define two separate firm outcomes $y_{ijot}=\{\text{Revenue}_{ijot}, \text{Employment}_{ijot}\}$.
	As mentioned previously, revenue and employment data are recorded at a lower frequency than session-level metrics, limiting the within-firm variation that can be utilized. To continue leveraging important spatial and temporal identifying variation from the first-stage, we average outcomes across firms within the same county-industry-year.\footnote{To leverage rich variation in categorization of businesses, we utilize ``industry x entity type" categories. However, for the sake of simplicity, we refer to each industry x entity type as industry in the empirical specifications. }  The average outcome is calculated as follows:
	\begin{equation}
		\frac{1}{n}\sum_{i=1}^n y_{ijot} = \bar{y}_{jot} \label{eq:avg_outcome}
	\end{equation}
	where $\bar{y}_{jot}$ is the average annual revenue (employment) for firms in the same industry $j$, located in county $o$, and apart of the NorCal SBDC Network in year $t$.
	
	This approach offers several advantages compared to a firm-level cross-sectional estimation. First, firms accumulate different amounts of consulting time with paired advisors each year. Depending on when annual revenue and employment information are collected from each firm, it may not account for the total accumulated consulting time shared between a firm and center, thus generating a misleading relationship between consulting time and performance outcomes. Second, leveraging county-year data in the second-stage preserves important identifying variation from the first-stage, since change in distance is determined by local spatial adjustments over time.\footnote{To assess the spatial structure of the identifying variation from the first-stage estimation, we compute Moran's I statistic, a standard measure of spatial autocorrelation (\citealt{anselin1988spatial}). We find a positive and statistically significant Moran’s I across a range of $k$-nearest neighbor specifications, indicating that geographically proximate firms experience positively correlated changes in distance. 
		Appendix Table \ref{tab:moran_results} reports numeric estimates.} Moreover, the second-stage structure accommodates the inclusion of industry-, county-, and year-fixed effects, which helps further account for possible endogeneity in the relationship between predicted firm-center consulting time and firm performance outcomes. The second-stage findings are robust to alternate aggregations, including the use of 5-digit ZIP code instead of county as the spatial aggregation unit (see Tables \ref{tab:alt_specs_secondstage_revenue} and \ref{tab:alt_specs_secondstage_employment}).
	
	The structure of this approach follows existing studies that leverage micro-level first-stage variation and aggregate predicted values to match coarser outcome data in the second stage. \cite{frankel1999does} estimate a bilateral trade equation using country-pair data and aggregate predicted trade flows to construct a country-level instrument for use in a cross-country growth regression. Similarly, \cite{keller2013link} exploit variation in trade at the city-pair level and aggregate fitted values at the city-level to study the relationship between trade and economic development. \cite{allcott2014gasoline} use detailed micro-level variation in gasoline prices and vehicle characteristics to construct predicted fuel economy measures that are then aggregated to match the level in which consumer outcomes are observed. 
	
	As a result of using industry-county-year variation in outcomes in the second-stage, we calculate average predicted contact time at the same level using firm-year predicted values generated from the first-stage:
	\begin{equation}
		\frac{1}{n}\sum_{i=1}^n \widehat{\text{ContactTime}}_{ijot} = \overline{\widehat{\text{ContactTime}}}_{jot} \label{eq:avg_contact_time}
	\end{equation}
	where $\overline{\widehat{\text{ContactTime}}}_{jot}$ is the average number of consulting hours for firms in industry $j$, county $o$, and year $t$ generated from the first-stage predicted firm-year contact time estimates.
	
	The second-stage estimating equation takes the following structure:
	\begin{equation} 
		\ln(\bar{y}_{jot})= \beta_0+ \beta_1\overline{\widehat{\text{ContactTime}}}_{jot}+X_{ot}\Delta+ \phi_j +\omega_o +\lambda_t+u_{jot} \label{eq:second_stage}
	\end{equation}
	where $\bar{y}_{jot}$ represents average firm-level outcomes in industry $j$ and county $o$ during year $t$, and $\phi_j$, $\omega_o$, and $\lambda_t$ indicate the industry-, county-, and year-fixed effects, respectively. We also include a set of economic indicators by county-year to account for possible endogeneity introduced by changing local economic conditions. $\beta_1$ measures the expected impact of changes in contact time on proportional changes in firm performance outcomes.
	
	\subsection{Threats to Identification and Instrument Validity}\label{threats}
\begin{figure}[h!]
		\centering
		\captionsetup{justification=centering}
		\caption{Count of Business-Center Pairings Based on Order of Geographic Proximity}
		\label{ranking}
		\vspace{-2mm}
		\includegraphics[width=0.65\textwidth]{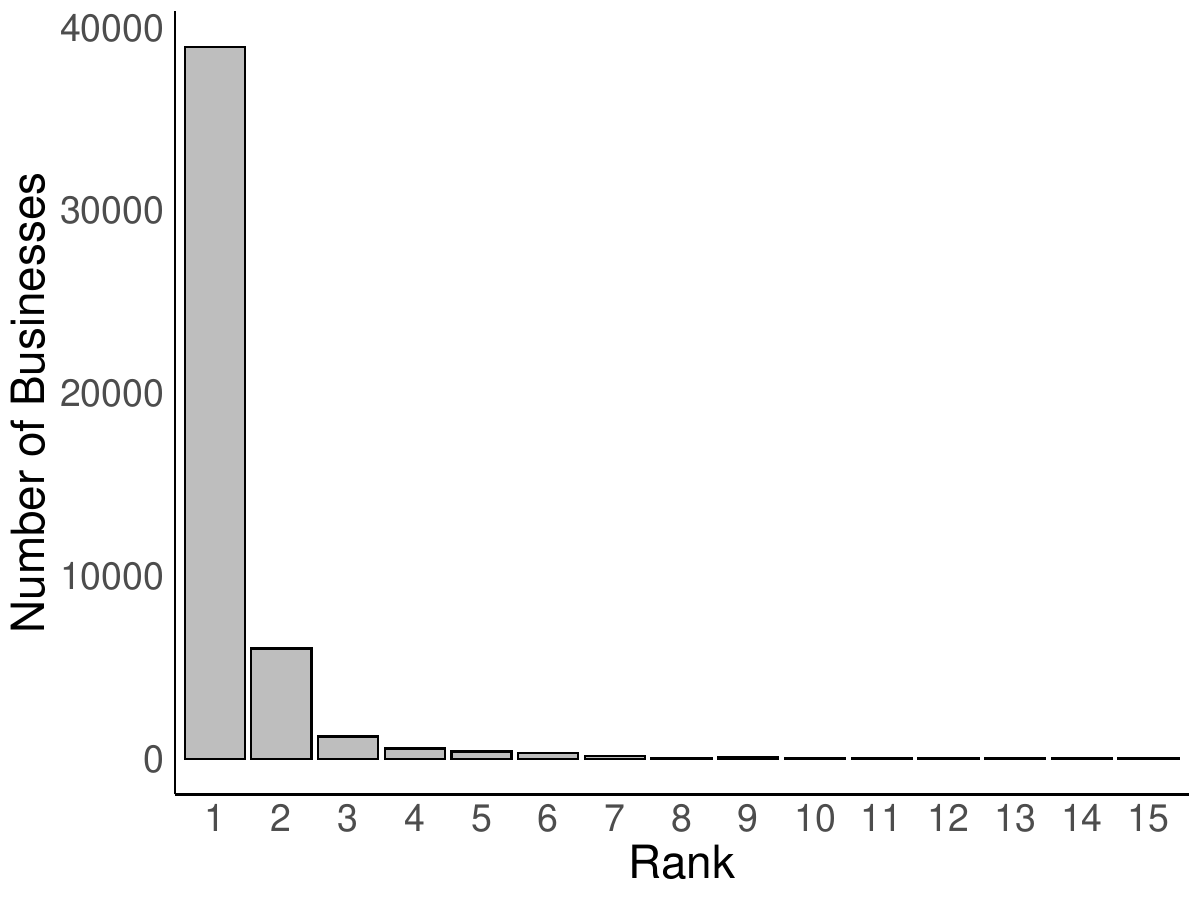} 
		\begin{tablenotes}
			\footnotesize
			\vspace{-2mm}
			\item \textit{Note}: For each business-year observation, we rank all potential centers from closest to furthest in distance and identify the center with which the business engaged. A rank of 1 indicates the business worked with its closest center, 2 indicates the second closest, and so on.
		\end{tablenotes}
	\end{figure}

	Here we discuss the validity of the IV approach in achieving identification. The instrument, change in distance, is expected to be predictive of consulting hours spent between a firm and corresponding SBDC. While we provide evidence of relevance in the first-stage estimation results in Table \ref{tab:iv_regression_summary}, Figure \ref{ranking} offers additional supporting context that firms generally choose to work with the center geographically closest to them.\footnote{While physical distance is certainly relevant in facilitating consulting time between a business and corresponding SBDC, it may also be important in terms of local knowledge and regulation a center can offer a business.} 
	
	The primary threat to identification is the existence of time-varying, unobserved firm- and location-specific shocks that may affect firm performance and are correlated with center re-locations, closures, and openings.\footnote{Time-invariant, location-specific shocks are accounted for with the inclusion of center and business fixed effects in the first-stage. Time-varying shocks affecting all units identically are controlled for with year fixed effects.} In other words, the validity of change in distance as an instrument is threatened if center adjustments systematically coincide with pre-adjustment firm performance as well as unobserved local economic upturns or downturns.\footnote{Note that the primary second-stage specification in equation \ref{eq:second_stage} controls for a set of time-varying, county-specific economic indicators.} For example, consider a center that relocates from an economically stagnant or declining region to one that is expanding, which facilitates a reduction in distance for some firms while they experience an improving economic environment. In this case, we are unable to identify to what extent changes in geographic proximity versus unobserved location-time specific changes in broader economic conditions generate improved business outcomes. It is important to note that unobserved changing economic conditions affecting regions differently is not a threat to identification per se. This is only concerning if region-specific changes in unobserved economic conditions are systematically correlated with patterns of center closures and openings. Figure \ref{centermovement} in the Appendix visually illustrates a hypothetical manifestation of this concern. 
	
	
	In light of these possible issues, we offer several pieces of supporting evidence to further validate the use of change in distance between firms and associated centers due to geographic shifts in the NorCal SBDC Network. First, we find no evidence that firm performance (as measured by annual revenue and employment) \textit{prior} to center openings and closures predicts the onset of such an event. Using a center-year panel consisting of all firms working with each specific center,\footnote{In the case of center openings, we use data in the years prior to the center opening from firms that are eventually matched with the newly-opened center.} we estimate the following linear probability model:
	
	\vspace{-8mm}
	\begin{align}
		\text{CenterEvent}_{ct} &= \pi_0 + \pi_1\text{AvgAnnualRev}_{c,t-1} + \pi_2\text{AvgAnnualRev}_{c,t-2} + \pi_3\text{AvgAnnualRev}_{c,t-3} + \nonumber \\ &\pi_4\text{AvgEmp}_{c,t-1} + \pi_5\text{AvgEmp}_{c,t-2} + \pi_6\text{AvgEmp}_{c,t-3} + \sigma_c + \delta_t + \psi_{ct} \label{eqn:centereventprediction}
	\end{align}
	
	where $\text{CenterEvent}_{ct} \in \{0,1\}$ indicates whether a specific center $c$ experiences a closure or opening in year $t$, $\text{AvgAnnualRev}_{c,t-h}$ and $\text{AvgEmp}_{c,t-h}$ represent the average annual revenue and employment, respectively, of firms associated with center $c$ and $h$ years prior to a closure or opening event, and $\sigma_c$ and $\delta_t$ represent center- and year-fixed effects, respectively, and control for time-invariant differences across centers and common temporal chocks. 
	
	\begin{table}[h!]
		\centering
		\caption{Lagged Firm Performance and Center Openings and Closures}
		\label{tab:event_prediction}
		\small
		\begin{tabular}{l*{3}{>{\centering\arraybackslash}p{0.17\textwidth}}} 
			\toprule
			& \multicolumn{3}{c}{Dependent Variable: Center Event Indicator} \\
			\cmidrule(lr){2-4}
			& \makecell{Openings \& \\ Closures} 
			& \makecell{Openings \\ Only} 
			& \makecell{Closures \\ Only} \\
			\midrule
			Avg. Revenue ($t-1$)
			& $-0.00018$ & $0.00002$ & $-0.00018$ \\
			& $(0.00035)$ & $(0.00019)$ & $(0.00031)$ \\
			
			Avg. Revenue ($t-2$)
			& $-0.00024$ & $-0.00032^{*}$ & $0.00000$ \\
			& $(0.00020)$ & $(0.00018)$ & $(0.00012)$ \\
			
			Avg. Revenue ($t-3$)
			& $-0.00240$ & $-0.00057$ & $-0.00160$ \\
			& $(0.00248)$ & $(0.00124)$ & $(0.00209)$ \\
			
			Avg. Employment ($t-1$)
			& $-0.00035$ & $-0.00390$ & $0.00272$ \\
			& $(0.00607)$ & $(0.00350)$ & $(0.00530)$ \\
			
			Avg. Employment ($t-2$)
			& $0.00186$ & $0.00153$ & $0.00234$ \\
			& $(0.00418)$ & $(0.00375)$ & $(0.00305)$ \\
			
			Avg. Employment ($t-3$)
			& $-0.00148$ & $-0.00239$ & $-0.00059$ \\
			& $(0.00326)$ & $(0.00327)$ & $(0.00215)$ \\
			
			\midrule
			Observations & $266$ & $266$ & $266$ \\
			Center FE & X & X & X \\
			Year FE & X & X & X \\
			Clustered SEs (Center) & X & X & X \\
			R$^2$ & $0.111$ & $0.165$ & $0.133$ \\
			\bottomrule
			\multicolumn{4}{l}{\textit{$^{*}p < 0.1$, $^{**}p < 0.05$, $^{***}p < 0.01$}} \\
		\end{tabular}
		\begin{tablenotes}
			\footnotesize
			\vspace{2mm}
			\item \textit{Note}: This table reports estimates from linear probability models where the dependent variable indicates whether a center experiences an opening or closure event in year $t$. Regressors include lagged average firm revenue (scaled in \$100,000s and in 2023 USD terms) and employment in the three years prior to the event. All specifications include center and year fixed effects, and standard errors are clustered at the center level.
		\end{tablenotes}
	\end{table}

	The findings are provided in Table \ref{tab:event_prediction}. The first specification, which considers both center closures and openings, shows no evidence that lagged average firm annual revenue or employment in the periods preceding a center opening or closure is associated with the probability of such an event occurring. Specifications (2) and (3) illustrate the findings separately with respect to center closures and openings, respectively.\footnote{There is a weak, but statistically significant, negative association between average firm annual revenue in period $t-2$ and the probability of a center opening in period $t$.} The findings are robust to differing time horizons of lagged firm performance outcomes.

	Next, we assess whether firms experiencing a center opening or closure exhibit differential pre-trends in annual revenue and employment compared to firms that do not. To do so, we estimate event-study specifications that trace out differences in average firm outcomes in the years leading up to a center opening or closure, relative to the year immediately preceding the event. Using the same center-year panel constructed for estimating equation \ref{eqn:centereventprediction}, we compare outcomes for treated centers in the pre-event period to those of control centers observed in the same calendar years. 
	The event-study specification takes the following structure:
	
	\vspace{-8mm}
	\begin{align}
		Y_{ct} = \sum_{k} \beta_k \cdot D^k_{ct} + \sigma_c + \delta_t + \varepsilon_{ct} \label{eqn:instrumentvalidityeventstudy}
	\end{align}
	
	where $Y_{ct}$ denotes average annual revenue (employment) for firms associated with center $c$ in year $t$, and $D^k_{ct}$ is an indicator equal to 1 if center $c$ is $k$ years relative to an opening or closure event in year $t$. The specification includes event-time indicators for $k \in \{-5,-4,-3,-2\}$, with $k = -1$ omitted as the reference period. $\beta_k$ therefore captures differences in outcomes between treated vs. untreated firms in year $k$ relative to the year immediately preceding a center opening or closure. The specification also includes center-fixed effects $\sigma_c$ and year-fixed effects $\delta_t$.
	
	\begin{figure}[h!]
		\centering
		\caption{Differences in Average Firm Performance Prior to Center Openings and Closures}
		\label{fig:eventstudyinstvalidity}
		
		\begin{subfigure}[t]{0.49\textwidth}
			\centering
			\caption*{Panel A: Annual Revenue}
			\includegraphics[width=\textwidth]{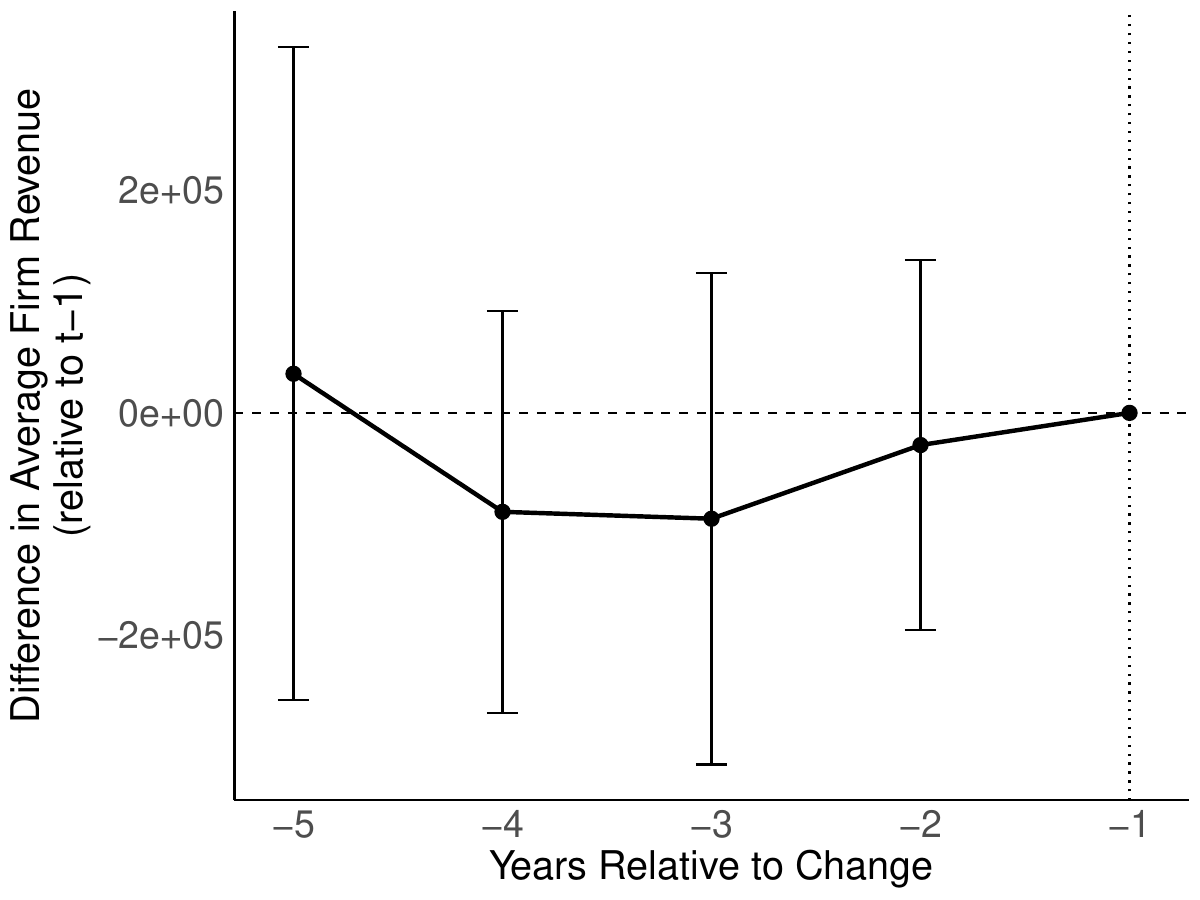}
		\end{subfigure}
		\hfill
		\begin{subfigure}[t]{0.49\textwidth}
			\centering
			\caption*{Panel B: Employment}
			\includegraphics[width=\textwidth]{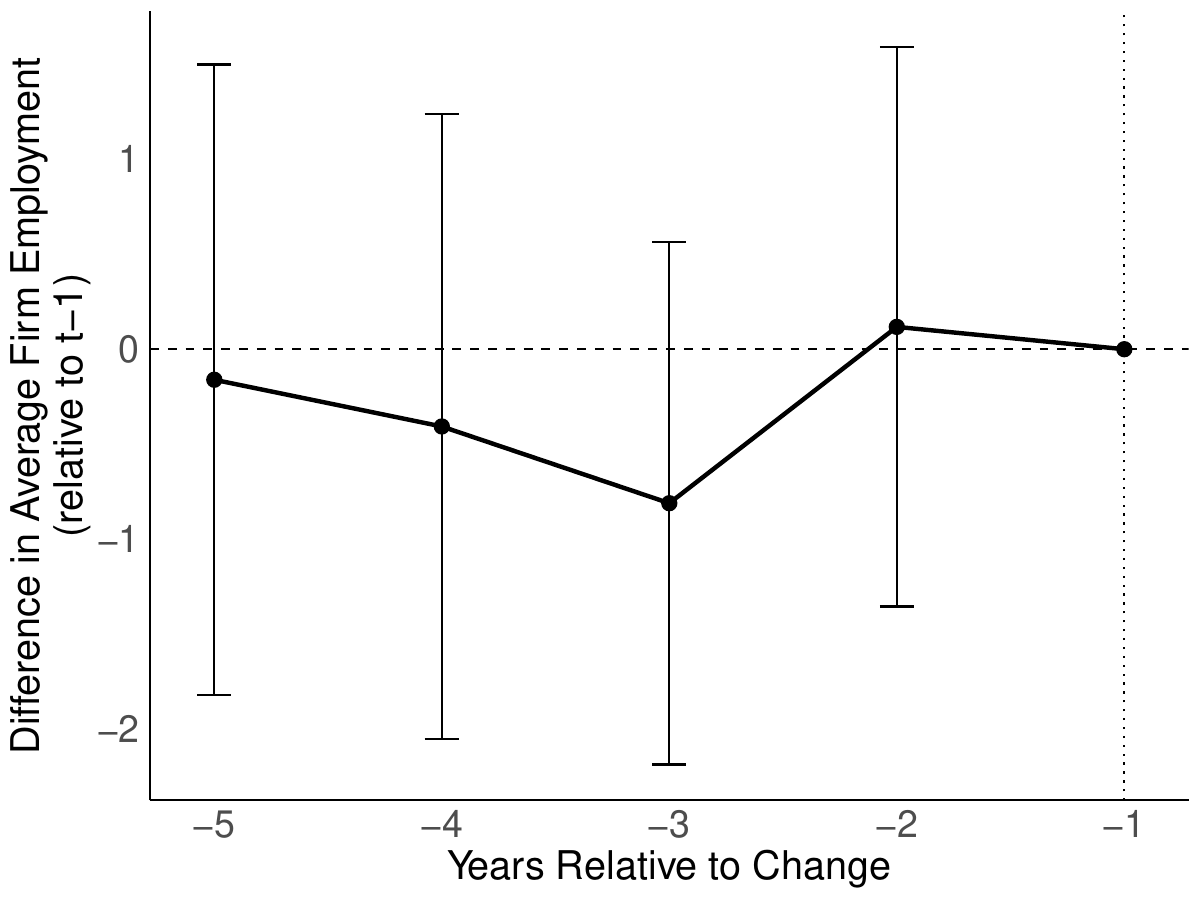}
		\end{subfigure}
		
		\begin{tablenotes}
			\footnotesize
			\item \textit{Note}: Points represent coefficient estimates from the event-study specification in equation \ref{eqn:instrumentvalidityeventstudy}. Confidence intervals are estimated using clustered-robust standard errors by center ID. 
		\end{tablenotes}
	\end{figure}
	
	As shown in Figure \ref{fig:eventstudyinstvalidity}, we find little evidence of systematic pre-trends in either annual revenue or employment in the years leading up to center openings or closures. Appendix Figure \ref{fig:eventstudyinstvalidity_Appendix} displays results for center closures and openings in separate estimations.

	
	\section{Results}\label{results}
	This section provides results from the empirical analysis. Table \ref{tab:iv_regression_summary} presents five specifications; specifications 1 and 2 feature estimates from an OLS specification examining the effect of consulting hours on annual revenue and employment averaged across businesses at the industry-county-year level. Specification 3 presents results from the first-stage estimation, measuring the effect of change in distance on predicted consulting hours a firm receives annually. Specifications 4 and 5 present the second-stage estimation results, examining the effect of contact time on average annual revenue and employment, respectively. Section \ref{section:robustness} shows robustness of estimates to important sample restrictions. Section \ref{section:heterogeneity} presents additional heterogeneity, estimating differential impacts of change in distance in relatively rural versus urban areas as well as based on the magnitude of change in distance experienced.
	
	\begin{table}[h!]
		\begin{center} 
			\caption{OLS and IV Estimation Results}
			\label{tab:iv_regression_summary}
			\footnotesize 
			\hspace*{-0.75cm}
			\begin{tabular}{l@{\hspace{4pt}}c@{\hspace{4pt}}cc@{\hspace{4pt}}c@{\hspace{4pt}}c} 
				\toprule
				& \multicolumn{2}{c}{OLS} & \multicolumn{1}{c}{First-Stage} & \multicolumn{2}{c}{Second-Stage} \\
				\cmidrule(lr){2-3} \cmidrule(lr){4-4} \cmidrule(lr){5-6} 
				& Log (Avg Rev) & Log (Avg Emp) & Contact Time & \hspace{1mm} Log (Avg Rev) & Log (Avg Emp) \\
				\midrule
				Avg Contact Time & $0.0066$ & $0.0070$ & & & \\ 
				& $(0.0147)$ & $(0.0045)$ & & & \\ 
				\addlinespace[0.5em] 
				$\Delta$ in Distance & & & $-0.0078^{***}$ & & \\ 
				& & & $(0.0020)$ & & \\ 
				\addlinespace[0.5em] 
				Avg Predicted Contact Time & & & & $0.051^{***}$ & $0.017^{***}$ \\ 
				& & & & $(0.017)$ & $(0.006)$ \\ 
				\midrule
				Observations & $2,382$ & $2,923$ & $91,236$ & $2,382$ & $2,923$ \\
				Firm-Fixed Effects &  &  & X &  &  \\
				Center-Fixed Effects &  &  & X &  &  \\
				Industry-Fixed Effects & X & X &  & X & X \\
				County-Fixed Effects & X & X &  & X & X \\
				Year-Fixed Effects & X & X & X & X & X \\
				County-Year Controls & X & X &  & X & X \\
				Clustered Robust SEs & X & X & X &  &  \\
				Bootstrapped SEs &  &  &  & X & X \\
				R$^2$ & $0.306$ & $0.310$ & $0.588$ & $0.308$ & $0.312$ \\
				F-Statistic &  &  & $15.9$ &  &  \\
				\bottomrule
				\multicolumn{6}{l}{\textit{$^{*}p < 0.1$, $^{**}p < 0.05$, $^{***}p < 0.01$}} \\
			\end{tabular}
		\end{center} 
		\begin{tablenotes}
			\footnotesize
			\vspace{-2mm}
			\item \textit{Note}: This table presents the coefficients (standard errors) of the OLS and IV first- and second-stage estimations. Information regarding fixed effects and estimation of standard errors is provided in the bottom panel of the table. For specifications (1) and (2), standard errors are clustered at the industry-county level, while in specifiation (3) they are clustered by center. When bootstrapping standard errors, we cluster at the industry x county level with 1,000 iterations.
		\end{tablenotes}
	\end{table}
	
	\subsection{First-Stage Results}
	The first-stage results present estimates of the impact of change in distance on number of consulting hours, leveraging panel data at the firm-year level. We find that for every one standard deviation increase in distance between a business and matched center (20 miles), predicted consulting time between them decreases by 0.15 hours each year. For the median firm, which engages in 2 hours of contact time annually, this represents an approximately 7.5\% decrease in utilization of advisory services. The estimate is highly statistically significant and has an F-statistic of 15.9, suggesting a moderately strong instrument. 
	
	Table \ref{tab:iv_regression_prep} provides six alternate specifications, including specifications employing (i) advisor preparation time and (ii) combined consulting and preparation time as outcome variables. Specification (5) in this table shows the effect of change in distance on preparation time; these values remain statistically significant but are approximately half the magnitude of the estimates in Table \ref{tab:iv_regression_summary}. While reasonable, we believe these results to be less relevant, as preparation time is logged by the advisor and does not measure actual interaction time between an advisor and firm. The last specification in Table \ref{tab:iv_regression_prep} shows the effect of change in distance on the sum of preparation and contact time; these results are consistent with (and moderately larger in magnitude than) the results in Table \ref{tab:iv_regression_summary}, reinforcing the relative importance of firm-advisor contact time vs. advisor preparation time.
	
	\subsection{Second-Stage Results}
	The second-stage results show the impact of changes in number of consulting hours on expected firm-level annual revenue and employment, estimated at the industry-county-year level. The primary specifications show average annual revenue increases approximately 5.1\% for every 1 additional consulting hour, while average employment increases 1.7\% for each additional consulting hour.\footnote{Due to the difference in structure of the residuals between the first- and second-stage estimations
		, we compute cluster bootstrapped standard errors by industry-county in the second-stage estimations (\citealt{freedman1984bootstrapping}; \citealt{davidson2008bootstrap}).} For the median firm in the sample, which has an annual revenue (employment) of \$18,200 (1.5 employees), one additional consulting hour generates approximately \$928 in expected additional annual revenue (in 2023 USD) and 0.025 expected additional hires, respectively.\footnote{Taken together, these estimates imply an expected annual wage of \$37,120 (in 2023 USD) for full-time employees, which may be reasonable in this setting.} Restricting the sample to firms experiencing a non-zero change in distance suggests even larger magnitude impacts for the median treated firm (Appendix Table \ref{tab:table2_summary_stats_treated}). Note that these estimates should be interpreted on the intensive margin, namely the effect of increases in consulting time for existing firms in the network. 
	
	Tables \ref{tab:alt_specs_secondstage_revenue} and \ref{tab:alt_specs_secondstage_employment} provide several alternate specifications featuring different aggregations and sets of fixed effects for the annual revenue and employment outcomes, respectively. Across these specifications, average annual revenue increases by approximately 3.6--5.2\% and average employment increases by 1.6--2.9\% for every 1 additional consulting hour, which corresponds to \$655--\$946 in additional annual revenue and 0.024--0.043 additional employees for the median firm in the sample.
	
	\subsection{Robustness}\label{section:robustness}
	
	This section assesses the sensitivity of the first- and second-stage results to two important considerations: the unique economic conditions associated with the post-COVID-19 period and the influence of a small number of extremely large changes in distance between firms and corresponding centers. The COVID-19 pandemic facilitated a substantial shift from in-person to virtual interactions, which is of particular relevance for this study environment. The continuous nature of the change-in-distance instrument may also introduce sensitivity to extremely large changes experienced by a relatively small number of firms. 
	
	\begin{table}[h!]
		\begin{center}
			\caption{First-Stage Results Excluding COVID-19 Periods and Change-in-Distance Outliers}
			\label{tab:firststage_robustness_outliers}
			\footnotesize
			
			\hspace*{-0.5cm}
			
			\begin{tabular}{l@{\hspace{31pt}}c@{\hspace{31pt}}c@{\hspace{31pt}}c@{\hspace{31pt}}c}
				\toprule
				
				& \makecell{(1) \\ Pre-COVID \\ Sample} 
				& \makecell{(2) \\ $|\Delta \text{Distance}|$ \\ $< 300$} 
				& \makecell{(3) \\ $|\Delta \text{Distance}|$ \\ $< 200$} 
				& \makecell{(4) \\ $|\Delta \text{Distance}|$ \\ $< 100$} \\
				
				\midrule
				
				$\Delta$ in Distance
				& $-0.0090^{***}$ 
				& $-0.0086^{***}$ 
				& $-0.0131^{***}$ 
				& $-0.0207^{***}$ \\
				
				& $(0.0014)$ 
				& $(0.0024)$ 
				& $(0.0041)$ 
				& $(0.0047)$ \\
				
				\midrule
				
				Observations 
				& $65{,}688$ 
				& $91{,}223$ 
				& $91{,}114$ 
				& $90{,}922$ \\
				
				Std. Dev. of Distance (Miles) 
				& $20.0$ 
				& $19.8$ 
				& $18.5$ 
				& $17.6$ \\
				
				Firm-Fixed Effects 
				& X & X & X & X \\
				
				Center-Fixed Effects 
				& X & X & X & X \\
				
				Year-Fixed Effects 
				& X & X & X & X \\
				
				County-Year Controls 
				&  &  &  &  \\
				
				Clustered Robust SEs 
				& X & X & X & X \\
				
				R$^2$ 
				& $0.606$ 
				& $0.588$ 
				& $0.588$ 
				& $0.588$ \\
				
				F-Statistic 
				& $43.5$ 
				& $12.7$ 
				& $10.3$ 
				& $19.3$ \\
				
				\bottomrule
				
				\multicolumn{5}{l}{\textit{$^{*}p < 0.1$, $^{**}p < 0.05$, $^{***}p < 0.01$}} \\
				
			\end{tabular}
			
		\end{center}
		
		\begin{tablenotes}
			\footnotesize
			\vspace{-2mm}
			
			\item \textit{Note}: This table reports robustness estimates for the primary first-stage specification presented in Column (3) of Table \ref{tab:iv_regression_summary}. Column (1) restricts the sample to the pre-COVID period (2006--2019). Columns (2)--(4) progressively omit clients experiencing large absolute changes in distance between firms and corresponding SBDC centers. Specifically, clients are excluded if they experience any absolute change in distance greater than 300, 200, or 100 miles, respectively. All specifications include firm, center, and year fixed effects, and standard errors clustered at the center level.
			
		\end{tablenotes}
	\end{table}
	
	Table \ref{tab:firststage_robustness_outliers} presents findings for the first-stage estimation that account for these considerations. Column (1) restricts the sample to the pre-COVID period (2006–2019); one can see the point estimate is slightly larger in magnitude than the estimate from the full sample, but not economically different. The increase in magnitude, albeit relatively small, is intuitive, given the COVID-19 pandemic likely reduced in-person interactions. It is also noteworthy that the F-Statistic nearly triples compared to the primary first-stage specification. Columns (2)--(4) progressively exclude firms experiencing absolute changes in distance greater than 300, 200, and 100 miles, respectively. Across all thresholds, the coefficient estimates remain negative and highly statistically significant. In fact, the magnitude of the estimated effect becomes significantly larger as more firms experiencing extreme changes-in-distance are removed, suggesting that moderate changes in proximity may have stronger marginal impacts on consulting intensity than the largest shifts. This finding is reinforced in Section \ref{section:heterogeneity_distancechangemagnitude}.
	
	\begin{table}[h!]
		\centering
		\caption{Second-Stage Robustness Excluding COVID-19 Periods and Change-in-Distance Outliers}
		\label{tab:secondstage_robustness}
		\scriptsize
		
		\begin{adjustbox}{center,margin*=-1.1cm 0cm}
			\begin{tabular}{
					l@{\hspace{8pt}}
					c@{\hspace{8pt}}
					c@{\hspace{8pt}}
					c@{\hspace{8pt}}
					c@{\hspace{14pt}}
					c@{\hspace{8pt}}
					c@{\hspace{8pt}}
					c@{\hspace{8pt}}
					c
				}
				\toprule
				
				& \multicolumn{4}{c}{Annual Revenue} & \multicolumn{4}{c}{Employment} \\
				\cmidrule(lr){2-5} \cmidrule(lr){6-9}
				
				& \makecell{(1) \\ Pre-COVID \\ Sample} 
				& \makecell{(2) \\ $|\Delta \text{Distance}|$ \\ $< 300$} 
				& \makecell{(3) \\ $|\Delta \text{Distance}|$ \\ $< 200$} 
				& \makecell{(4) \\ $|\Delta \text{Distance}|$ \\ $< 100$} 
				& \makecell{(5) \\ Pre-COVID \\ Sample} 
				& \makecell{(6) \\ $|\Delta \text{Distance}|$ \\ $< 300$} 
				& \makecell{(7) \\ $|\Delta \text{Distance}|$ \\ $< 200$} 
				& \makecell{(8) \\ $|\Delta \text{Distance}|$ \\ $< 100$}  \\
				
				\midrule
				
				{Avg. Predicted Contact Time}
				& $0.051^{***}$ 
				& $0.050^{***}$ 
				& $0.052^{***}$ 
				& $0.052^{***}$
				& $0.016^{**}$ 
				& $0.017^{***}$ 
				& $0.017^{***}$ 
				& $0.017^{***}$ \\
				
				& $(0.019)$ 
				& $(0.017)$ 
				& $(0.017)$ 
				& $(0.016)$
				& $(0.006)$ 
				& $(0.006)$ 
				& $(0.006)$ 
				& $(0.006)$ \\
				
				\midrule
				
				Observations 
				& $1{,}683$ 
				& $2{,}381$ 
				& $2{,}379$ 
				& $2{,}374$
				& $2{,}226$ 
				& $2{,}922$ 
				& $2{,}920$ 
				& $2{,}914$ \\
				
				Industry Fixed Effects 
				& X & X & X & X & X & X & X & X \\
				
				County Fixed Effects 
				& X & X & X & X & X & X & X & X \\
				
				Year Fixed Effects 
				& X & X & X & X & X & X & X & X \\
				
				County-Year Controls 
				& X & X & X & X & X & X & X & X \\
				
				Bootstrapped SEs 
				& X & X & X & X & X & X & X & X \\
				
				R$^2$ 
				& $0.314$ 
				& $0.301$ 
				& $0.302$ 
				& $0.303$
				& $0.317$ 
				& $0.312$ 
				& $0.311$ 
				& $0.311$ \\
				
				\bottomrule
				
				\multicolumn{9}{l}{\textit{$^{*}p < 0.1$, $^{**}p < 0.05$, $^{***}p < 0.01$}} \\
				
			\end{tabular}
		\end{adjustbox}
		
		\begin{tablenotes}
			\footnotesize
			\vspace{2mm}
			
			\item \textit{Note}: This table reports robustness estimates for the primary second-stage specifications for annual revenue and employment presented in Columns (4) and (5) of Table \ref{tab:iv_regression_summary}, respectively. Columns 1--4 use annual revenue as the outcome and Columns 5--8 use employment. Columns (1) and (5) restrict the sample to the pre-COVID period (2006--2019). Columns (2)--(4) and (6)--(8) progressively omit treated firms experiencing large absolute changes in distance with corresponding centers. Specifically, clients are excluded if they experience any absolute change in distance greater than 300, 200, or 100 miles, respectively. All specifications use the primary industry-county-year aggregation with county-year controls and bootstrapped standard errors clustered at the industry-county level.
			
		\end{tablenotes}
	\end{table}
	
	Table \ref{tab:secondstage_robustness} presents analogous findings for the second-stage estimates. Columns (1)--(4) depict specifications using annual revenue as the outcome, and Columns (5)--(8) use employment. 
	One can see that across all specifications, the estimates are nearly identical to the primary results in Table \ref{tab:iv_regression_summary}.
	
	\subsection{Heterogeneity}\label{section:heterogeneity}
	
	\subsubsection{Urban vs. Rural Firms}
	
	The first avenue of heterogeneity we explore is the extent to which change in distance facilitates differential effects on firm utilization of SBDC advisory services depending on the urbanicity of local areas where firms and associated centers reside.\footnote{Urbanicity, which takes on a value between 0-100\%, is defined according to the 2010 Decennial Census and measures the extent to which a geographic area is considered urban vs. rural. For each 5-digit ZIP code in our sample, we obtain the 2010 Census measure of urbanicity. Appendix Figure \ref{fig:urbanicity_hist} shows the distribution of urbanicity separately for all firms and treated firms in the sample.} While we do not observe whether specific sessions are held virtually or in-person, according to the NorCal SBDC administrative office, centers and firms in more rural areas tend to rely much more heavily on in-person meetings than centers and firms in heavily urban areas.\footnote{According to the NorCal SBDC administrative office, meetings tend to be 80\% virtual vs. 20\% in-person for centers in predominantly urban areas, while the breakdown is closer to 50/50 for centers in more rural areas.} As a result, one might expect that changes in distance are more impactful for firms and centers in more rural areas. Of course, distance may not only be associated with transportation and time costs incurred by meeting with center advisors in person; proximity may be relevant in terms of local knowledge that can be shared by regionally embedded advisors. Exploring differences in impacts across urban vs. rural environments offers additional support for the validity of the IV approach in the primary analysis.
	
	To explore this, we estimate the following modified first-stage specification.
	
	\vspace{-10mm}
	\begin{equation}
		\text{ContactTime}_{ijoct}= \alpha_0+\alpha_1\Delta\text{Distance}_{ijoct}+\alpha_2(\Delta\text{Distance x Urbanicity \%})_{ijoct}+\gamma_i+\sigma_c+\delta_t+\varepsilon_{ijoct} \label{eqn:urbanicity_interacted}
	\end{equation}
	
	Where $\alpha_2$ represents the differential effect of change in distance on contact time between advisors and firms depending on the urbanicity of a firm's 5-digit ZIP code. 
	
	\begin{figure}[h!]
		\centering
		\caption{First-Stage IV Estimates by Urbanicity Level}
		\label{urbanicity_graph}
		\vspace{-4mm}
		\includegraphics[width=0.9\textwidth]{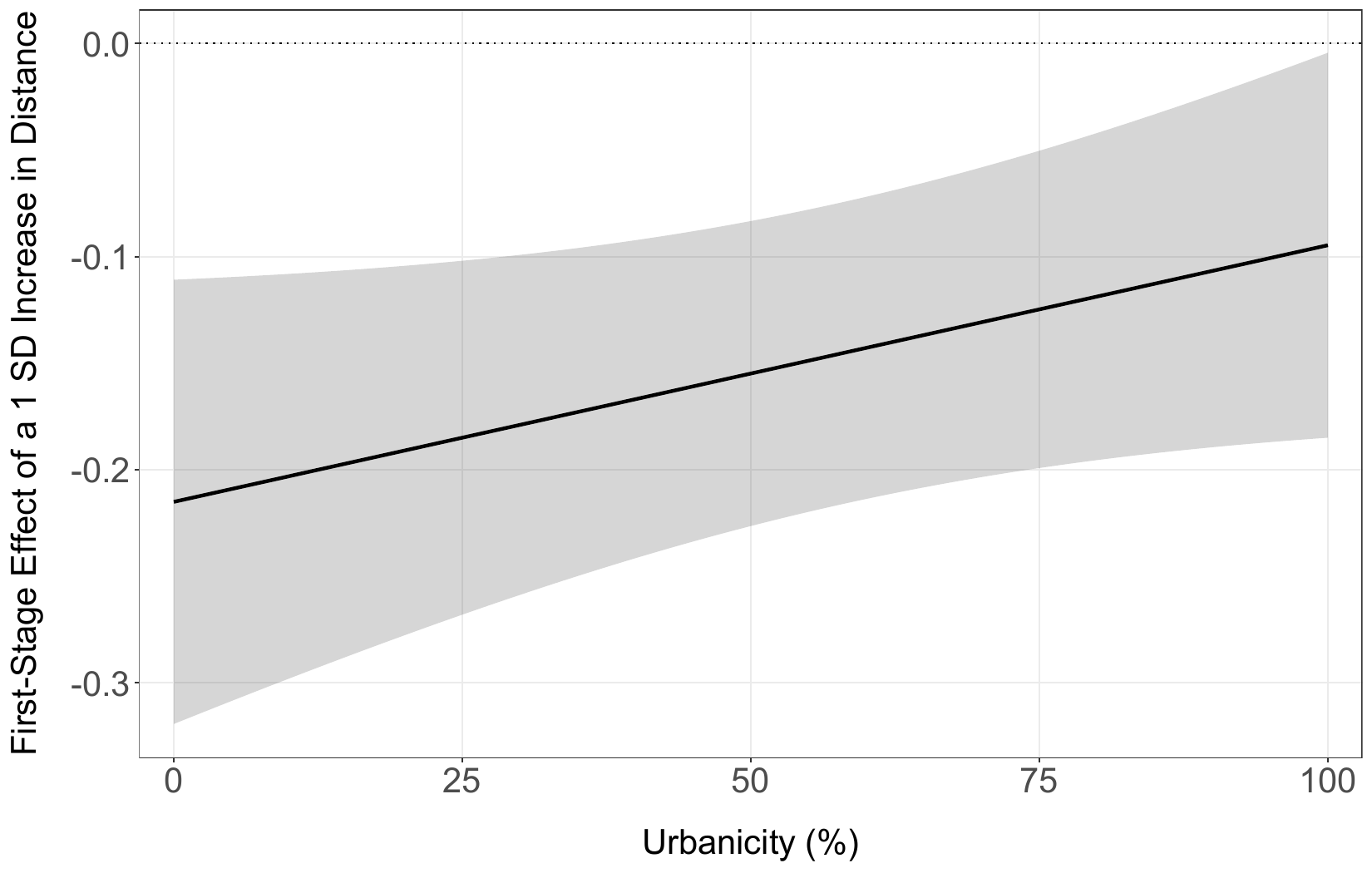} 
		\begin{tablenotes}
			\footnotesize
			\item \textit{Note}: Points represent scaled coefficient estimates from the interacted first-stage specification seen in Equation \ref{eqn:urbanicity_interacted} (scaled by a one standard deviation change in distance, which corresponds to 20 miles). Standard errors and associated 95\% confidence intervals are estimated using the delta method. As indicated in specification (1) of Table \ref{tab:iv_regression_percurban}, the coefficient on the interaction term is significant at the 10\% level.
		\end{tablenotes}
	\end{figure}
	
	To illustrate this effect, Figure \ref{urbanicity_graph} plots the estimated impact of a one standard deviation increase in distance (20 miles) on annual contact time as a function of urbanicity.\footnote{Standard errors and associated 95\% confidence intervals are constructed using the delta method.} The findings suggest that urbanicity mediates the impact of change in distance on contact time between firms and advisors. That is, for firms in completely rural areas, a 1 SD (20 mile) increase in distance from their associated advisory center causes a 0.22 hour expected reduction in annual contact time with advisors ($\approx$11\% for the median firm). In contrast, in completely urban areas, the same increase in distance results in a smaller, but still economically meaningful, reduction of about 0.09 hours ($\approx$5\% for the median firm). Importantly, the effect remains statistically significant at the 5\% level even in fully urban areas, suggesting that geographic proximity continues to matter despite the greater prevalence of virtual interactions, and potentially reflecting the value of localized expertise provided by SBDC advisors.

	Appendix Table \ref{tab:iv_regression_percurban} offers two additional specifications using advisor preparation time and the sum of contact and preparation time as outcomes. The findings offer little evidence that urbanicity differentially affects the relationship between distance and advisor preparation time. This is intuitive, given the time advisors spend preparing for meetings is less likely to depend on whether consulting interactions occur in-person or virtually.
	
	

	\subsubsection{Magnitude of Change in Distance}\label{section:heterogeneity_distancechangemagnitude}
	The primary first-stage results depict the marginal impact of a change in distance between firms and centers on expected contact time, leveraging the entire sample of firm-center observations. However, there is substantial variation among treated firms with respect to the magnitude of changes in distances experienced. The absolute change in distance experienced by a firm may lead to differently-sized marginal effects on expected contact time. This may be due to physical travel considerations when meeting with center advisors, center and advisor time and scheduling constraints, and possible diminishing returns to advisory services. 
	
\begin{figure}[h!]
		\centering
		\caption{First-Stage IV Estimates by Change-in-Distance Bin}
		\label{heterogeneity_graph}
		\begin{subfigure}[t]{0.49\textwidth}
			\centering
			\caption*{Panel A: Marginal Effects}
			\includegraphics[width=\textwidth]{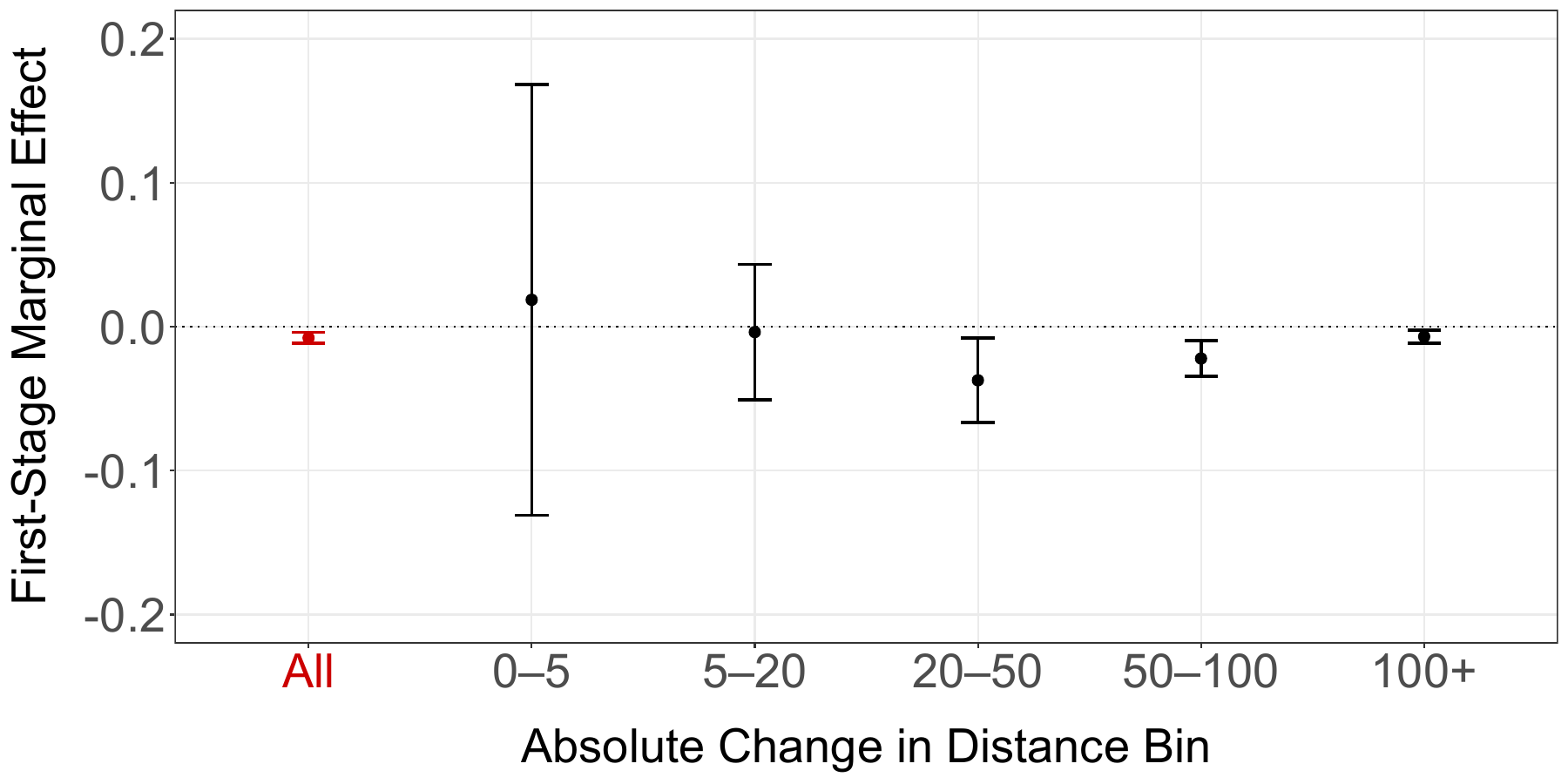}
		\end{subfigure}
		\hfill
		\begin{subfigure}[t]{0.49\textwidth}
			\centering
			\caption*{Panel B: Cumulative Effects}
			\includegraphics[width=\textwidth]{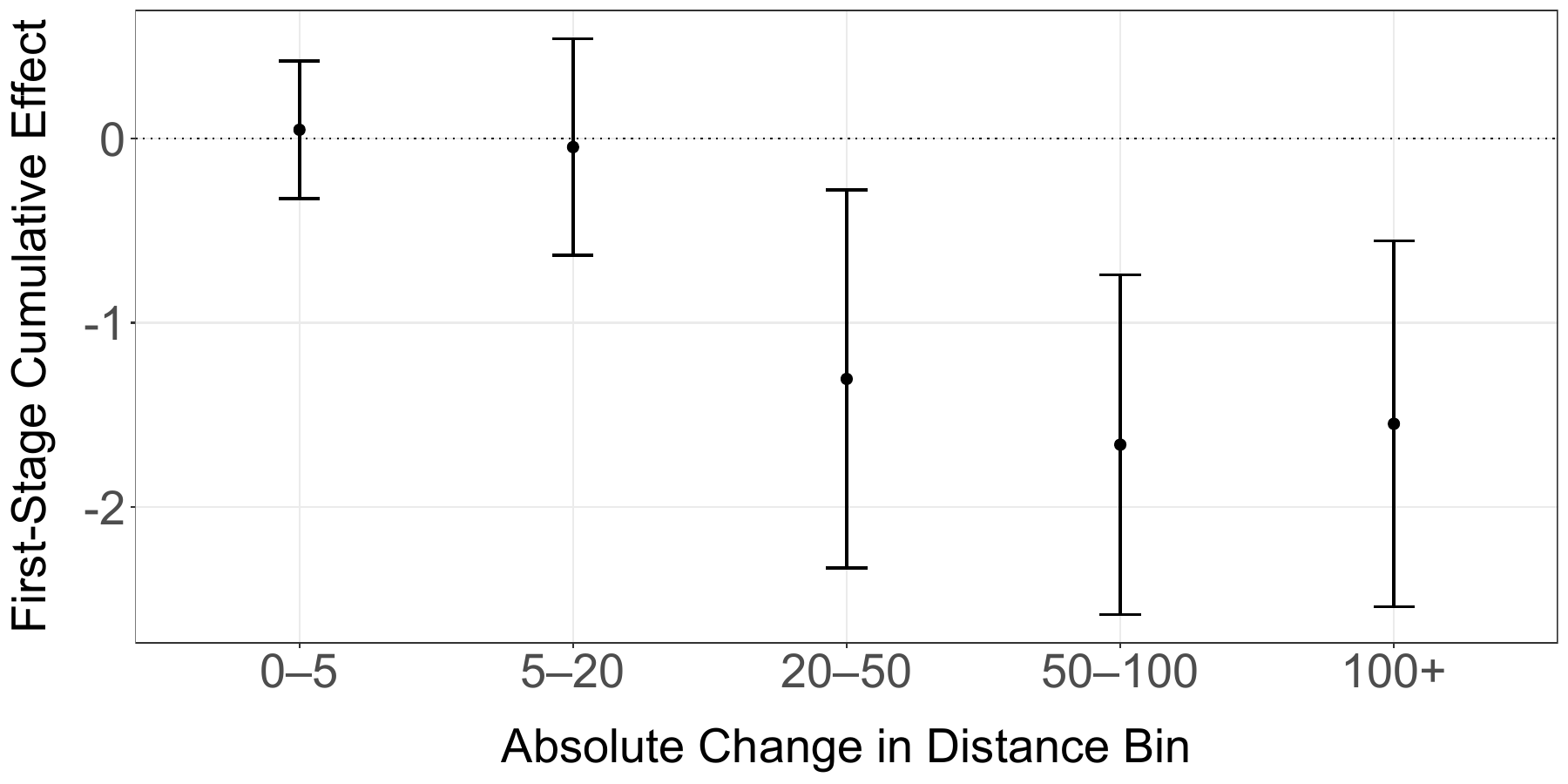}
		\end{subfigure}
		\begin{tablenotes}
			\footnotesize
			\item \textit{Note}: Panel A shows coefficient estimates from the first-stage specification for each change-in-distance bin grouping treated businesses. Confidence intervals are estimated using clustered-robust standard errors. Treated clients with multiple change-in-distance instances are assigned to the bin corresponding to their largest (in magnitude) $\Delta$ in distance. Panel B scales the marginal effect estimates by the midpoint of each absolute change-in-distance bin grouping; the 100+ bin scales by the midpoint of the minimum (100) and maximum (347.34) absolute change in distance values within the bin.
		\end{tablenotes}
	\end{figure}

	To test for such heterogeneity, we subset the data into several change-in-distance bins and estimate separate IV specifications for each. Panel A of Figure \ref{heterogeneity_graph} depicts the marginal effects associated with the first-stage specification, while Panel B depicts the corresponding cumulative effects.\footnote{A more detailed set of numeric results is contained in Table \ref{tab:firststage_heterogeneity}.} Each bin contains observations of treated businesses that experience a change in distance with their corresponding center falling within the bounds of the bin, as well as observations of all businesses never experiencing a change in distance. 
	Thus, bins simply differ by the composition of treated businesses they include. The ``All" estimation in Panel A refers to the primary specification presented in column 3 of Table \ref{tab:iv_regression_summary}.
	
	Panel A indicates there is substantial heterogeneity in the marginal effect of change in distance on predicted annual contact time based on absolute change in distance. The estimates indicate a one-mile change in distance has the largest marginal impact for firms experiencing \textit{moderate} shifts in distance, while both smaller and larger changes yield attenuated effects (and are statistically insignificant for the smallest changes). Panel B complements this by plotting the corresponding cumulative effects, evaluated at the midpoint of each absolute change-in-distance bin.\footnote{The 100+ bin scales the associated marginal effect estimate by the midpoint of the minimum (100) and maximum (347.34) absolute change in distance values within the bin.} These estimates imply that firms experiencing larger overall distance adjustments face substantially greater cumulative changes in annual consulting time.
	
	Taken together, these patterns are intuitive and highlight two distinct mechanisms. First, firms facing small changes in distance are typically already located close to corresponding centers and may be operating near an optimal level of engagement, subject to advisor capacity or time constraints. Consequently, marginal adjustments in distance have limited impact on annual consulting time. Moreover, small changes preserve the existing geographic context of firm–advisor interactions, maintaining access to region-specific knowledge and networks. 
	
	Second, and in contrast, moderate changes generate the largest marginal responses, consistent with firms moving away from this locally-optimal range and encountering increasing frictions in engagement. For the largest changes, while the marginal effect is smaller, Panel B shows that the total impact on annual consulting time is nevertheless substantial, reflecting the scale of the geographic shift. The cumulative estimates suggest that moderate to large adjustments in geographic proximity may effectively move firms outside of their original advisory ecosystem, reducing the relevance of expertise offered. Overall, the results indicate that geographic proximity affects engagement not only through incremental time and travel costs, but also through nonlinear effects tied to the preservation---or loss---of locally embedded advisor knowledge.
	
	
	\begin{figure}[h!]
		\centering
		\caption{Second-Stage Estimates by Change-in-Distance Bin}
		\label{fig:secondstage_graphs}
		
		\begin{subfigure}[t]{0.49\textwidth}
			\centering
			\caption*{Panel A: Annual Revenue}
			\includegraphics[width=\textwidth]{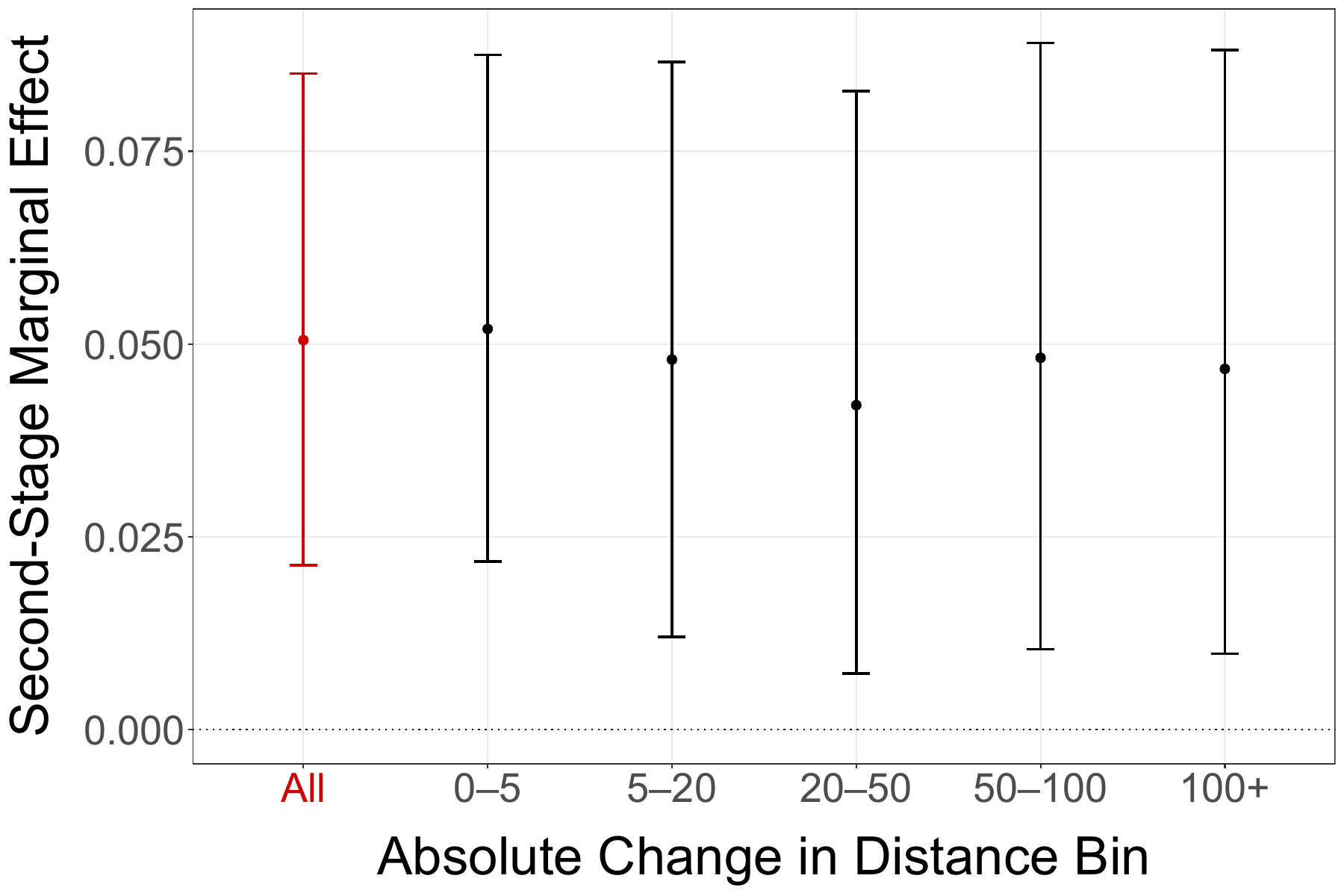}
		\end{subfigure}
		\hfill
		\begin{subfigure}[t]{0.49\textwidth}
			\centering
			\caption*{Panel B: Employment}
			\includegraphics[width=\textwidth]{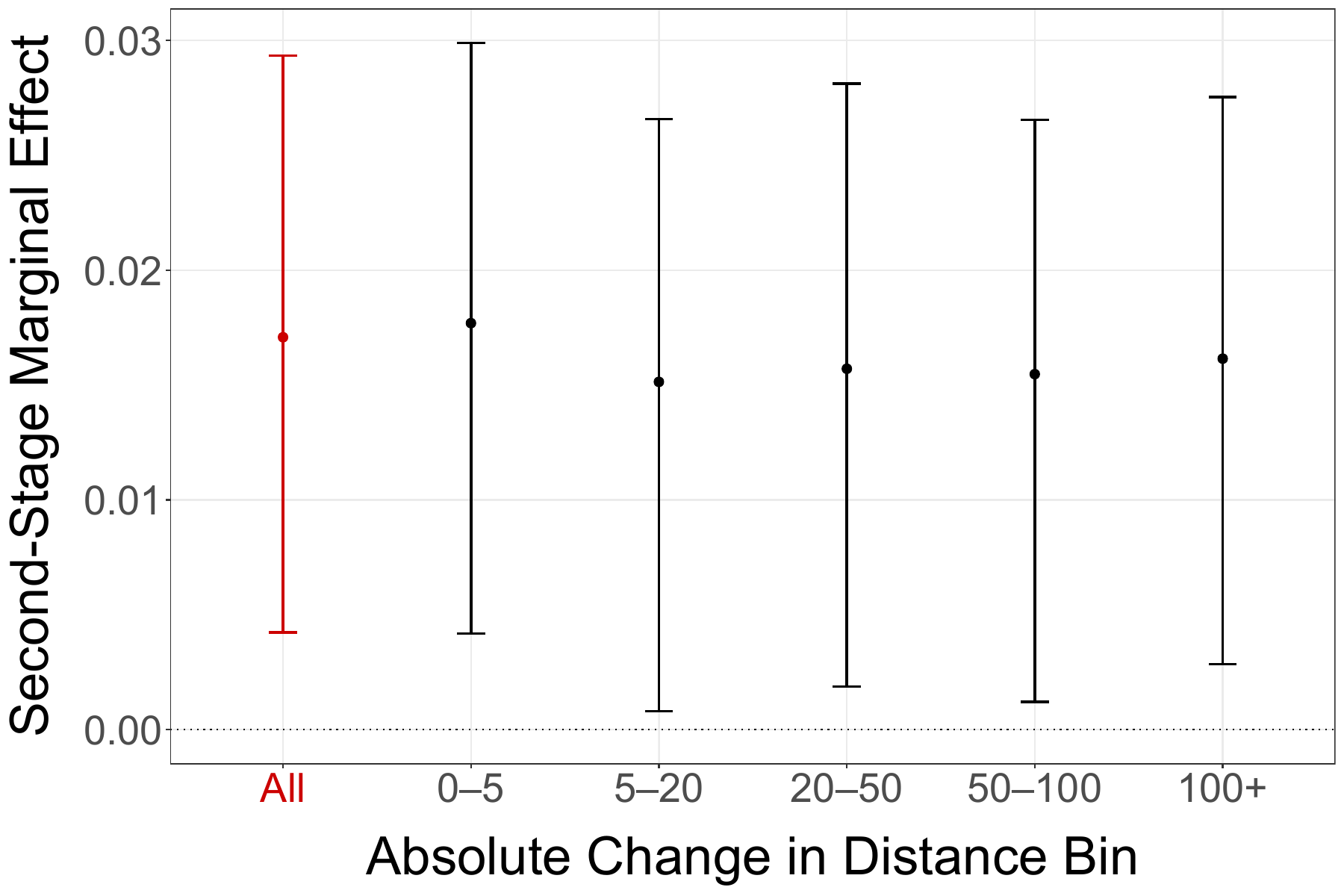}
		\end{subfigure}
		
		\begin{tablenotes}
			\footnotesize
			\item \textit{Note}: Points represent coefficient estimates from the second-stage specification for each change-in-distance bin, using predicted contact time from the first-stage regressions. Confidence intervals are estimated using bootstrapped standard errors. Note the difference in y-axis scales for each panel. Treated clients with multiple change-in-distance instances are assigned to the bin corresponding to their \underline{largest} (in magnitude) $\Delta$ in distance.
		\end{tablenotes}
	\end{figure}

	Figure \ref{fig:secondstage_graphs} depicts the second-stage results for annual revenue (Panel A) and employment (Panel B) under each change-in-distance bin first-stage specification. The point estimates and significance are stable and similar to the primary estimation results in columns 4 and 5 of Table \ref{tab:iv_regression_summary}. This suggests the effect of contact time on firm annual revenue and employment is not sensitive to the source of variation in change-in-distance from the first-stage estimation.

	\section{Discussion}
	Our findings show that closer geographic proximity increases consulting time between firms and advisors, which in turn leads to higher annual revenue and employment for firms working with the NorCal SBDC Network. These results highlight that physical proximity remains an important determinant of firm-advisor engagement, even in an increasingly digital environment. We also find the effects of proximity are more pronounced in relatively rural areas, consistent with greater reliance on in-person interactions. 
	
	At the same time, the importance of distance likely reflects more than time and travel costs alone, pointing to the value of localized knowledge embedded in center-firm interactions. The urban–rural results support this interpretation: even in more urban areas, where virtual interactions are more common, changes in distance continue to affect consulting time. Moreover, the finding that moderate-to-large changes in distance generate the largest marginal and cumulative impacts on consulting time suggests that such shifts may disrupt access to locally-embedded centers and advisors
	, whereas small changes are less likely to alter the underlying informational environment. A broader literature similarly emphasizes the role of geographic proximity in facilitating knowledge transmission. \cite{jaffe1993geographic} examines the geographic locations of patent development and citations of those patents, finding that citations are more concentrated in areas proximal to where patents are developed. \cite{jarafigueroa:etal2018} finds that access to location-specific knowledge significantly improves the growth and survival of pioneer firms.\footnote{\cite{jarafigueroa:etal2018} define pioneer firms as ``firms operating in an industry that was not present in a region."} Our findings indicate that physical access may also represent informational access and relevant advisory expertise.\footnote{
		Such expertise is also critical in creating a more even playing field between smaller and larger firms operating in heavily bureaucratic policy environments. Large firms tend to have more internal resources to navigate ``red tape" as well as identify and take advantage of favorable policy loopholes (\citealt{welsh:white1981}).}

	The second primary set of findings suggests that increased consulting time with advisors leads to improved firm performance. Specifically, each additional hour of consulting annually generates an approximately 3.6--5.2\% and 1.6--2.9\% increase in average firm annual revenue and employment, respectively. Given that firms typically engage in relatively few consulting hours annually, even modest increases in engagement 
	can improve business performance. These effects may reflect both direct benefits to firms, such as improved planning, financial management, and regulatory guidance, as well as complementary, indirect benefits. For instance, more consulting time may facilitate introductions to separate professional networks or financing opportunities
	. More broadly, additional engagement with advisors may facilitate a stronger economic environment for small businesses owners and entrepreneurs to exchange ideas, leverage resources, and find support within the local community.
	
	From a policy perspective, these findings suggest public advisory services can significantly impact entrepreneurship and small business growth, particularly for businesses that may be less competitive in accessing private resources. Additionally, the size of the impacts we estimate suggest these advisory services are cost-effective. The 2023-24 operating budget for the NorCal SBDC network was \$10.5 million, and according to our data, delivered  59,408 total hours of advisor preparation and consulting time to clients. This suggests each advisor-hour costed \$230.60 on average (in 2023 USD), while our findings indicate that each consulting hour delivered an additional \$655--\$946 in expected additional annual revenue (in 2023 USD) and 0.024--0.043 expected additional hires. In fact, due to the SBDC network funding model requiring matching donor funds for each federal dollar, the taxpayer funded cost for each advisor-hour was \$115.30 on average.
	
	The costs associated with maintaining the network of SBDCs appear to deliver economically meaningful benefits in the form of higher revenues and employment of participating small businesses, suggesting public investments in advisory services may ultimately benefit the broader economy through tax revenue and job creation. 
	In fact, this study likely underestimates the economic impacts of these investments, since it does not consider the external benefits SBDC clients may provide to businesses that do not work directly with SBDCs. The findings also indicate the importance of both a dense network of centers and locating centers in under-served areas. A denser center network facilitates more geographically localized expertise, allowing businesses to select into the most topically relevant center. Increased accessibility of centers, particularly in under-served areas, is expected to enhance and create less costly interactions between centers and businesses. 
	
	\section{Conclusion}
	The findings in this study reflect the importance of both a dense network of small business advisory centers and locating centers in under-served areas. A denser center network facilitates closer geographic proximity to firms and more localized expertise.
	Increased engagement between centers and advisors, particularly in under-served areas, is expected to enhance firm performance in a cost-effective manner.

	This work contributes to a relatively sparse literature examining the impact of publicly funded business advisory services levied in the US and wealthier economies more generally. 
	Several avenues for future research could further our understanding of the effectiveness of SBDCs and publicly funded business advisory services. First, richer longitudinal data on firm outcomes would enable a more comprehensive assessment of impacts across a broader set of economic dimensions, including costs, financing activity, and capital access. Tracking firms beyond their tenure in the SBDC network would also allow for a clearer evaluation of longer-run effects. Second, exploring heterogeneity across industries could help identify where advisory services generate the greatest returns. Finally, examining broader local economic impacts would shed light on potential spillovers, including benefits to non-participating firms and the surrounding business environment.

	\clearpage
	
	\bibliographystyle{aea}
	\bibliography{references2}

	\clearpage
	
	\appendix
	\section{Appendix}
	\renewcommand{\thefigure}{A.\arabic{figure}}
	\renewcommand{\thetable}{A.\arabic{table}}
	\renewcommand{\theequation}{A.\arabic{equation}}
	\setcounter{figure}{0}
	\setcounter{table}{0}
	\setcounter{equation}{0}
	
	\subsection{Additional Information About the NorCal SBDC Network}\label{AdditionalDetailSBDC}
	
	This section provides additional detail regarding the NorCal SBDC network and services it provides. Specifically, we detail the process by which a business owner joins the network and services available to firms, the format of consulting sessions, and additional detail regarding how the advisor-firm relationship evolves over time. These institutional details offer important context for interpreting the empirical analysis in the paper.

	\subsubsection{Joining the NorCal SBDC Network}
	
	This section describes the process by which business owners enter the NorCal SBDC network and come to engage with a specific center. The modal SBDC client is an inexperienced small business owner in the early stages of building a business, often lacking critical experience and management expertise. In the initial stages of building a business, owners typically take on the duties of accountant, marketer, lawyer, and product developer simultaneously. External help is highly valuable at this stage, yet new business owners (particularly ones that select into working with the SBDC network) typically lack sufficient funds to hire employees or contract a paid consultant. 
	
	The process by which each business initiates their engagement with the SBDC network begins with an online form detailing basic information about the business. An on-boarding agent at the network's Lead Center then conducts an initial 30-minute meeting to assess the businesses stage of development and needs. The purpose of the meeting is twofold. First, the agent wants to assess whether the stage of the business and associated needs are appropriate for the services SBDCs provide.\footnote{Advisors will not brainstorm with clients on a “potential” idea.}  
	Second, the agent wants to match the business owner with the most suitable advisor at a local center.\footnote{Advisors can and often do specialize in multiple topics. The matching process aims to align the business owner's needs with the area of expertise that an advisor offers, such as marketing, financial planning, or regulatory navigation.} This on-boarding structure is reflected in the data: many businesses initially register contact time with the Lead Center before being reassigned to a local center, which motivates our decision to omit contact time associated with the Lead Center (see Section \ref{Data}).
	
	Once a business owner is on-boarded into the network and assigned a center and advisor, there are several types of services they can access. The first is individual advisory services, which are the primary focus of this paper. The structure of individual advisory services is detailed in the next subsection. The second is training and workshops. SBDCs offer general classes on topics like bookkeeping or developing a marketing plan, designed specifically for the types of early-stage business owners described previously. These group workshops can also facilitate additional interactions with SBDC advisors. Third is access to the center's referral network. Individual SBDCs actively engage with local chambers of commerce, government offices, and other nonprofit organizations to market their services, and while these sources can refer businesses to the SBDC network itself, they can also serve as resources to existing SBDC clientele. Finally, SBDCs occasionally host professional networking social events, facilitating additional interactions among SBDC clients and with the local business community. 
	

	\subsubsection{Structure of Consulting Sessions and the Advisor-Firm Relationship}
	
	Before describing the specific consulting relationship between SBDC firms and advisors, it is useful to understand the composition of the advisor pool at a typical center. Centers tend to feature two generalist advisors that can consult across a broad range of topics, alongside several additional advisors with more specialized backgrounds.\footnote{NorCal SBDCs typically employ an average of 10 advisors, though this can range from 5 to 15. Beyond center-level advisors, the NorCal network as a whole maintains a small set of highly specialized advisors, such as a cybersecurity expert, who can be utilized by any center as needed.} The subcontract model used by the NorCal SBDC Network gives center directors considerable flexibility in shaping this advisor pool.\footnote{As noted in Section \ref{SBDC_Background}, contracted advisors are atypical of SBDC networks nationally, where the majority of advisors are full-time employees.} Directors actively manage the composition of their advisors over time, recruiting individuals to match the specific needs of their local client base. For instance, a director who frequently serves prospective food truck owners can bring on an advisor with previous food truck experience. At the same time, center directors across the network share a common preference: they look for local members of the community who have themselves started a business. 
	
	Despite varying backgrounds and expertise, advisors across the network share a common approach to consulting, grounded in a philosophy of teaching rather than doing. Advisors aim to provide the necessary tools that allow business owners to build components independently. 
	In addition to this shared philosophy among advisors, the SBDC network provides a standardized toolkit and set of resources that advisors can pair with their own personal experiences when working with clients. Advisors routinely share information with one another to refine their approaches and improve the quality of services delivered to business owners.
	
	The first meeting in an advisor-client relationship typically follows a consistent structure across pairings, mirroring the initial on-boarding process. Its primary purpose is to establish a scope of work between the business and advisor, identifying the central questions the business owner faces and determining the organization of subsequent sessions to address them. After this initial meeting, the trajectory of the advisor-client relationship varies substantially from one pairing to the next, as advisors tailor sessions to the specific, evolving needs of each business. Thus, sessions are typically structured in ``bunches," moving the business owner from one plateau to the next. Each session typically lasts no longer than one hour and occurs periodically. 

	There is considerable variation in the frequency of consulting sessions across firms, and the advisor-client relationship can end for several reasons. One common occurrence is that business owners simply become too busy running day-to-day operations to maintain regular sessions. Advisors, too, face time constraints, as most work for the center on a part-time basis. Sessions may also end when the business owner has accomplished goals associated with their engagement in the network. Through the advisory process, owners often build their own professional networks facilitated by SBDC introductions to local contacts and service providers, and eventually outgrow the program or gain sufficient resources to pay for private services (e.g. an accountant). 
	Third, though rarely, the relationship can end because the specific business is headed towards failure, and the advisor determines there is no productive future in continuing.\footnote{The extreme rarity of advisor-initiated terminations suggests this channel is unlikely to introduce substantial selection bias in the data.} Generally, and in congruence with the SBDC model, an advisor's goal is to guide the business to a point where it no longer needs SBDC advisory resources.

	\clearpage
	
	\subsection{Additional Tables and Figures}
	
	\renewcommand{\cellset}{\renewcommand{\arraystretch}{0.65}}
	\begin{center}
		\scriptsize
		\setlength{\tabcolsep}{2.9pt}
		\renewcommand{\arraystretch}{0.65}
		
		\captionof{table}{Center-Specific Descriptive Statistics}
		\label{tab:center_characteristics_compact}
		\hspace*{-2.1cm}
		\begin{tabular}{cccccccccccccc}
			\toprule
			\makecell{Center \\ ID} & \makecell{First \\ Year} & \makecell{Last \\ Year} & \makecell{\# \\ Firms} & \makecell{\# \\ Firm-Years} & \makecell{Contact \\ Time (Hrs)} & \makecell{Prep. \\ Time (Hrs)} & \makecell{Tenure \\ (Yrs)} & \makecell{Firm Age \\ (Yrs)} & \makecell{Avg. Annual \\ Revenue (\$)} & \makecell{Median Annual \\ Revenue (\$)} & \makecell{Employment} & \makecell{\# Annual \\ Sessions} & \makecell{Distance \\ (Miles)} \\
			\midrule
			1 & 2006 & 2023 & 3,829 & 6,103 & 4.35 & 1.38 & 1.61 & 8.68 & 354,142.50 & 3,863.21 & 4.76 & 3.35 & 9.55 \\
			2 & 2016 & 2023 & 5,497 & 8,726 & 3.95 & 0.82 & 1.76 & 9.34 & 162,397.30 & 32,638.38 & 3.52 & 3.48 & 5.94 \\
			3 & 2006 & 2009 & 1,681 & 1,919 & 2.86 & 1.45 & 1.36 & 9.02 & 584,720.18 & 12,319.06 & 2.25 & 2.23 & 60.51 \\
			4 & 2010 & 2015 & 2,659 & 3,352 & 3.76 & 1.98 & 1.71 & 8.44 & 138,150.96 & 5,238.30 & 3.90 & 3.26 & 9.43 \\
			5 & 2022 & 2023 & 132 & 171 & 6.74 & 3.49 & 2.21 & 7.67 & 3,017,020.09 & 450,184.50 & 13.64 & 5.41 & 9.10 \\
			6 & 2015 & 2015 & 488 & 488 & 5.90 & 1.79 & 2.57 & 10.14 & 371,182.27 & 371,182.27 & 5.50 & 4.99 & 7.53 \\
			7 & 2013 & 2016 & 301 & 389 & 3.99 & 1.38 & 1.88 & 8.32 & 160,688.84 & 104,796.32 & 1.10 & 4.40 & 14.08 \\
			8 & 2019 & 2021 & 1,351 & 1,723 & 2.77 & 1.04 & 1.53 & 7.71 & 403,186.34 & 335,500.00 & 8.62 & 3.12 & 19.88 \\
			9 & 2017 & 2019 & 319 & 407 & 3.37 & 1.11 & 1.59 & 7.96 & 0.00 & 0.00 & 2.00 & 4.00 & 11.72 \\
			10 & 2022 & 2023 & 65 & 71 & 4.04 & 2.16 & 1.29 & 8.51 & 132,934.20 & 74,843.17 & 4.25 & 3.77 & 10.45 \\
			11 & 2022 & 2023 & 356 & 381 & 3.91 & 2.26 & 1.80 & 9.07 & 965,256.14 & 156,000.00 & 8.36 & 3.54 & 14.94 \\
			12 & 2019 & 2022 & 2,527 & 3,784 & 2.83 & 1.49 & 1.89 & 9.23 & 903,059.20 & 130,391.72 & 4.97 & 3.55 & 3.04 \\
			13 & 2023 & 2023 & 332 & 332 & 2.96 & 1.18 & 2.53 & 10.56 & 600,887.19 & 145,654.98 & 4.95 & 3.42 & 3.37 \\
			14 & 2016 & 2018 & 824 & 1,115 & 5.37 & 1.98 & 2.66 & 11.42 & 708,553.19 & 63,943.83 & 4.19 & 4.82 & 2.88 \\
			15 & 2013 & 2015 & 627 & 781 & 5.60 & 2.22 & 2.33 & 10.88 & 376,231.57 & 43,215.97 & 3.68 & 4.28 & 3.82 \\
			16 & 2006 & 2012 & 2,830 & 3,884 & 5.67 & 1.86 & 1.56 & 8.73 & 662,986.30 & 0.00 & 5.28 & 3.77 & 7.99 \\
			17 & 2009 & 2015 & 2,403 & 2,951 & 4.17 & 1.81 & 1.41 & 8.36 & 1,307,212.62 & 23,394.73 & 4.65 & 2.88 & 24.44 \\
			18 & 2006 & 2008 & 777 & 926 & 3.51 & 2.64 & 1.40 & 7.42 & 11,301.32 & 0.00 & 1.17 & 2.66 & 44.66 \\
			19 & 2016 & 2017 & 847 & 916 & 3.92 & 1.18 & 1.45 & 8.09 & 31,778.25 & 24,228.30 & 1.12 & 2.69 & 14.41 \\
			20 & 2018 & 2021 & 1,143 & 1,622 & 1.57 & 1.15 & 1.60 & 8.25 & 263,550.92 & 263,550.92 & 1.75 & 2.74 & 11.83 \\
			21 & 2010 & 2017 & 3,927 & 5,910 & 2.58 & 1.44 & 1.72 & 8.67 & 77,249.77 & 2,790.82 & 2.00 & 3.56 & 31.38 \\
			22 & 2013 & 2015 & 542 & 690 & 4.23 & 1.54 & 2.29 & 9.95 & 166,639.84 & 2,795.83 & 2.21 & 3.85 & 6.40 \\
			23 & 2016 & 2023 & 2,749 & 4,330 & 3.84 & 1.16 & 1.72 & 9.13 & 422,673.29 & 58,447.40 & 3.46 & 3.74 & 6.97 \\
			24 & 2006 & 2009 & 1,736 & 2,069 & 3.27 & 1.58 & 1.48 & 8.21 & 46,409.08 & 36,359.47 & 3.33 & 2.49 & 177.08 \\
			25 & 2018 & 2019 & 920 & 1,043 & 2.15 & 0.94 & 1.80 & 9.74 & 17,500.00 & 17,500.00 & 1.00 & 2.97 & 20.51 \\
			26 & 2008 & 2016 & 1,764 & 2,550 & 3.13 & 1.55 & 1.63 & 8.56 & 203,511.72 & 0.00 & 2.57 & 3.04 & 13.31 \\
			27 & 2006 & 2007 & 202 & 237 & 2.75 & 1.21 & 1.92 & 8.85 & 172,456.02 & 523.83 & 3.31 & 3.21 & 17.37 \\
			28 & 2017 & 2017 & 135 & 135 & 3.23 & 0.62 & 2.06 & 9.18 & 374,079.77 & 0.00 & 3.57 & 2.13 & 9.59 \\
			29 & 2019 & 2023 & 633 & 951 & 3.77 & 1.33 & 2.23 & 10.64 & 4,330,985.63 & 40,069.55 & 6.34 & 4.17 & 11.37 \\
			30 & 2018 & 2018 & 99 & 99 & 3.64 & 0.99 & 2.54 & 10.97 & 288,105.83 & 14,575.87 & 3.21 & 3.61 & 10.36 \\
			31 & 2013 & 2021 & 3,240 & 5,041 & 3.92 & 1.21 & 1.96 & 9.44 & 439,859.75 & 75,000.00 & 3.22 & 3.48 & 21.39 \\
			32 & 2006 & 2012 & 4,351 & 6,078 & 4.02 & 1.70 & 1.67 & 8.91 & 828,451.39 & 56,844.00 & 3.80 & 3.23 & 44.70 \\
			33 & 2022 & 2023 & 219 & 249 & 3.37 & 1.30 & 2.32 & 9.85 & 570,711.22 & 316,530.44 & 4.50 & 3.45 & 9.57 \\
			34 & 2022 & 2023 & 273 & 289 & 2.59 & 0.77 & 2.15 & 8.98 & 182,444.87 & 99,025.97 & 1.83 & 2.81 & 44.32 \\
			35 & 2022 & 2023 & 3 & 3 & 3.33 & 0.67 & 1.00 & 5.33 & 150,000.00 & 150,000.00 & 2.00 & 2.00 & 67.79 \\
			36 & 2018 & 2021 & 2,779 & 3,585 & 4.39 & 1.31 & 1.44 & 8.27 & 556,782.46 & 82,500.00 & 2.87 & 3.98 & 19.30 \\
			37 & 2022 & 2023 & 373 & 461 & 3.00 & 1.02 & 2.31 & 9.62 & 434,199.94 & 57,883.63 & 2.96 & 3.14 & 13.27 \\
			38 & 2018 & 2023 & 1,775 & 2,560 & 3.47 & 1.05 & 1.50 & 8.59 & 762,916.83 & 125,085.44 & 5.42 & 3.18 & 42.65 \\
			39 & 2019 & 2023 & 806 & 1,177 & 4.29 & 1.91 & 1.78 & 9.11 & 286,323.34 & 66,250.72 & 4.10 & 3.86 & 27.82 \\
			40 & 2007 & 2018 & 1,067 & 1,456 & 4.71 & 2.24 & 1.62 & 8.43 & 129,185.10 & 0.00 & 2.68 & 3.85 & 24.57 \\
			41 & 2018 & 2023 & 1,691 & 3,130 & 2.56 & 1.29 & 1.86 & 9.52 & 1,293,528.90 & 149,326.81 & 4.85 & 3.69 & 14.55 \\
			42 & 2018 & 2020 & 583 & 732 & 2.97 & 0.87 & 1.69 & 8.84 & 510,368.37 & 90,285.34 & 3.07 & 3.05 & 13.48 \\
			43 & 2021 & 2023 & 561 & 702 & 2.18 & 0.82 & 1.82 & 9.01 & 459,495.29 & 110,000.14 & 3.07 & 2.65 & 15.20 \\
			44 & 2006 & 2022 & 3,580 & 5,733 & 3.49 & 1.77 & 1.64 & 8.78 & 291,329.87 & 9,601.66 & 3.26 & 2.90 & 13.69 \\
			46 & 2023 & 2023 & 61 & 61 & 2.07 & 1.21 & 3.02 & 11.15 & 468,645.29 & 102,403.28 & 2.69 & 2.34 & 10.39 \\
			48 & 2018 & 2023 & 289 & 466 & 4.08 & 1.99 & 1.70 & 9.03 & 340,550.64 & 6,101.27 & 2.00 & 4.41 & 44.97 \\
			49 & 2006 & 2015 & 922 & 1,630 & 4.98 & 1.70 & 1.84 & 9.72 & 76,015.71 & 0.00 & 0.78 & 3.45 & 14.29 \\
			50 & 2016 & 2017 & 115 & 119 & 2.22 & 0.72 & 1.70 & 8.47 & 68,632.50 & 1,270.83 & 1.68 & 2.28 & 22.49 \\
			\bottomrule
		\end{tabular}
		
		\vspace{-2mm}
		
		\begin{tablenotes}
			\footnotesize
			\singlespacing
			\textit{Note}: This table reports center-level descriptive statistics for associated client firms. Each row corresponds to a distinct center. Center IDs 45 and 47 refer to the NorCal SBDC lead center (which consists of two separate center IDs based on network organizational shifts) and are excluded.
		\end{tablenotes}
		
	\end{center}
	
	\clearpage
	
	\begin{table}[h!]
		\centering
		\footnotesize
		\caption{Summary Statistics of County-Year US Census CBP Characteristics}
		\label{tab:cbp_summary_stats}
		\hspace*{-0.5cm}
		\begin{threeparttable}
			\begin{tabular}{lclllll}
				\toprule
				Variable & \makecell{Number of \\ Observations} & Mean & SD & Median & Min & Max \\
				\midrule
				
				\multicolumn{7}{l}{\textit{County-Year CBP Characteristics}} \\
				\midrule
				
				Total Establishments 
				& 828 & 7,480 & 10,800 & 3,100 & 154 & 34,200 \\
				
				
				Annual Payroll per Establishment (\$1000s)
				& 828 & 605 & 524 & 482 & 203 & 1,460 \\
				
				\midrule
				\multicolumn{7}{l}{\textit{Establishment Size Shares}} \\
				\midrule
				
				1--4 Employees
				& 828 & 0.570 & 0.061 & 0.554 & 0.499 & 0.686 \\
				
				5--9 Employees
				& 828 & 0.192 & 0.019 & 0.194 & 0.165 & 0.216 \\
				
				10--19 Employees
				& 828 & 0.123 & 0.022 & 0.127 & 0.081 & 0.148 \\
				
				20--49 Employees
				& 828 & 0.077 & 0.022 & 0.082 & 0.031 & 0.105 \\
				
				50--99 Employees
				& 828 & 0.022 & 0.010 & 0.023 & 0.000 & 0.035 \\
				
				100--249 Employees
				& 828 & 0.011 & 0.005 & 0.011 & 0.000 & 0.019 \\
				
				250--499 Employees
				& 828 & 0.003 & 0.002 & 0.002 & 0.000 & 0.005 \\
				
				500--999 Employees
				& 828 & 0.001 & 0.002 & 0.001 & 0.000 & 0.002 \\
				
				1,000+ Employees
				& 828 & 0.0004 & 0.0005 & 0.0003 & 0.000 & 0.001 \\
				
				\bottomrule
			\end{tabular}
			\begin{tablenotes}[flushleft]
				\footnotesize
				\singlespacing
				\vspace{-4mm}
				\item \textit{Note}: This table reports summary statistics for county-year observations from the US Census County Business Patterns (CBP) data across Northern California counties from 2006--2023. Establishment size shares are calculated as the proportion of total establishments within each county-year belonging to the corresponding employment-size category.
			\end{tablenotes}
		\end{threeparttable}
	\end{table}
	
	\clearpage
	
	\begin{figure}[h!]
		\centering
		\caption{Firm Performance Outcomes by Length of Tenure with SBDC Network}
		\label{fig:firm_rev_emp_violinstats}
		
		\begin{subfigure}[t]{0.49\textwidth}
			\centering
			\caption*{Panel A: Annual Revenue}
			\includegraphics[width=\textwidth]{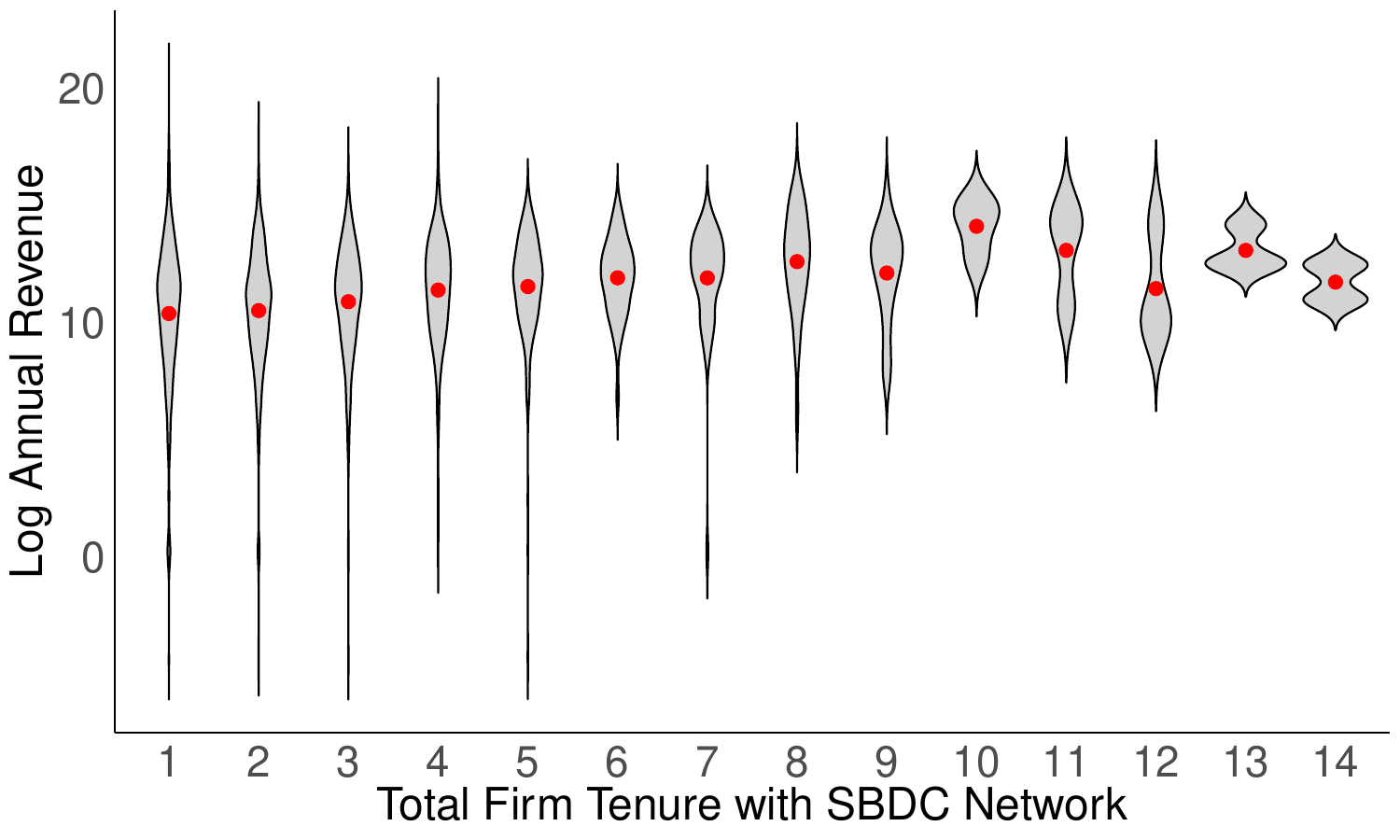}
		\end{subfigure}
		\hfill
		\begin{subfigure}[t]{0.49\textwidth}
			\centering
			\caption*{Panel B: Employment}
			\includegraphics[width=\textwidth]{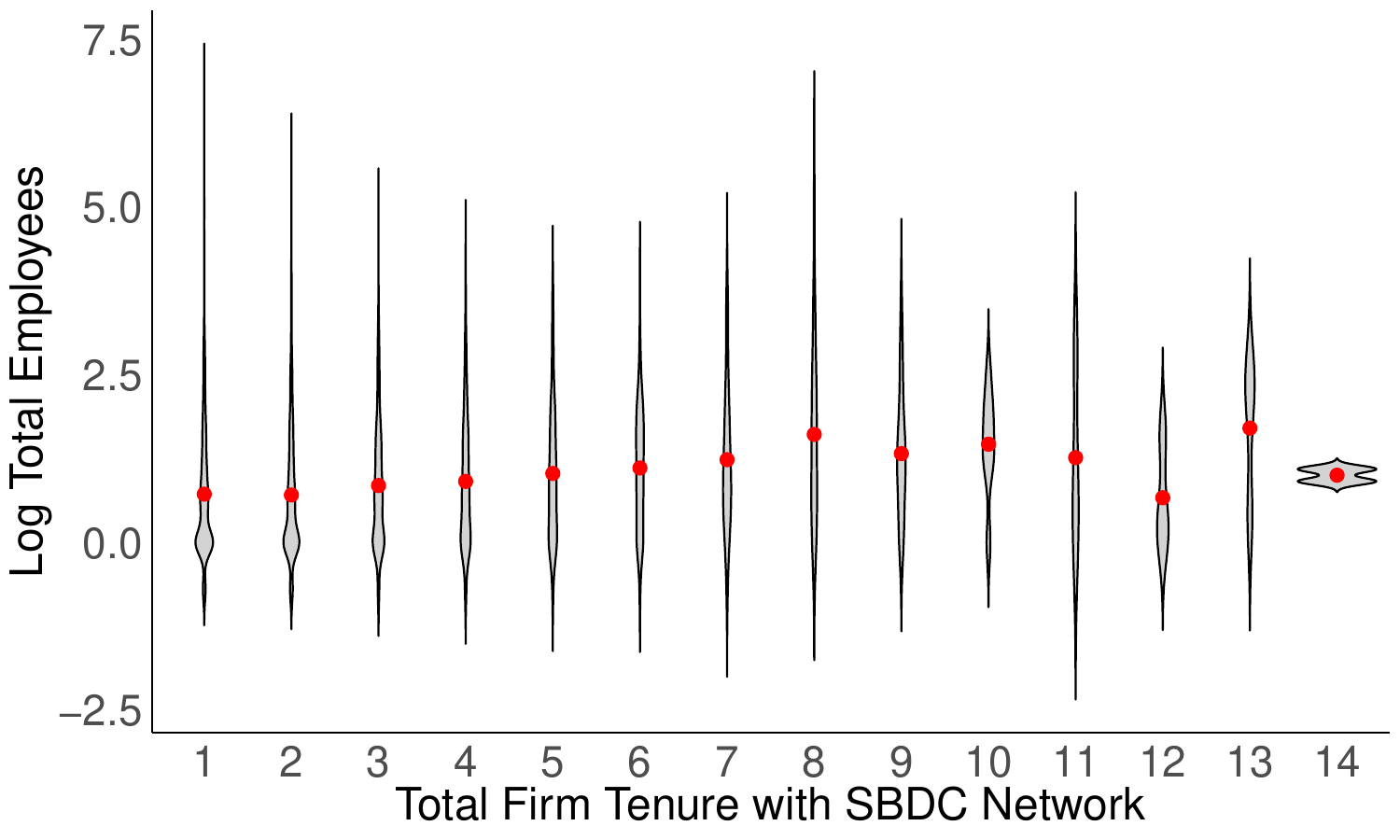}
		\end{subfigure}
		
		\begin{tablenotes}
			\footnotesize
			\item \textit{Note}: The red dots and associated shading represent the average and distribution of performance outcome values, respectively, conditional on firm tenure with the NorCal SBDC Network. 
		\end{tablenotes}
	\end{figure}
	
	\begin{figure}[h!]
		\centering
		\caption{Tenure Year of Recorded Firm Performance Outcomes by Length of Tenure in SBDC Network}
		\label{fig:firm_rev_emp_tenureyearofrecord}
		
		\begin{subfigure}[t]{0.49\textwidth}
			\centering
			\caption*{Panel A: Annual Revenue}
			\includegraphics[width=\textwidth]{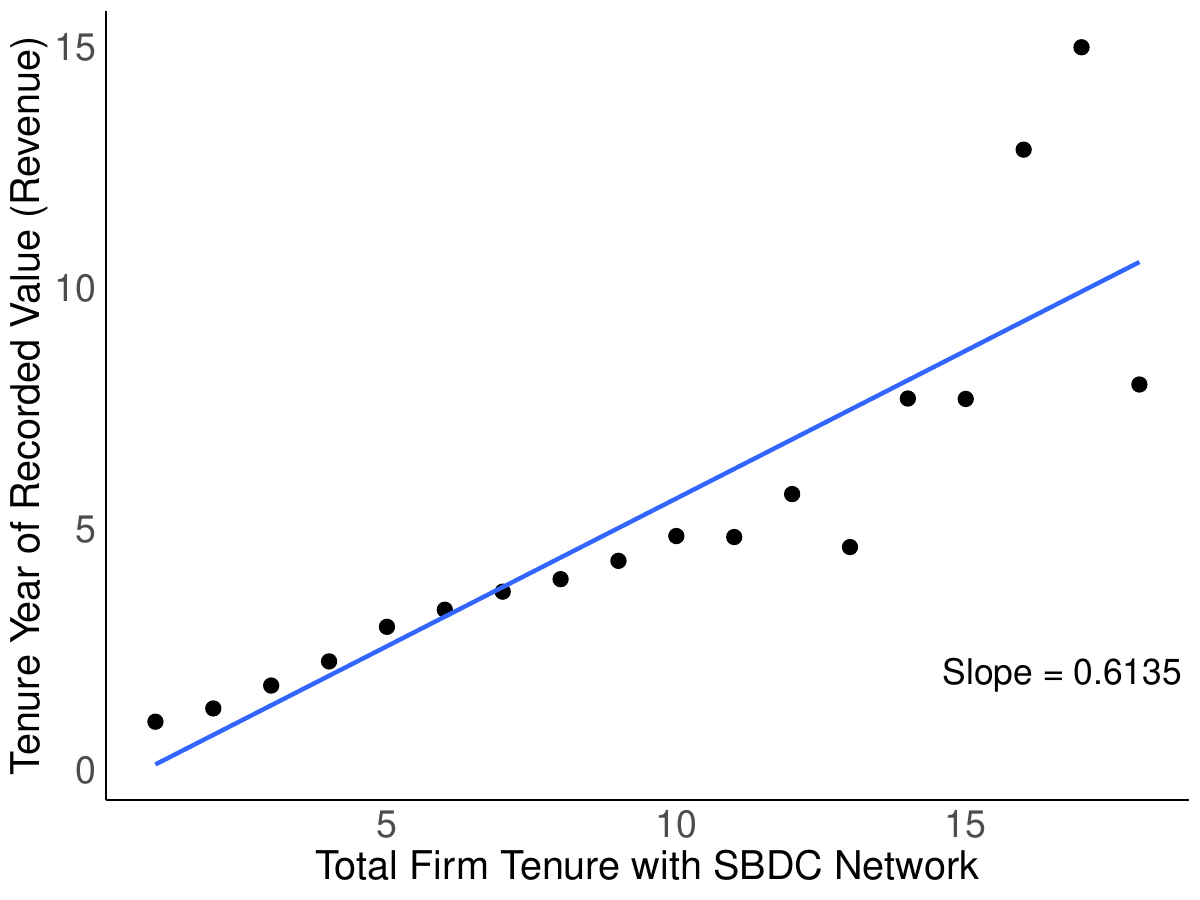}
		\end{subfigure}
		\hfill
		\begin{subfigure}[t]{0.49\textwidth}
			\centering
			\caption*{Panel B: Employment}
			\includegraphics[width=\textwidth]{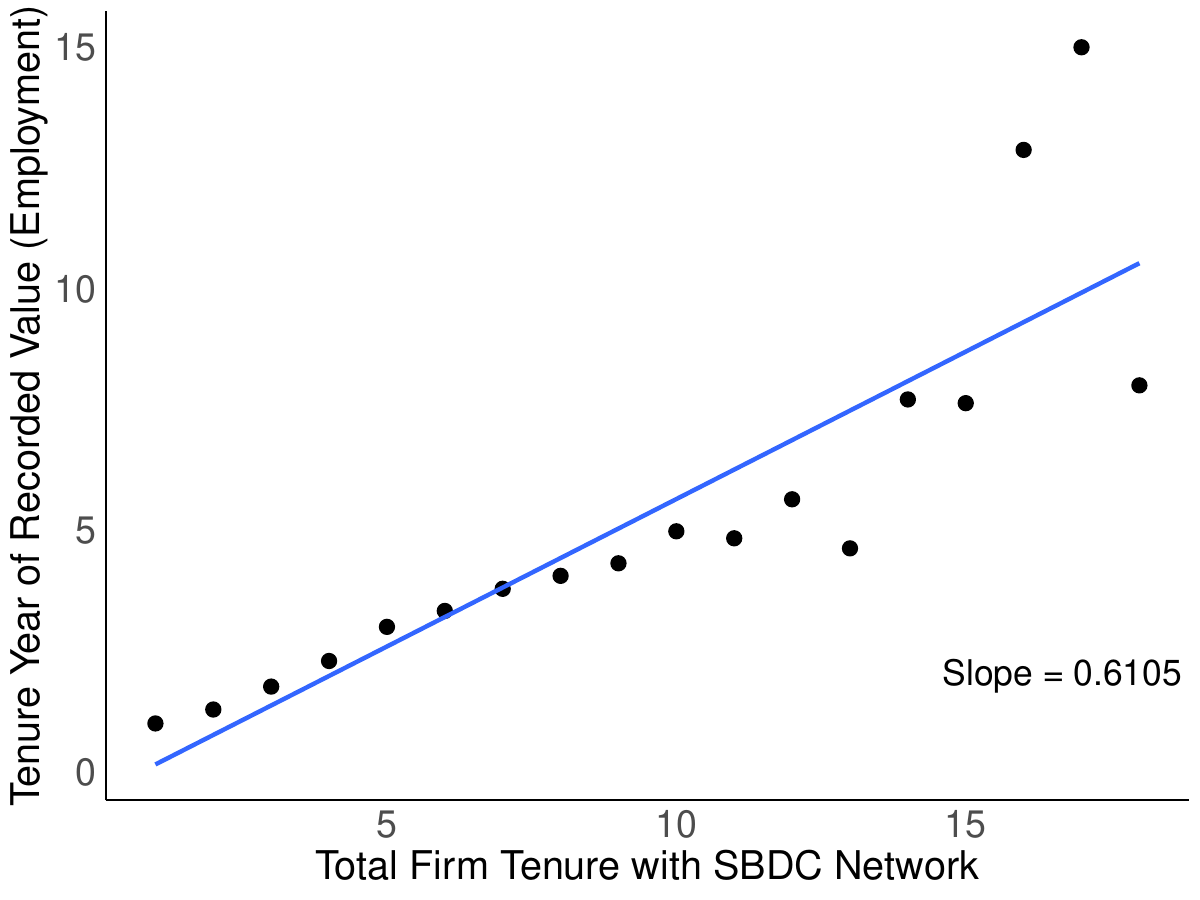}
		\end{subfigure}
		
		\begin{tablenotes}
			\footnotesize
			\item \textit{Note}: The black dots represent a binscatter of average values of the tenure year in which annual revenue (panel A) and employment (panel B) are recorded vs. the total firm tenure with the network. The blue line is a best-fit line through the binscatter points.
		\end{tablenotes}
	\end{figure}
	
	\clearpage
	
	\begin{figure}[!h]
		\centering
		\captionsetup{justification=centering}
		\caption{Active SBDC Centers}
		\label{activecenters}
		\vspace{-2mm}
		\includegraphics[width=0.65\textwidth]{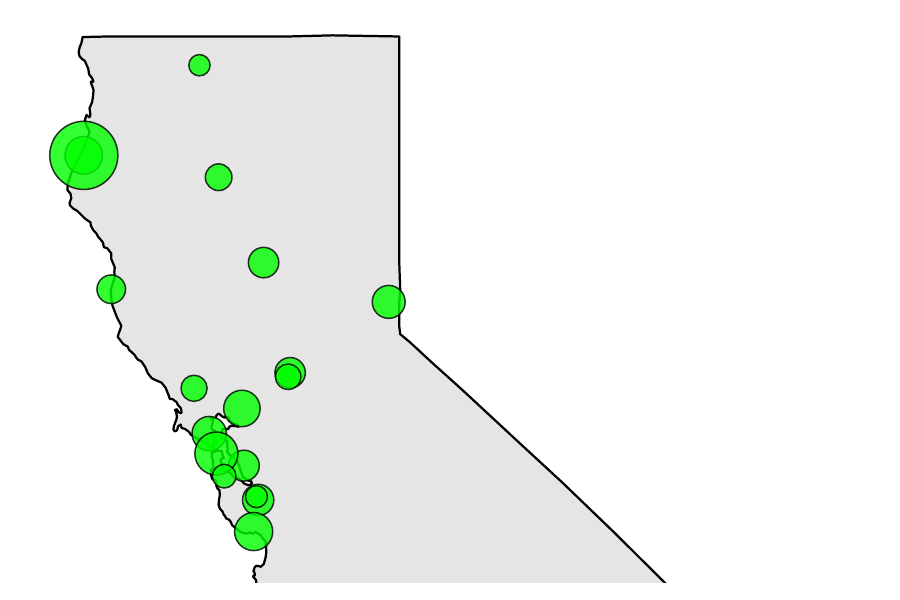}
		\vspace{-5mm}
		\begin{center}
			Active Centers: 19
		\end{center}
		\begin{tablenotes}
			\footnotesize
			\vspace{-2mm}
			\item \textit{Note}: This figure maps the number of active centers in the NorCal SBDC Network in 2023. Each center is identified by a green circle. The larger the circle, the greater number of clients that the center has served. 
		\end{tablenotes}
	\end{figure}
	
	\begin{table}[h!]
		\centering
		\caption{Moran's I Estimates Across $k$-Nearest Neighbor Specifications}
		\label{tab:moran_results}
		\small
		\setlength{\tabcolsep}{10pt}       
		\renewcommand{\arraystretch}{1.1}
		\begin{tabular}{lccccc}
			\toprule
			\textbf{$k$} & \textbf{Moran's I} & \textbf{Expected I} & \textbf{Variance} & \textbf{$p$-value} & \textbf{$N$} \\
			\midrule
			10  & 0.0154 & -0.0003 & $4.24 \times 10^{-5}$ & 0.0079        & 3{,}409 \\
			20  & 0.0349 & -0.0003 & $2.27 \times 10^{-5}$ & $8.38 \times 10^{-14}$ & 3{,}409 \\
			50  & 0.0456 & -0.0003 & $9.68 \times 10^{-6}$ & $1.77 \times 10^{-49}$ & 3{,}409 \\
			100 & 0.0414 & -0.0003 & $4.87 \times 10^{-6}$ & $8.90 \times 10^{-80}$ & 3{,}409 \\
			\bottomrule
		\end{tabular}
		\begin{flushleft}
			\footnotesize
			\textit{Note:} This table reports Moran's I statistics for firm-year changes in distance using $k$-nearest neighbor spatial weights matrices. The statistic measures the degree of spatial autocorrelation in change in distance. Expected I and the associated variance correspond to the null hypothesis of spatial randomness. $p$-values are derived from the standard normal approximation of Moran's I under this null. $N$ corresponds to the number of unique firm-year change in distance values witnessed in the sample. Positive and statistically significant values of Moran's I indicate spatial autocorrelation in change in distance among geographically proximate firms.
		\end{flushleft}
	\end{table}
	
	\clearpage
	
	\begin{figure}[h!]
		\centering
		\captionsetup{justification=centering}
		\caption{Place-Specific Shocks Correlated with Center Movement}
		\label{centermovement}
		\vspace{-3mm}
		\includegraphics[width=\textwidth]{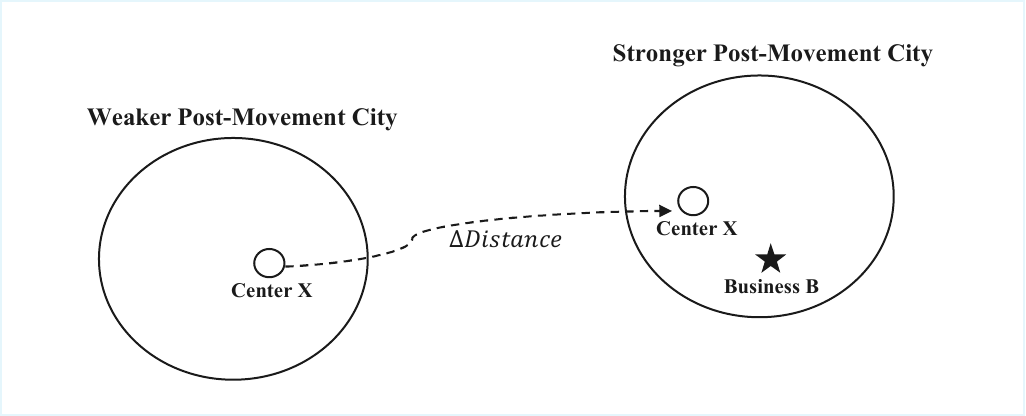} 
		\begin{tablenotes}
			\footnotesize
			\vspace{-2mm}
			\item \textit{Note}: This figure displays the movement of Center X from a city that is economically declining to one that is improving. Business B, located in the economically improving city, experiences both a decrease in distance to Center X and a stronger post-movement economic environment compared to its peers.
		\end{tablenotes}
	\end{figure}
	
	\begin{figure}[h!]
		\centering
		\caption{Differences in Average Firm Performance Prior to Center Openings (Top) and Closures (Bottom)}
		\label{fig:eventstudyinstvalidity_Appendix}
		
		\begin{subfigure}[t]{0.49\textwidth}
			\centering
			\caption*{Panel A: Annual Revenue (Openings)}
			\includegraphics[width=\textwidth]{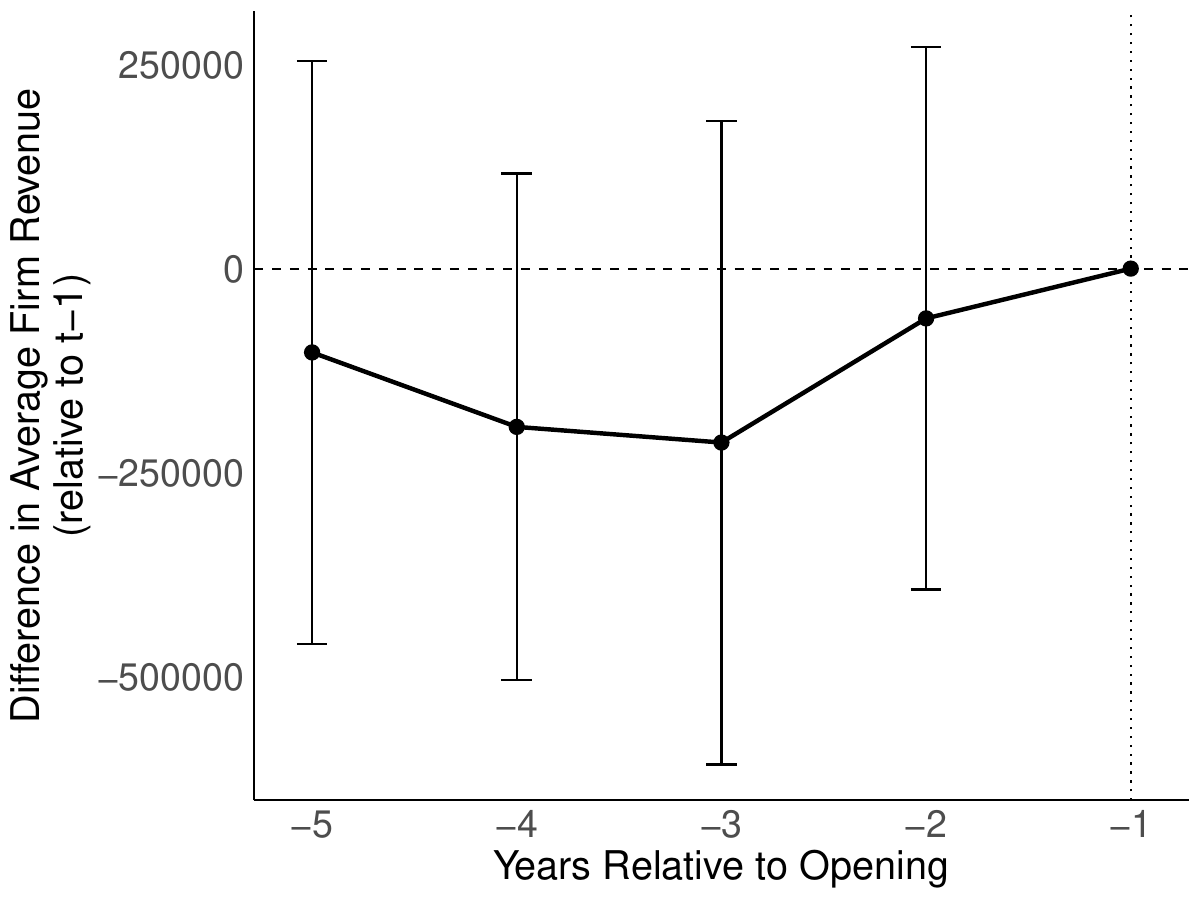}
		\end{subfigure}
		\hfill
		\begin{subfigure}[t]{0.49\textwidth}
			\centering
			\caption*{Panel B: Employment (Openings)}
			\includegraphics[width=\textwidth]{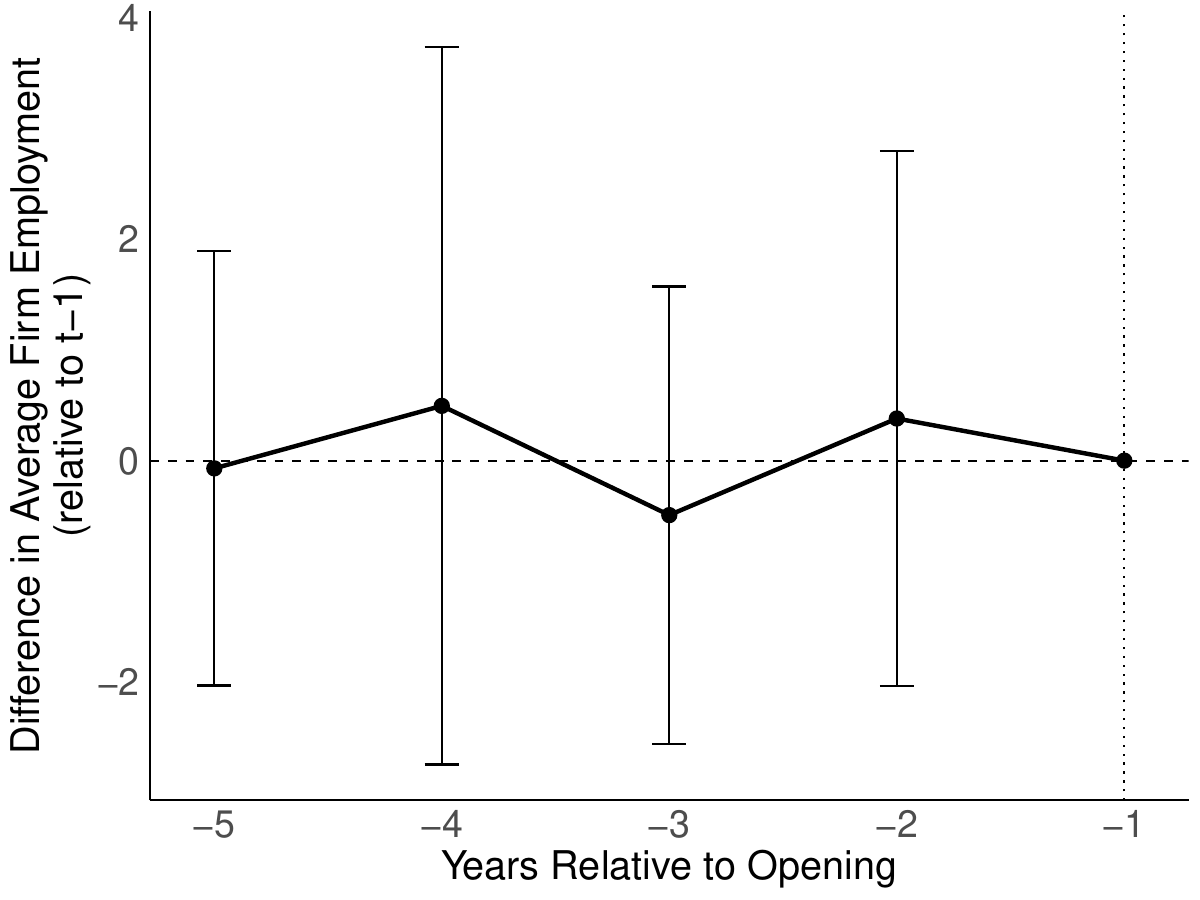}
		\end{subfigure}
		
		\vspace{3mm}
		
		\begin{subfigure}[t]{0.49\textwidth}
			\centering
			\caption*{Panel C: Annual Revenue (Closures)}
			\includegraphics[width=\textwidth]{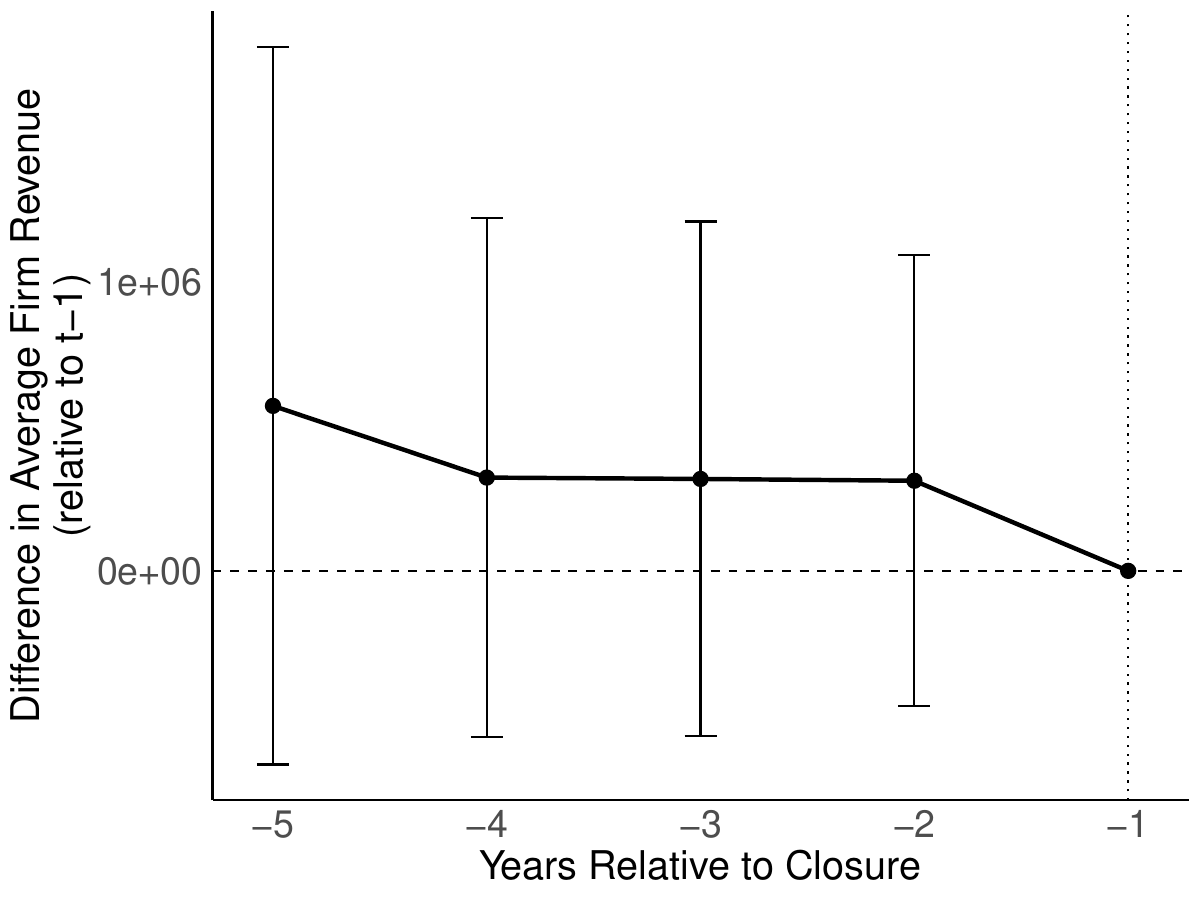}
		\end{subfigure}
		\hfill
		\begin{subfigure}[t]{0.49\textwidth}
			\centering
			\caption*{Panel D: Employment (Closures)}
			\includegraphics[width=\textwidth]{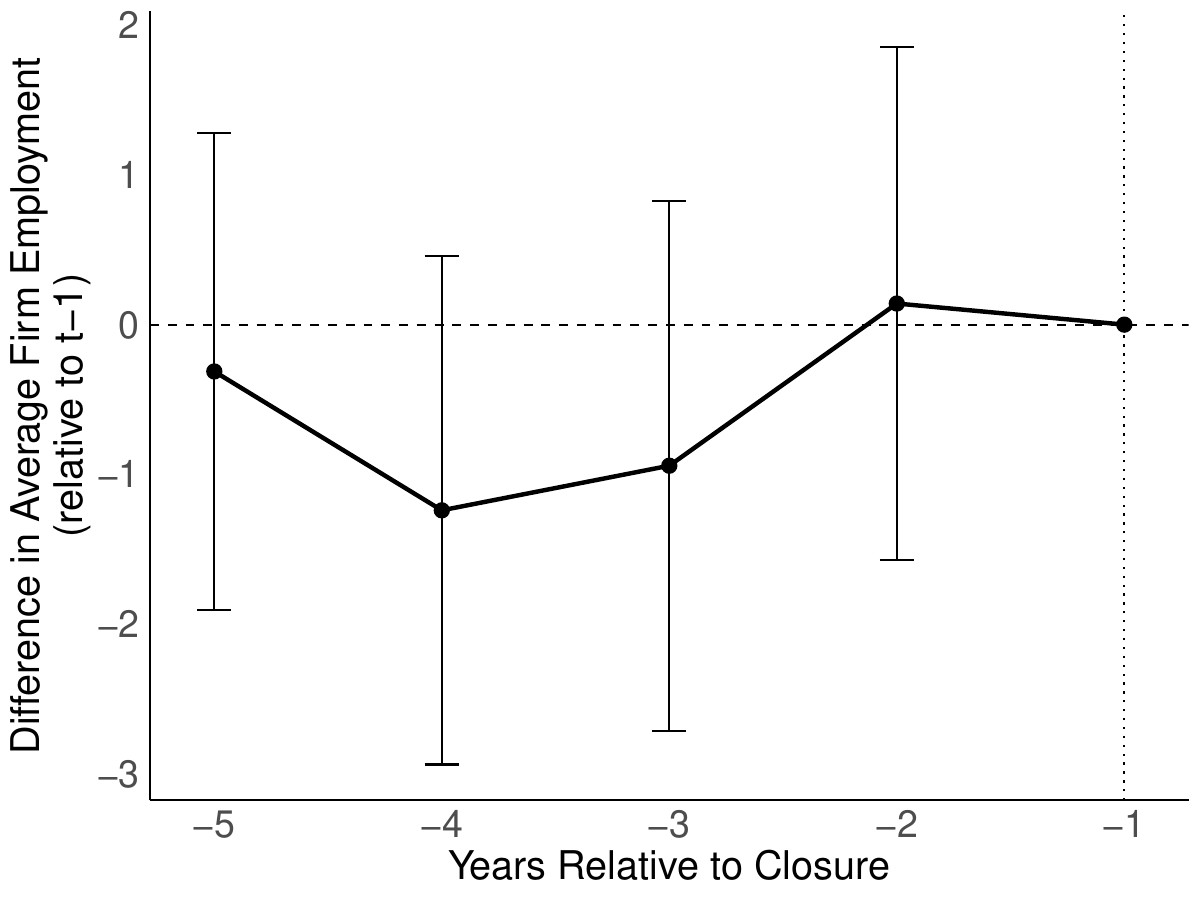}
		\end{subfigure}
		
		\begin{tablenotes}
			\footnotesize
			\item \textit{Note}: Points represent coefficient estimates from the event-study specification in equation \ref{eqn:instrumentvalidityeventstudy}. Confidence intervals are estimated using clustered-robust standard errors by center ID. 
		\end{tablenotes}
	\end{figure}

	\clearpage
	
	\begin{table}[h!] 
		\centering
		\caption{Additional First-Stage Estimation Results}
		\label{tab:iv_regression_prep}
		\footnotesize
		\begin{tabular}{l c c c c c c}
			\toprule
			& \multicolumn{6}{c}{First-Stage} \\
			\cmidrule(lr){2-7}
			& \makecell{Contact\\Time} & \makecell{Contact\\Time} & \makecell{Contact\\Time} & \makecell{Contact\\Time} & \makecell{Preparation\\Time} & \makecell{Preparation + \\ Contact\\Time} \\
			\midrule
			$\Delta$ in Distance
			& $-0.0078^{***}$ & $-0.0038^{***}$ & $-0.0058^{**}$ & $-0.0040^{**}$ & $-0.0037^{***}$ & $-0.0115^{***}$ \\
			& $(0.0020)$ & $(0.0013)$ & $(0.0024)$ & $(0.0016)$ & $(0.0009)$ & $(0.0025)$ \\
			\midrule
			Observations & $91,236$ & $91,236$ & $47,614$ & $47,614$ & $83,113$ & $91,832$ \\
			Firm FE & X &  & X &  & X & X \\
			Industry FE &  & X &  & X &  &  \\
			Center FE & X & X & X & X & X & X \\
			Year FE & X & X & X & X & X & X \\
			County-Year Controls &  &  & X & X &  &  \\
			Clustered Robust SEs & X & X & X & X & X & X \\
			R$^2$ & $0.587$ & $0.021$ & $0.571$ & $0.059$ & $0.614$ & $0.591$ \\
			F-Statistic & $15.9$ & $9.0$ & $4.5$ & $2.6$ & $16.8$ & $20.9$ \\
			\bottomrule
			\multicolumn{7}{l}{\textit{$^{*}p < 0.1$, $^{**}p < 0.05$, $^{***}p < 0.01$}} \\
		\end{tabular}%
		\begin{tablenotes}
			\footnotesize
			\vspace{2mm}
			\item \textit{Note}: This table presents the coefficients (standard errors in parentheses) of the primary first-stage estimation results (specification 1), alternate specifications with different sets of fixed-effects and inclusion of county-year economic indicator controls (specifications 2-4), and specifications where the outcome variable is preparation time (specification 5) and combined preparation and contact time (specification 6). Column headers indicate the dependent variable for each specification. Information regarding fixed effects and estimation of standard errors is provided in the bottom panel of the table. 
		\end{tablenotes}
	\end{table}
	
	\begin{table}[htbp]
		\centering
		\caption{Additional Second-Stage Estimation Results (Annual Revenue)}
		\label{tab:alt_specs_secondstage_revenue}
		\footnotesize
		\hspace*{-1cm}
		\begin{tabular}{lcccccccc}
			\toprule
			& \multicolumn{8}{c}{Second-Stage Specifications (Dependent Variable: Annual Revenue)} \\
			\cmidrule(lr){2-9}
			& \makecell{county \\ industry \\ year} 
			& \makecell{zip5 \\ industry \\ year}
			& \makecell{industry \\ year} 
			& \makecell{county \\ industry \\ year} 
			& \makecell{county \\ industry \\ year} 
			& \makecell{center \\ county \\ industry \\ year} 
			& \makecell{center \\ county \\ industry \\ year} 
			& \makecell{center \\ county \\ industry \\ year} \\
			\midrule
			Avg Predicted Contact Time
			& $0.051^{***}$ & $0.042^{***}$ & $0.043$ & $0.052^{***}$ & $0.039^{**}$ & $0.043^{**}$ & $0.043^{**}$ & $0.036^{*}$ \\
			& $(0.017)$ & $(0.010)$ & $(0.045)$ & $(0.017)$ & $(0.018)$ & $(0.017)$ & $(0.017)$ & $(0.019)$ \\
			\midrule
			Observations & $2,382$ & $3,504$ & $980$ & $2,382$ & $2,382$ & $2,393$ & $2,393$ & $2,393$ \\
			Industry Fixed Effects & X & X & X & X & X & X & X & X \\
			County Fixed Effects & X &  &  & X &  & X & X &  \\
			ZIP5 Fixed Effects &  & X &  &  &  &  &  &  \\
			Year Fixed Effects & X & X & X & X &  & X & X &  \\
			County-Year Fixed Effects &  &  &  &  & X &  &  & X \\
			Center Fixed Effects &  &  &  &  &  & X & X & X \\
			County-Year Controls & X &  &  &  &  & X &  &  \\
			Bootstrapped SEs & X & X & X & X & X & X & X & X \\
			R$^2$ & $0.308$ & $0.362$ & $0.475$ & $0.308$ & $0.403$ & $0.328$ & $0.327$ & $0.407$ \\
			\bottomrule
			\multicolumn{9}{l}{\textit{$^{*}p < 0.1$, $^{**}p < 0.05$, $^{***}p < 0.01$}} \\
		\end{tabular}
	\end{table}
	
	\begin{table}[htbp]
		\centering
		\caption{Additional Second-Stage Estimation Results (Employment)}
		\label{tab:alt_specs_secondstage_employment}
		\footnotesize
		\hspace*{-1cm}
		\begin{tabular}{lcccccccc}
			\toprule
			& \multicolumn{8}{c}{Second-Stage Specifications (Dependent Variable: Employment)} \\
			\cmidrule(lr){2-9}
			& \makecell{county \\ industry \\ year} 
			& \makecell{zip5 \\ industry \\ year}
			& \makecell{industry \\ year} 
			& \makecell{county \\ industry \\ year} 
			& \makecell{county \\ industry \\ year} 
			& \makecell{center \\ county \\ industry \\ year} 
			& \makecell{center \\ county \\ industry \\ year} 
			& \makecell{center \\ county \\ industry \\ year} \\
			\midrule
			Avg Predicted Contact Time
			& $0.017^{***}$ & $0.017^{***}$ & $0.029^{*}$ & $0.017^{***}$ & $0.016^{**}$ & $0.018^{***}$ & $0.018^{***}$ & $0.016^{**}$ \\
			& $(0.006)$ & $(0.004)$ & $(0.015)$ & $(0.006)$ & $(0.007)$ & $(0.007)$ & $(0.007)$ & $(0.007)$ \\
			\midrule
			Observations & $2,923$ & $4,515$ & $1,115$ & $2,923$ & $2,923$ & $2,942$ & $2,942$ & $2,942$ \\
			Industry Fixed Effects & X & X & X & X & X & X & X & X \\
			County Fixed Effects & X &  &  & X &  & X & X &  \\
			ZIP5 Fixed Effects &  & X &  &  &  &  &  &  \\
			Year Fixed Effects & X & X & X & X &  & X & X &  \\
			County-Year Fixed Effects &  &  &  &  & X &  &  & X \\
			Center Fixed Effects &  &  &  &  &  & X & X & X \\
			County-Year Controls & X &  &  &  &  & X &  &  \\
			Bootstrapped SEs & X & X & X & X & X & X & X & X \\
			R$^2$ & $0.312$ & $0.359$ & $0.476$ & $0.311$ & $0.375$ & $0.318$ & $0.318$ & $0.383$ \\
			\bottomrule
			\multicolumn{9}{l}{\textit{$^{*}p < 0.1$, $^{**}p < 0.05$, $^{***}p < 0.01$}} \\
		\end{tabular}
	\end{table}
	
	\clearpage
	
	\begin{table}[t]
		\centering
		\caption{Summary Statistics of \underline{Treated} Client Firms in the NorCal SBDC Network}
		\label{tab:table2_summary_stats_treated}
		\hspace*{-0.75cm}
		\begin{threeparttable}
			\begin{tabular}{lllllll}
				\toprule
				Variable & \makecell{Number of \\ Observations} & Mean & SD & Median & \makecell{5\textsuperscript{th}\\Percentile} & \makecell{95\textsuperscript{th}\\Percentile}\\
				\midrule
				\# of Sessions & 9,813 & 5.15 & 5.94 & 3 & 1 & 16 \\
				Contact Time (hours) & 9,813 & 5.14 & 7.40 & 2.5 & 0.25 & 18.5 \\
				Preparation Time (hours) & 9,813 & 1.94 & 3.21 & 1.0 & 0 & 7 \\
				Distance (miles) & 9,813 & 15.9 & 28.5 & 7.22 & 0.78 & 56.5 \\
				$\Delta$ in Distance (miles)$^*$ & 3,409 & 0.33 & 43.1 & -0.17 & -47.7 & 43.0 \\
				\% Urban & 9,642 & 88.8 & 24.5 & 100.0 & 0 & 100 \\
				Firm Tenure w/ Network (years) & 2,865 & 4.85 & 3.27 & 4 & 2 & 12 \\
				Firm Age (years) & 5,999 & 10.40 & 13.10 & 6 & 0 & 36 \\
				Annual Revenue (\$1000s)$^{+}$ & 1,247 & 548.2 & 3,813.2 & 59.8 & 0.0 & 1,703.8 \\
				Employees$^\dagger$ & 1,200 & 3.98 & 8.45 & 2.0 & 0 & 14 \\
				\bottomrule
			\end{tabular}
			\begin{tablenotes}[flushleft]
				\footnotesize
				\singlespacing
				\vspace{-4mm}
				\item \textit{Note}: These statistics are restricted to treated businesses, defined as firms that experienced a non-zero change in distance resulting from a center closure, opening, or location change within the NorCal SBDC Network. The sample omits all businesses that engaged with the NorCal SBDC lead center. The lead center is often responsible for on-boarding businesses into the network, thus logging contact time not associated with the primary business advisory services.\\
				$^*$The number of observations for the ``$\Delta$ in Distance" variable consists of every unique change in business-center pairing resulting from center openings, closures, or re-locations. Comparing this value to the number of treated businesses indicates that some firms experienced multiple changes in center pairing over time. \\
				$^{+}$Annual revenue is reported in 2023 USD terms and displayed in thousands of dollars. \\
				$^\dagger$Full-time and part-time employees are documented separately. We count each part-time employee as 0.5 full-time employees.
			\end{tablenotes}
		\end{threeparttable}
	\end{table}
	
	\clearpage
	
	\begin{figure}[h!]
		\centering
		\caption{Distribution of Firm Urbanicity}
		\label{fig:urbanicity_hist}
		
		\begin{subfigure}[t]{0.49\textwidth}
			\centering
			\caption*{Panel A: All Firms}
			\includegraphics[width=\textwidth]{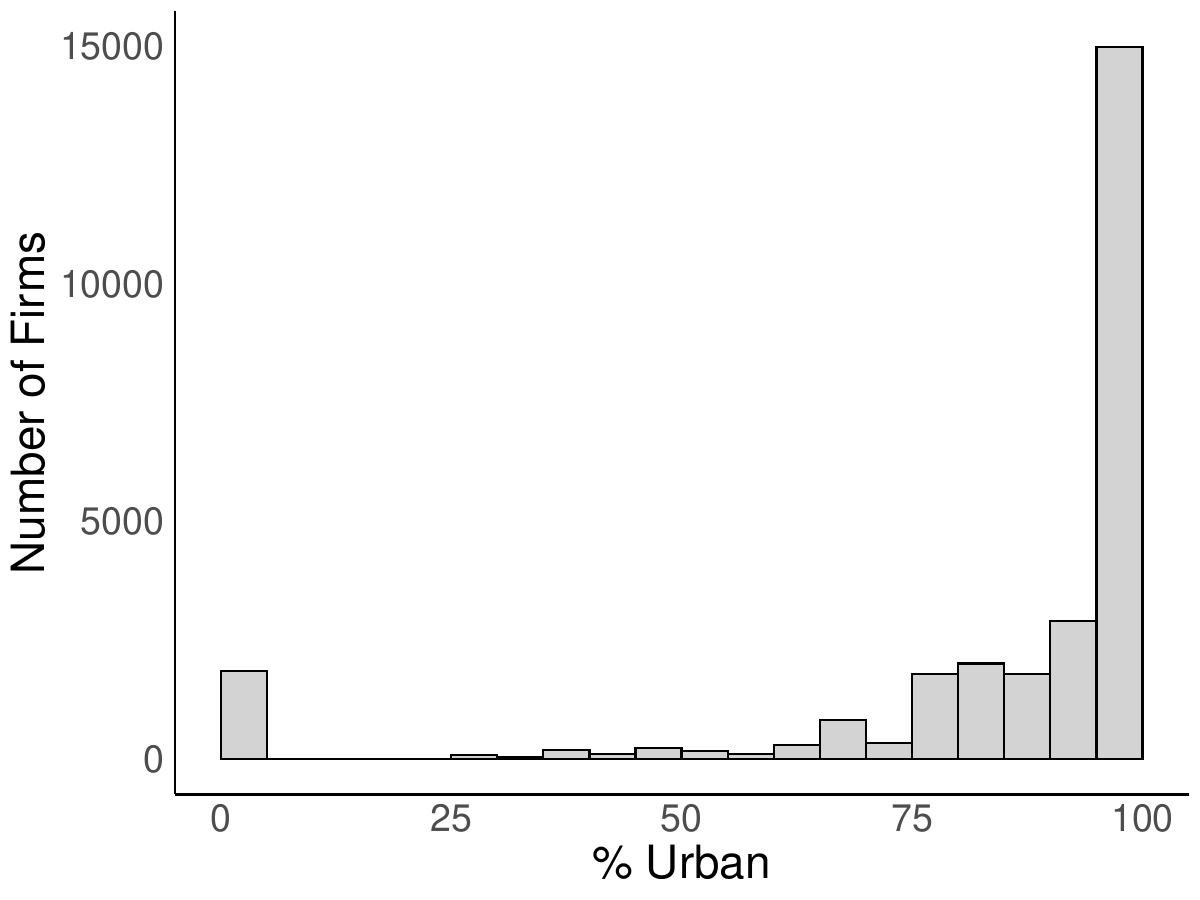}
		\end{subfigure}
		\hfill
		\begin{subfigure}[t]{0.49\textwidth}
			\centering
			\caption*{Panel B: Treated Firms}
			\includegraphics[width=\textwidth]{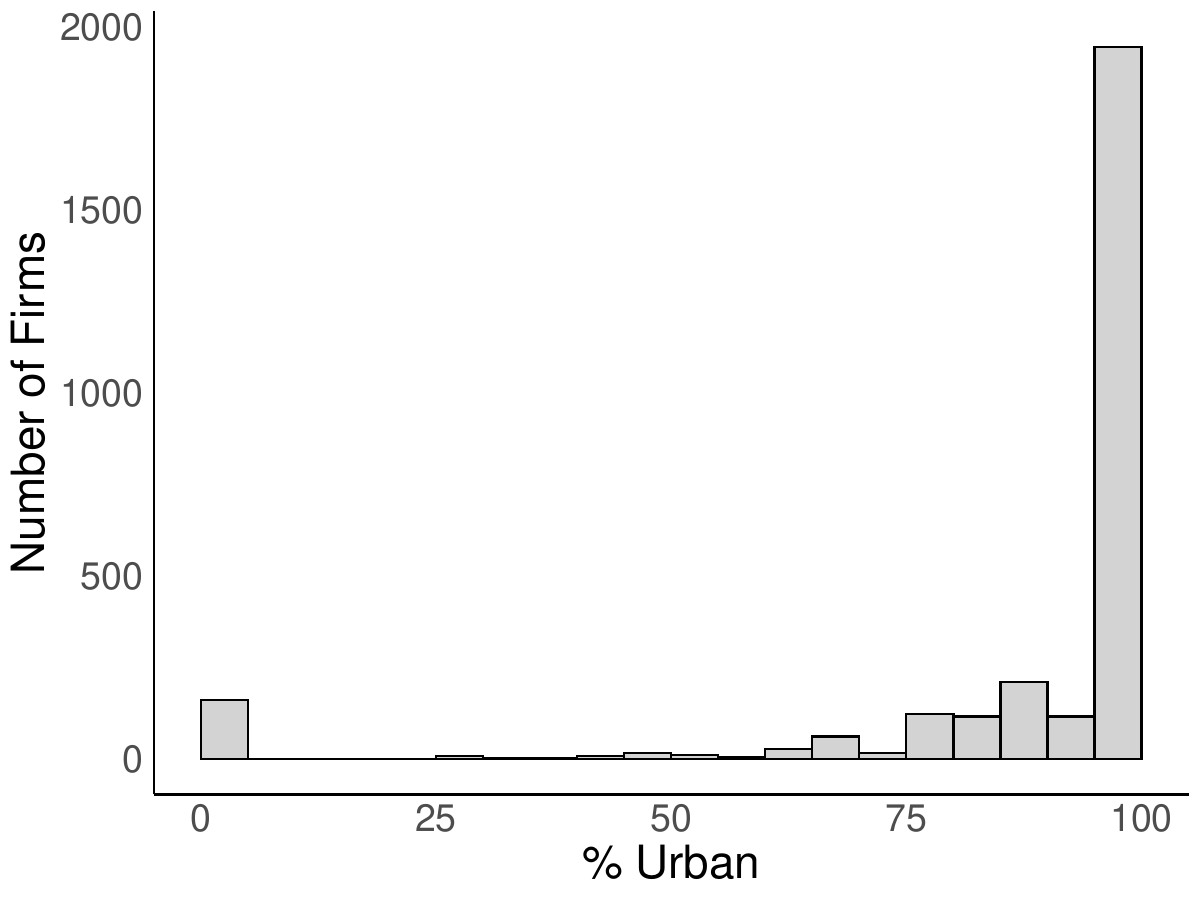}
		\end{subfigure}
		
	\end{figure}

	\begin{table}[h!] 
		\centering
		\caption{First-Stage Estimation Results with Urbanicity Heterogeneity}
		\label{tab:iv_regression_percurban}
		\small
		\begin{tabular}{l c c c}
			\toprule
			& \multicolumn{3}{c}{Dependent Variable} \\
			\cmidrule(lr){2-4}
			& \makecell{Contact\\Time} 
			& \makecell{Preparation\\Time} 
			& \makecell{Preparation + \\ Contact\\Time} \\
			\midrule
			$\Delta$ in Distance
			& $-0.0108^{***}$ & $-0.0012$ & $-0.0121^{***}$ \\
			& $(0.0027)$ & $(0.0009)$ & $(0.0031)$ \\
			$\Delta$ in Distance $\times$ \% Urban
			& $0.00006^{*}$ & $-0.00002$ & $0.00005$ \\
			& $(0.00003)$ & $(0.00002)$ & $(0.00004)$ \\
			\midrule
			Observations & $46,226$ & $42,195$ & $46,726$ \\
			Firm FE & X & X & X \\
			Industry FE &  &  &  \\
			Center FE & X & X & X \\
			Year FE & X & X & X \\
			County-Year Controls &  &  &  \\
			Clustered Robust SEs & X & X & X \\
			R$^2$ & $0.574$ & $0.572$ & $0.566$ \\
			F-Statistic & $5.45$ & $0.51$ & $4.98$ \\
			\bottomrule
			\multicolumn{4}{l}{\textit{$^{*}p < 0.1$, $^{**}p < 0.05$, $^{***}p < 0.01$}} \\
		\end{tabular}%
		\begin{tablenotes}
			\footnotesize
			\vspace{2mm}
			\item \textit{Note}: This table presents first-stage estimates including an interaction between change in distance and local urbanicity (\% Urban). All specifications include firm, center, and year fixed effects. Standard errors are clustered at the center level.
		\end{tablenotes}
	\end{table}
	
	\begin{table}[h!]
		\centering
		\caption{First-Stage Estimates by Change-in-Distance Bin}
		\label{tab:firststage_heterogeneity}
		\footnotesize
		\setlength{\tabcolsep}{1.5pt}       
		\renewcommand{\arraystretch}{1}   
		\begin{tabular}{lccccccccc}
			\toprule
			\textbf{Bin} & \textbf{Coefficient} & \textbf{95\% CI} & \makecell{\textbf{Cumulative} \\ \textbf{Effect}} & \makecell{\textbf{Cumulative} \\ \textbf{95\% CI}} & \textbf{$R^2$} & \textbf{$F$-stat} & \textbf{$N$} & 
			\makecell{\textbf{\# Treated}\\\textbf{Clients}} & 
			\makecell{\textbf{\# Total}\\\textbf{Clients}} \\
			\midrule
			All          & -0.0078 & [-0.0117, -0.0040] & ---      & ---         & 0.588 & 15.93 & 91{,}236 & 2{,}865 & 58{,}549 \\
			0--5         &  0.0186 & [-0.1310,  0.1683] &  0.0466  & [-0.3275,  0.4207]  & 0.593 & 0.06  & 87{,}436 & 1{,}778 & 57{,}462 \\
			5--20        & -0.0038 & [-0.0508,  0.0432] & -0.0474  & [-0.6351,  0.5402]  & 0.602 & 0.03  & 83{,}044 &   464   & 56{,}148 \\
			20--50       & -0.0373 & [-0.0666, -0.0080] & -1.3046  & [-2.3301, -0.2790]  & 0.601 & 6.22  & 82{,}780 &   391   & 56{,}075 \\
			50--100      & -0.0221 & [-0.0344, -0.0099] & -1.6610  & [-2.5821, -0.7399]  & 0.602 & 12.49 & 82{,}084 &   154   & 55{,}838 \\
			$\geq$ 100   & -0.0069 & [-0.0114, -0.0025] & -1.5479  & [-2.5402, -0.5557]  & 0.603 & 9.35  & 81{,}788 &    78   & 55{,}762 \\
			\bottomrule
		\end{tabular}
		\begin{flushleft}
			\footnotesize
			\textit{Notes:} Estimates are generated from the first-stage specification for each change-in-distance bin grouping treated businesses. Confidence intervals are estimated using clustered-robust standard errors. The column “\# Treated Clients” counts the number of businesses with at least one non-zero change in distance in each bin; treated clients with multiple change-in-distance instances are assigned to the bin corresponding to their \underline{largest} (in magnitude) change-in-distance. There are 55,684 clients that never change distance and are included in every specification. Cumulative effects scale the marginal effect by the midpoint of each bin; for the $\geq$100 bin, the midpoint is defined as the average of 100 and the maximum observed absolute change in distance.
		\end{flushleft}
	\end{table}

	\clearpage
	
	\begin{table}[htbp]
		\centering
		\caption{Second-Stage Estimates by First-Stage Change-in-Distance Bin Specifications (Annual Revenue)}
		\label{tab:heterogeneity_secondstage_revenue}
		\begin{tabular}{lcccc}
			\toprule
			\textbf{Bin} & \textbf{Coefficient} & \textbf{95\% CI} & \textbf{$R^2$} & \textbf{$N$} \\
			\midrule
			All          & 0.0505 & [0.0213, 0.0851] & 0.308 & 2{,}382 \\
			0--5         & 0.0520 & [0.0218, 0.0875] & 0.313 & 2{,}286 \\
			5--20        & 0.0480 & [0.0120, 0.0866] & 0.317 & 2{,}092 \\
			20--50       & 0.0421 & [0.0073, 0.0828] & 0.317 & 2{,}053 \\
			50--100      & 0.0482 & [0.0104, 0.0891] & 0.320 & 2{,}042 \\
			$\geq$ 100   & 0.0468 & [0.0098, 0.0881] & 0.319 & 2{,}041 \\
			\bottomrule
		\end{tabular}
		\begin{flushleft}
			\footnotesize
			\textit{Notes:} Estimates are generated for the second-stage specification examining annual revenue, using predicted contact time estimates generated from the first-stage specifications in Table \ref{tab:firststage_heterogeneity} for each corresponding change-in-distance bin. Confidence intervals are estimated using bootstrapped standard errors. 
		\end{flushleft}
	\end{table}

	\begin{table}[htbp]
		\centering
		\caption{Second-Stage Estimates by First-Stage Change-in-Distance Bin Specifications (Total Employment)}
		\label{tab:heterogeneity_secondstage_employment}
		\begin{tabular}{lcccc}
			\toprule
			\textbf{Bin} & \textbf{Coefficient} & \textbf{95\% CI} & \textbf{$R^2$} & \textbf{$N$} \\
			\midrule
			All          & 0.0171 & [0.0042, 0.0293] & 0.312 & 2{,}923 \\
			0--5         & 0.0177 & [0.0042, 0.0299] & 0.309 & 2{,}826 \\
			5--20        & 0.0151 & [0.0008, 0.0266] & 0.319 & 2{,}624 \\
			20--50       & 0.0157 & [0.0019, 0.0281] & 0.318 & 2{,}582 \\
			50--100      & 0.0155 & [0.0012, 0.0265] & 0.318 & 2{,}573 \\
			$\geq$ 100   & 0.0161 & [0.0028, 0.0275] & 0.318 & 2{,}571 \\
			\bottomrule
		\end{tabular}
		\begin{flushleft}
			\footnotesize
			\textit{Notes:} Estimates are generated for the second-stage specification examining total employment, using predicted contact time estimates from the first-stage specifications in Table \ref{tab:firststage_heterogeneity} for each corresponding $\Delta$ in distance bin. Confidence intervals are estimated using bootstrapped standard errors. 
		\end{flushleft}
	\end{table}

	\end{document}